\documentclass[11pt]{article}

\usepackage{indentfirst}
\usepackage{amsmath}
\usepackage{amssymb}
\usepackage{amsfonts}
\usepackage{mathtools}
\usepackage{geometry}
\usepackage{hyperref}
\usepackage[dvipsnames]{xcolor}
\hypersetup{bookmarksnumbered, colorlinks=true}
\usepackage[pdftex]{graphicx}
\usepackage{subfigure}
\usepackage{float}
\usepackage{cite}
\usepackage{tikz}
\usetikzlibrary{matrix, calc}

\hypersetup
{
    colorlinks = true,
    urlcolor = blue,
    linkcolor = red,
    citecolor = green
}
\graphicspath{{./Figs/}}

\begin{document}
\begin{titlepage}
\begin{center}
    {\Large\bf Bound on Lyapunov exponent for a charged particle \\ in Kerr-Sen-AdS Black Hole}
    \\ \vspace{1.5cm}
    {$\mbox{Hocheol Lee}$}\footnote{\it email: insaying@dongguk.edu}, \quad
    {$\mbox{Bogeun Gwak}$}\footnote{\it email: rasenis@dgu.ac.kr}
    \\ \vspace{0.5cm}
    {\small \it Department of Physics, Dongguk University, Seoul 04620, Korea}
    \\ \vspace{1.0cm}
\end{center}

\begin{center}
\begin{abstract}
    We investigate the upper bound of the Lyapunov exponent for a charged particle in the Gibbons--Maeda--Garfinkle--Horowitz--Strominger (GMGHS)--AdS and Kerr--Sen--AdS black hole backgrounds, which originate from the low-energy effective actions of heterotic string theory and gauged supergravity. We analyze the Lyapunov exponent near the unstable orbit to examine possible violations of the bound. Our results indicate that the bound is sensitive to the signs and magnitudes of the charges, the angular momentum of the particle, the black hole spin, and the negative cosmological constant. The violations are pronounced in the extremal or near-extremal regime. Numerical analysis supports the analytical predictions and highlights the interplay between the string-inspired black hole and the charged particle.
\end{abstract}
\end{center}
\end{titlepage}

\setcounter{page}{1}
\section{Introduction} \label{intro}
    Understanding the properties of black holes is pivotal in probing fundamental aspects of gravity and quantum theories. Typically, black holes govern geodesic motion, determine the causal structure of spacetime, and lead to phenomena such as frame dragging, gravitational redshift, and strong lensing---exemplified by the black hole shadow \cite{EventHorizonTelescope:2019dse}. These classical aspects also encompass the role of black holes as major sources of gravitational waves \cite{LIGOScientific:2016aoc} generated from their dynamical evolution during mergers. Direct gravitational wave detection has confirmed general relativity and opened a new observational window onto the strong-field regime of gravity. Moreover, black holes are governed by black hole mechanics, following principles analogous to those of thermodynamics and supporting their interpretation as thermodynamic systems. From a quantum perspective, Hawking radiation \cite{Hawking:1975vcx} reveals that black holes emit thermal radiation due to quantum effects near the event horizon, indicating that they cannot be considered purely classical objects. This thermal behavior is consistent with the Bekenstein--Hawking entropy \cite{Bekenstein:1973ur}, which assigns an entropy to a black hole that is proportional to the area of its event horizon, further reinforcing its thermodynamic nature. This insight raises fundamental questions concerning unitarity and information loss, emphasizing the need for a consistent quantum theory of gravity to describe black hole dynamics.

    The connection between gravity and quantum theory is exemplified by the Anti-de Sitter (AdS)\slash conformal field theory (CFT) correspondence, proposed by Maldacena \cite{Maldacena:1997re}, which posits a duality between a gravitational theory in asymptotically AdS spacetime and a CFT defined on the boundary of the AdS spacetime. Within this framework, black holes in AdS spacetime correspond to thermal states in the boundary conformal field theory, enabling strongly coupled quantum systems to be studied through classical gravitational dynamics. The AdS/CFT correspondence has been extended to other geometric and physical frameworks. The de Sitter (dS)/CFT correspondence \cite{Strominger:2001pn, Strominger:2001gp, McInnes:2001zw, Anninos:2011ui} proposes a duality between quantum gravity in dS spacetime and a Euclidean conformal field theory defined at future infinity of the dS geometry, providing a potential framework to describe cosmological evolution via duality. The Kerr/CFT correspondence \cite{Guica:2008mu, Matsuo:2009sj, Castro:2010fd, Compere:2012jk} proposes that the near-horizon geometry of extremal Kerr black holes is dual to a geometry described by a two-dimensional chiral CFT, offering insights into the microscopic origin of black hole entropy. Similarly, the Reissner--Nordström (RN)/CFT correspondence \cite{Garousi:2009zx} proposes a duality between the near-horizon region of extremal Reissner--Nordstr\"om black holes and a region defined by CFT. These dualities have been applied to diverse physical systems. In AdS/quantum chromodynamics (QCD) \cite{Sakai:2004cn, Erlich:2005qh, Karch:2006pv}, phenomena such as confinement and chiral symmetry breaking are modeled by constructing effective gravitational theories grounded in string theory. Similarly, in AdS/condensed matter theory (CMT) \cite{Gubser:2008px, Hartnoll:2008vx, Hartnoll:2008kx}, the duality is used to study strongly correlated condensed matter systems such as high-temperature superconductors and quantum phase transitions.
    
    Moreover, the AdS/CFT correspondence has been applied to the study of quantum information theory. The study by Ryu and Takayanagi \cite{Ryu:2006bv}, along with its covariant generalization \cite{Hubeny:2007xt}, represents a major development and provides a geometric interpretation of entanglement entropy in quantum field theories through the area of minimal surfaces in the bulk geometry. This approach has been extended to other quantum information concepts, such as holographic complexity \cite{Sekino:2008he}. Additionally, the study of quantum chaos has benefited considerably from this duality, leading to the concept of fast scramblers, which are systems that thermalize information at the maximal rate permitted by quantum mechanics; black holes are the fastest scramblers \cite{Sekino:2008he}. Furthermore, investigations have established connections to the Sachdev--Ye--Kitaev model \cite{Maldacena:2016hyu}, a solvable quantum system with maximal chaos that captures key features of the duality. Among these developments, the Lyapunov exponent has garnered attention. Maldacena, Shenker, and Stanford \cite{Maldacena:2015waa} derived an upper bound on the Lyapunov exponent $\lambda$, given by $\lambda \leq 2\pi T$, where $T$ denotes the temperature of the system.

    The universal upper bound on the Lyapunov exponent in quantum chaos has sparked renewed interest in the classical Lyapunov exponent. Hashimoto and Tanahashi \cite{Hashimoto:2016dfz} extended the bound on the Lyapunov exponent---originally formulated for quantum systems with temperature $T$---to Hawking temperature $T_\mathrm{H}$. They revealed that the bound $\lambda \leq 2 \pi T_\mathrm{H}$ holds for a particle orbiting a static, spherically symmetric black hole in the presence of a Maxwell field or a massless scalar field. Extending the bound to black holes has prompted extensive investigations of Lyapunov exponents for particles in various gravitational backgrounds, including static and spherically symmetric black holes \cite{Hashimoto:2022kfv, Gao:2022ybw, Guo:2022kio, Jeong:2023hom, Das:2024iuf, Gallo:2024wju, Gogoi:2024akv, Lei:2024qpu, Shukla:2024tkw}, rotating black holes \cite{Kan:2021blg, Yu:2022tlr, Prihadi:2023tvr, Prihadi:2023qmk, Giataganas:2024hil}, torus-like black holes \cite{Yin:2022mjv}, Einstein--Euler--Heisenberg black holes \cite{Chen:2022tbb}, Einstein--Gauss--Bonnet black holes \cite{Xie:2023tjc}, Einstein--Maxwell--Chern--Simons black holes \cite{Yu:2023spr}, Einstein--Maxwell--Dilaton black holes \cite{Kumara:2024obd}, Taub--NUT black holes \cite{Chen:2023wph}, $f(R)$ black holes \cite{Das:2024iuf}, $R$-charged black holes \cite{Awal:2025irl}, acoustic black holes \cite{Singh:2024qfw}, AdS/dS black holes \cite{Gwak:2022xje, Song:2022lhf, Park:2023lfc, R:2025gok}, higher-dimensional AdS black holes \cite{Han:2023ckr}, and black branes \cite{Lei:2023jqv, Dutta:2023yhx, Dutta:2024rta}.

    In investigating Lyapunov exponents, we focus on the Kerr--Sen black hole---a charged, rotating solution obtained from the low-energy effective action of heterotic string theory. This solution incorporates the dilaton and antisymmetric two-form field. The additional fields modify the spacetime geometry, providing a framework for examining the influence of string-inspired theory on classical black holes. The Kerr--Sen solution builds on an earlier study by Gibbons and Maeda \cite{Gibbons:1987ps}, who obtained a static, spherically symmetric, electrically and/or magnetically charged black hole solution in the context of the Einstein--Maxwell--Dilaton theory, incorporating a 2-form and a $( D - 2 )$-form field. Garfinkle, Horowitz, and Strominger \cite{Garfinkle:1990qj} derived a similar solution from the low-energy effective action of heterotic string theory, focusing on magnetic charge and omitting the antisymmetric three-form field. Hassan and Sen \cite{Hassan:1991mq}, through an alternative approach, extended the solution by generating electric and magnetic charges from the magnetically charged solution via the Hassan--Sen transformation. Based on previous studies, Sen \cite{Sen:1992ua} used the Hassan--Sen transformation on the Kerr metric to construct a rotating and charged black hole solution known as the Kerr--Sen black hole. Extending the Kerr--Sen black hole to an asymptotically AdS spacetime enables the incorporation of ideas motivated by the AdS/CFT correspondence. This extension can be achieved within the framework of gauged supergravity \cite{Chong:2004na, Chow:2013gba}, where a negative cosmological constant naturally emerges from gauge couplings. The setup enables a systematic analysis of the influence of string-inspired theories and AdS curvature on chaotic dynamics and the bound on chaos.

    In this work, we investigate the Lyapunov exponent of a charged particle in the Gibbons--Maeda--Garfinkle--Horowitz--Strominger (GMGHS)--AdS and Kerr--Sen--AdS backgrounds, focusing on the local maximum of the effective potential associated with the unstable orbit. Hashimoto and Tanahashi \cite{Hashimoto:2016dfz} demonstrated that the bound on the Lyapunov exponent holds when the angular momentum of the particle is neglected. However, subsequent studies of Kerr--Newman black holes \cite{Kan:2021blg} revealed that the angular momenta of the black hole and particle are crucial in violating the bound. This investigation was further extended to Kerr--Newman--AdS \cite{Gwak:2022xje} and Kerr--Newman--dS black holes \cite{Park:2023lfc}. Following on these findings, our analysis extends to string-inspired theories on classical black holes employing Kerr--Sen--AdS black holes. We derive the effective potential governing the radial motion and compute the Lyapunov exponent as a function of black hole and particle parameters. Within this framework, we demonstrate the equivalence between the Lyapunov exponent obtained from the Jacobi matrix and that derived from the effective Lagrangian. We analyze the violation of the bound on chaos by comparing the Lyapunov exponent to the surface gravity. Our analysis indicates that the violation is sensitive to the relative sign of the black hole and particle charges, the alignment between the black hole spin and particle angular momentum, and the magnitude of the negative cosmological constant.
    
    The remainder of this paper is organized as follows. Section \ref{section_review} reviews the Kerr--Sen--AdS black hole solution and the formulation of the Lyapunov exponent. Section \ref{section_lyapunov_exponent} presents the derivation of the effective potential for a charged particle and the computation of the Lyapunov exponent in the Kerr--Sen--AdS background. Section \ref{section_result} presents an analysis of the violation of the bound on chaos in the GMGHS--AdS and Kerr--Sen--AdS geometries with a detailed discussion of parameter dependence. Section \ref{section_conclusion} concludes the study with a summary of the major results and possible directions for future research.

\section{Review of Kerr--Sen AdS Black Hole and Lyapunov Exponent} \label{section_review}
    Here, we provide a brief overview of the Kerr--Sen--AdS black hole and the Lyapunov exponent, which are essential in our analysis of dynamical instability and chaotic behavior near black holes.

\subsection{Kerr--Sen--AdS Black Hole}
    In heterotic string theories, the Neveu--Schwarz--Neveu--Schwarz (NS--NS) or bosonic sector of the four-dimensional effective action in the string frame is given by
\begin{equation}
    \bar{S} = \int \mathrm{d}^4x \sqrt{-\bar{g}} e^{-\phi} \left[ \bar{R} + \partial_\mu \phi \partial^\mu \phi - F_{\mu \nu} F^{\mu \nu} - \frac{1}{12} H_{\mu \nu \rho} H^{\mu \nu \rho} \right],
\end{equation}
    where the bar denotes the string frame, $R$ is the Ricci scalar, $\phi$ is the dilaton scalar field, $F_{\mu \nu}$ is the $U(1)$ Maxwell field strength with $A_{\mu}$, and $H_{\mu \nu \rho}$ is the field strength of the Kalb--Ramond NS--NS \textit{B}-field with the gauge Chern--Simons term, which are written as
\begin{eqnarray}
    F_{\mu \nu} &=& \nabla_\mu A_\nu - \nabla_\nu A_\mu \;\; = \;\; \partial_\mu A_\nu - \partial_\nu A_\mu,
    \\
    H_{\alpha \beta \gamma} &=& 3 \nabla_{[\alpha} B_{\beta \gamma]} - \frac{3}{4} A_{[\alpha} F_{\beta \gamma]} \nonumber
    \\
    &=& \partial_\alpha B_{\beta \gamma} + \partial_\beta B_{\gamma \alpha} + \partial_\gamma B_{\alpha \beta} - \frac{1}{4} \left( A_\alpha F_{\beta \gamma} + A_\beta F_{\gamma \alpha} + A_\gamma F_{\alpha \beta} \right).
\end{eqnarray}
    Applying the following conformal transformation,
\begin{equation}
    \mathrm{d}s_\mathrm{E}^2 = e^{-\phi} \mathrm{d}\bar{s}^2,
\end{equation}
    and the effective action in the Einstein frame is obtained as
\begin{equation}
    S_\mathrm{E} = \int \mathrm{d}^4x \sqrt{-g} \left[ R - \frac{1}{2} \partial_\mu \phi \partial^\mu \phi - e^{-\phi} F_{\mu \nu} F^{\mu \nu} - \frac{e^{- 2 \phi}}{12} H_{\mu \nu \rho} H^{\mu \nu \rho} \right].
    \label{effective_action}
\end{equation}
    Modifying the definition of the three-form field strength is necessary to incorporate a non-zero cosmological constant into the theory and obtain asymptotically AdS solutions. Particularly, the cosmological constant breaks the conformal invariance of the low-energy effective string action, and a dualization procedure is introduced to restore consistency. This involves redefining the three-form field $H_{\mu \nu \rho}$ in terms of a one-form pseudo-scalar field $\chi$, which is often referred to as the axion field in the dual formulation. The modified definition is expressed as
\begin{equation}
    H = \mathrm{d}B - \frac{1}{4} A \wedge F \equiv - e^{2 \phi} \star \mathrm{d}\chi,
\end{equation}
    where $\star$ denotes the Hodge star operator. Introducing the $\chi$ field enables the consistent incorporation of the effect of a negative cosmological constant into the theory by capturing the dynamics of the dualized three-form field. This reformulation enables a well-defined action compatible with AdS asymptotics, which is valuable for analyzing black hole solutions in string-inspired theories, where dilaton and axion fields are crucial in low-energy effective dynamics. With this setup, the four-dimensional effective Lagrangian takes the form \cite{Wu:2020cgf, Ali:2023ppg}
\begin{eqnarray}
    \mathcal{L}_\mathrm{E} &=& \sqrt{-g} \left\{ R - \frac{1}{2} \partial_\mu \phi \partial^\mu \phi - \frac{e^{2 \phi}}{2} \partial_\mu \chi \partial^\mu \chi - e^{-\phi} F_{\mu \nu} F^{\mu \nu} - \frac{\Lambda}{3} \left[ 4 + e^{-\phi} + e^{\phi} \left( 1 + \chi^2 \right) \right] \right\} \nonumber
    \\
    && \quad + \chi F_{\mu \nu} \tilde{F}^{\mu \nu},
\end{eqnarray}
    where $\tilde{F}^{\mu \nu} = \frac{1}{2} \epsilon^{\mu \nu \rho \sigma} F_{\rho \sigma}$ is the dual tensor of $F^{\mu \nu}$, and $\epsilon^{\mu \nu \rho \sigma}$ is the Levi-Civita tensor. The Kerr--Sen--AdS metric in the Einstein frame, expressed in four-dimensional Boyer--Lindquist coordinates, is given by
\begin{equation}
    \mathrm{d}s_\mathrm{E}^2 = - \frac{\Delta_r}{\rho^2} \left( \mathrm{d}t - \frac{a \sin^2\theta}{\Xi} \mathrm{d}\varphi \right)^2 + \frac{\rho^2}{\Delta_r} \mathrm{d}r^2 + \frac{\rho^2}{\Delta_\theta} \mathrm{d}\theta^2 + \frac{\Delta_\theta \sin^2\theta}{\rho^2} \left( a dt - \frac{r^2 + 2 b r + a^2}{\Xi} \mathrm{d}\varphi \right)^2,
    \label{metric}
\end{equation}
    where
\begin{eqnarray}
    \rho^2 &=& r^2 + 2 b r + a^2 \cos^2\theta, \qquad \Xi \;\; = \;\; 1 + \frac{\Lambda}{3} a^2, \nonumber
    \\
    \Delta_r &=& \left( r^2 + 2 b r + a^2 \right) \left[ 1 - \frac{\Lambda}{3} \left( r^2 + 2 b r \right) \right] - 2 M r, \nonumber
    \\
    \Delta_\theta &=& 1 + \frac{\Lambda}{3} a^2 \cos^2\theta. \nonumber
\end{eqnarray}
    The Kerr--Sen--AdS metric is characterized by the mass $M$, angular momentum $J$, and electric charge $Q$. The spin parameter is defined as $a = J/M$, and the dilatonic scalar charge as $b = Q^2 / ( 2 M )$. In the absence of a cosmological constant $(\Lambda = 0)$ and vanishing spin parameter $(a = 0)$, the metric reduces to the electrically charged GMGHS solution via a radial coordinate shift. The electromagnetic potential, assuming the absence of magnetic charge, is given by
\begin{equation}
    A = - \frac{Q r}{\rho^2} \left( \mathrm{d}t - \frac{a \sin^2\theta}{\Xi} \mathrm{d}\varphi \right).
\end{equation}
    
    In the Kerr--Sen--AdS background \eqref{metric}, the angular velocity of an observer at spatial infinity is given by
\begin{equation}
    \Omega_\infty = \left. - \frac{g_{t\varphi}}{g_{\varphi\varphi}} \right|_{r \to \infty} = \frac{a \Lambda}{3}.
\end{equation}
    We define the frame of an asymptotically static observer by performing a coordinate transformation \cite{Hawking:1998kw,Gwak:2018akg,Gwak:2021tcl} that eliminates the angular velocity at spatial infinity,
\begin{equation}
    t \to T, \qquad \varphi \to \Phi + \frac{a \Lambda}{3} T.
\end{equation}
    After the coordinate transformation, the Kerr-Sen-AdS metric takes the form
\begin{eqnarray}
    \mathrm{d}s_\mathrm{E}^2 &=& - \frac{\Delta_r}{\Xi^2 \rho^2} \left( \Delta_\theta \mathrm{d}T - a \sin^2\theta \mathrm{d}\Phi \right)^2 + \frac{\rho^2}{\Delta_r} \mathrm{d}r^2 + \frac{\rho^2}{\Delta_\theta} \mathrm{d}\theta^2 \nonumber
    \\
    && \quad + \frac{\Delta_\theta \sin^2\theta}{\Xi^2 \rho^2} \left\{ a \left[ 1 - \frac{\Lambda}{3} \left( r^2 + 2 b r \right) \right] \mathrm{d}T - \left( r^2 + 2 b r + a^2 \right) \mathrm{d}\Phi \right\}^2
\end{eqnarray}
    and the corresponding potential becomes
\begin{equation}
    A =  - \frac{Q r}{\Xi \rho^2} \left( \Delta_\theta \mathrm{d}T - a \sin^2\theta \mathrm{d}\Phi \right).
\end{equation}
    The Hawking temperature remains invariant under this transformation, which is given by
\begin{equation}
    T = \frac{r_+ \left[ 1 - \frac{a^2}{r_+^2} - \frac{\Lambda}{3} a^2 - \frac{\Lambda}{3} \left( r_+ + 2 b \right) \left( 3 r_+ + 2 b \right) \right]}{4 \pi \left( r_+^2 + 2 b r_+ + a^2 \right)},
\end{equation}
    where $r_+$ denotes the outer horizon of the Kerr--Sen--AdS black hole. The surface gravity
\begin{equation}
    \kappa = 2 \pi T = \frac{r_+ \left[ 1 - \frac{a^2}{r_+^2} - \frac{\Lambda}{3} a^2 - \frac{\Lambda}{3} \left( r_+ + 2 b \right) \left( 3 r_+ + 2 b \right) \right]}{2 \left( r_+^2 + 2 b r_+ + a^2 \right)},
    \label{surface_gravity}
\end{equation}
    and the corresponding bound on the Lyapunov exponent \cite{Maldacena:2015waa} is \cite{Maldacena:2015waa} is
\begin{equation}
    \lambda \leq \kappa.
\end{equation}
    For convenience, we adopt the squared form of this condition, $\lambda^2 \leq \kappa^2$.

\subsection{Lyapunov Exponent}
    Computing the Lyapunov exponent is essential for analyzing the dynamical stability of particle trajectories. This quantity characterizes the exponential divergence of nearby trajectories, which provides a measure of chaotic behavior. We begin by deriving the relevant equations of motion for a probe particle, expressed as follows:
   \begin{equation}
    \frac{\mathrm{d}}{\mathrm{d}\tau} X^i = F_i(X^j).
\end{equation}
    The corresponding linearized form is expressed as
\begin{equation}
    \frac{\mathrm{d}}{\mathrm{d}\tau} \delta X_i = J_{ij} \delta X_j,
    \label{eom_linear}
\end{equation}
    where $X_i$ are generalized coordinates, and $F_i$ are functions of these coordinates. The components of the Jacobian matrix  $J_{ij}$ are defined as
    \begin{equation}
    J_{ij} = \left. \frac{\partial F_i}{\partial X_j} \right|_{X = X_i}.
\end{equation}

    We consider the generalized Lagrangian of the particle to express the dynamics in terms of the classical phase space variables $X_i = (r, p)$ and the radial motion of a particle near the unstable equilibrium point $r_0$
\begin{equation}
    \mathcal{L} = \frac{1}{2} K(r) \dot{r}^2 - V(r),
\end{equation}
    where the radial-dependent kinetic term $K(r)$ reflects the influence of the nontrivial background geometry, and the dot denotes the derivatives with respect to the geodesic parameter. From the Euler--Lagrangian equation,
\begin{eqnarray}
    F_1 &=& \dot{r} \;\;\:\: = \;\; \frac{p}{K},
    \\
    F_2 &=& \dot{p}_r \;\; = \;\; \frac{1}{2} K' \dot{r}^2 - V',
\end{eqnarray} 
    where the prime indicates the derivative relative to the radius. At the unstable orbit, $\left. \dot{r} \right|_{r = r_0} = \left. p \right|_{r = r_0} = 0$, $V'(r_0) = 0$, and $V''(r_0) < 0$. We analyze fluctuations around $r_0$ by considering small deviations $\delta r$ and $\delta p$ as
\begin{eqnarray}
    r &=& r_0 + \delta r,
    \\
    p &=& \delta p.
\end{eqnarray}
    Consequently, Eq. \eqref{eom_linear} reduces to
\begin{equation}
    \frac{\mathrm{d}}{\mathrm{d}\tau}
    \begin{pmatrix} 
        \delta r
        \\
        \delta p
    \end{pmatrix}
    =
    \begin{pmatrix} 
        0 & \frac{1}{K}
        \\
        - V'' & 0
    \end{pmatrix}
    \begin{pmatrix} 
        \delta r
        \\
        \delta p
    \end{pmatrix}.
\end{equation}
    The local Lyapunov exponent is determined by the eigenvalue of the Jacobian matrix evaluated at the unstable radial equilibrium point $r_0$,
\begin{equation}
    \lambda^2 = - \frac{V''(r_0)}{K(r_0)}.
    \label{lyapunov_exponent}
\end{equation}
    The Lyapunov exponents defined relative to proper time are commonly used to study particle geodesic motion. Additionally, the Lyapunov exponents in coordinate time have been extensively studied \cite{Cardoso:2008bp, Pradhan:2012rkk, Pradhan:2013bli, Pradhan:2014tva, Lei:2020clg, Lei:2021koj, Chen:2025xqc, Ali:2025znb}. In this study, we apply a static gauge condition (Eq. \eqref{lagrangian_particle}), aligning the geodesic parameter with coordinate time.

\section{Lyapunov Exponent in the Kerr--Sen--AdS Black Hole} \label{section_lyapunov_exponent}
    We investigate the motion of a probe particle in the background of a Kerr--Sen--AdS black hole. Near a local extremum of the effective potential, the dynamics of the particle can be represented by those of an (inverse) harmonic oscillator. This enables the computation of the Lyapunov exponent for the effective Lagrangian, characterizing the instability of the orbit.

    We systematically analyze the particle's motion, starting from the Polyakov-type action for an electrically charged particle in a curved spacetime. The action is given by
\begin{eqnarray}
    S_\mathrm{p} &=& \int \mathrm{d}\tau \mathcal{L}_\mathrm{p},
    \label{action}
    \\
    \mathcal{L}_\mathrm{p} &=& \frac{1}{2 e(s)} \left( \frac{\mathrm{d}X}{\mathrm{d}s} \right)^2 - \frac{e(s)}{2} m^2 + q A_{\mu} \frac{\mathrm{d}X^\mu}{\mathrm{d}s},
    \label{lagrangian_particle}
\end{eqnarray}
    where $m$ and $q$ are the mass and electric charge of the particle, respectively, $e$ is the auxiliary field, $s$ parametrizes the geodesic of the particle, and $X$ denotes the position of spacetime where $X = \{ T(s), \, r(s), \, \theta(s), \, \Phi(s) \}$. By integrating out the auxiliary field, the action \eqref{action} becomes equivalent to the Nambu--Goto--type action. We adopt the static gauge $s = T$ and restrict the motion to the equatorial plane, {\it i.e.}, $\theta = \pi/2$. Consequently, the metric functions are simplified as $\rho^2 = r^2$ and $\Delta_\theta = 1$. Applying these to the metric, the Lagrangian \eqref{lagrangian_particle} for equatorial motion is expressed as
\begin{eqnarray}
    \mathcal{L}_\mathrm{p} &=& \frac{1}{2 e \Xi^2 \left( r^2 + 2 b r \right)} \left\{ \frac{\Xi^2 \left( r^2 + 2 b r \right)^2}{\Delta_r} \dot{r}^2 + \left( 1 - \frac{\Lambda}{3} \left( r^2 + 2 b r \right) \right)^2 - \Delta_r \right. \nonumber
    \\
    && \quad \left. + \left[ \left( r^2 + 2 b r + a^2 \right)^2 - a^2 \Delta_r \right] \dot{\Phi}^2 \right. \nonumber
    \\
    && \quad \left. + 2 a \left[ \frac{\Lambda}{3} \left( r^2 + 2 b r \right) \left( r^2 + 2 b r + a^2 \right) - \left( r^2 + 2 b r + a^2 - \Delta_r \right) \right] \dot{\Phi} \right\} \nonumber
    \\
    && - \frac{e}{2} m^2 - \frac{Q q}{\Xi \left( r + 2 b \right)} \left( 1 - a \dot{\Phi} \right).
    \label{reduced_lagrangian_particle}
\end{eqnarray}
    The reduced Lagrangian \eqref{reduced_lagrangian_particle} is independent of $\Phi$; therefore, the corresponding conserved quantity, the angular momentum of the particle, is given by
\begin{equation}
    L \! = \! \frac{\partial \mathcal{L}_\mathrm{p}}{\partial \dot{\Phi}} \! = \! \frac{a \left\{ \Delta_r \! - \! \left( r^2 \! + \! 2 b r \! + \! a^2 \right) \left[ 1 \! - \! \frac{\Lambda}{3} \left( r^2 \! + \! 2 b r \right) \right] \! + \! e \Xi Q q r \right\} \! + \! \left[ \left( r^2 \! + \! 2 b r \! + \! a^2 \right)^2 \! - \! a^2 \Delta_r \right] \dot{\Phi}}{e \Xi^2 \left( r^2 + 2 b r \right)}.
\end{equation}
    In addition, we obtain the constraint $\dot{X}^2 = - e^2 m^2$ from the equation of motion for the auxiliary field $e$, leading to
\begin{equation}
    e^2 = \frac{r^2 + 2 b r}{\alpha(r)} \left( \Delta_r - \frac{\beta(r)}{\Delta_r} \dot{r}^2 \right),
\end{equation}
    where
\begin{eqnarray}
    \alpha(r) &=& m^2 \beta(r) + \frac{r}{r + 2 b} \left[ L \Xi \left( r + 2 b \right) - Q q a \right]^2,
    \\
    \beta(r) &=& \left( r^2 + 2 b r + a^2 \right)^2 - a^2 \Delta_r.
\end{eqnarray}
    By incorporating the conserved angular momentum $L$ and the auxiliary field $e$, we derive the effective Lagrangian
\begin{equation}
    \mathcal{L}_\mathrm{eff} = \mathcal{L}_\mathrm{p} - L \dot{\Phi}.
\end{equation}
    Owing to the conservation of angular momentum, the effective Lagrangian no longer depends explicitly on $\Phi$. Consequently, the dynamics are reduced to a one-dimensional system solely governed by the radial coordinate $r$. We investigate the sensitivity of the probe particle's motion near an unstable local extremum of the effective potential, assuming a perturbatively small initial velocity. In this non-relativistic regime, where $\dot{r}^2 \ll 1$, the effective Lagrangian is simplified to
\begin{equation}
    \mathcal{L}_\mathrm{eff} = \frac{1}{2} K(r) \dot{r}^2 - V_\mathrm{eff}(r) + \mathcal{O}(\dot{r}^4),
    \label{effective_lagrangian}
\end{equation}
    where
\begin{eqnarray}
    K(r) &=& \sqrt{\frac{\left( r^2 + 2 b r \right) \alpha(r)}{\Delta_r^3}},
    \\
     V_\mathrm{eff}(r) &=& \frac{1}{\beta(r)} \left[ a \left( r^2 + 2 b r + a^2 - \Delta_r \right) \left( L \Xi - \frac{Q q a}{r + 2 b} \right) + \sqrt{\left( r^2 + 2 b r \right) \Delta_r \alpha(r)} \right] \nonumber
     \\
     && \quad + \frac{Q q}{r + 2 b} - \frac{\Lambda}{3} L a.
     \label{potential}
\end{eqnarray}
    We evaluate the maximum Lyapunov exponent, focusing on the motion of the particle near a local extremum of the effective potential. The position of the extremum is denoted by $r_0$, determined by the condition $V'(r_0) = 0$. Considering a small perturbation around this point, $r = r_0 + \epsilon$, the effective Lagrangian can be approximated as
\begin{equation}
    \mathcal{L}_\mathrm{eff} = \frac{1}{2} K(r_0) \left( \dot{\epsilon}^2 + \lambda^2 \epsilon^2 \right),
\end{equation}
    where constant terms have been omitted. In this approximation, the coefficient of the second-order term determines the squared Lyapunov exponent:
\begin{equation}
    \lambda^2 = - \frac{V_\mathrm{eff}''(r_0)}{K(r_0)},
\end{equation}
    which is consistent with the expression given in Eq. \eqref{lyapunov_exponent}.

\section{Bound on Chaos in GMGHS--AdS and Kerr--Sen--AdS Black Holes} \label{section_result}
    We investigated the bound on the Lyapunov exponent for massless and massive particles near unstable orbits in the GMGHS--AdS and Kerr--Sen--AdS spacetimes. We analyzed the structure of the effective potential governing particle motion, identifying the unstable radial equilibrium points $r_0$, which correspond to chaotic dynamics at a local maximum of the potential. Specifically, we examine limiting cases, such as a vanishing cosmological constant, a large negative cosmological constant, and the near-horizon limit, where the potential enables more tractable analytic expressions. From this analysis, we derived criteria for the emergence of chaos based on the behavior of the Lyapunov exponent at the equilibrium points.

\subsection{GMGHS--AdS Black Hole}
    We first examined the GMGHS--AdS black hole, which corresponds to the non-rotating limit $( a = 0 )$ of the Kerr--Sen--AdS black hole. In this limit, the effective potential \eqref{potential} reduces to
\begin{equation}
    \left. V_\mathrm{eff}(r) \right|_{a = 0} = \frac{Q q r + \sqrt{\left[ L^2 + m^2 \left( r^2 + 2 b r \right) \right] \Delta_r}}{r^2 + 2 b r}.
\end{equation}
    The condition $V'(r) = 0$ at the local extremum $r = r_0$ determines the electric charge of the particle $q$ as
\begin{equation}
    \left. q \right|_{a = 0} = \left\{ \frac{r_0 + 2 b}{2 Q r_0} \frac{\Delta_r'(r_0)}{\Delta_r(r_0)} - \frac{r_0 + b}{Q r_0^2} \left[ 1 + \frac{L^2}{L^2 + m^2 (r_0^2 + 2 b r_0)} \right] \right\} \sqrt{\left[ L^2 + m^2 \left( r_0^2 + 2 b r_0 \right) \right] \Delta_r(r_0)}.
\end{equation}
    Subsequently, the squared Lyapunov exponent $\lambda^2$ is reduced to
\begin{eqnarray}
    \left. \lambda^2 \right|_{a = 0} &=& - \frac{\Delta_r(r_0)}{2 r_0^3 (r_0 + 2 b)^2} \left[ r_0 \Delta''(r_0) - \frac{2 \left( L^2 + m^2 b r_0 \right)}{L^2 + m^2 \left( r_0^2 + 2 b r_0 \right)} \Delta_r'(r_0) \right] + \frac{\left( \Delta'(r_0) \right)^2}{4 r_0^2 \left( r_0 + 2 b \right)^2} \nonumber
    \\
    && \quad - \frac{2 L^4 + 3 L^2 \left( r_0^2 + 2 b r_0 \right) + m^4 r_0^2 \left( 2 b r_0 + 3 b^2 \right)}{r_0^4 \left( r_0 + 2 b \right)^2 \left[ L^2 + m^2 \left( r_0^2 + 2 b r_0 \right) \right]^2} \Delta^2 (r_0).
\end{eqnarray}

    \noindent \textbf{\textit{Vanishing cosmological constant}}---We consider the case of a vanishing cosmological constant $( \Lambda = 0 )$. In this scenario, the squared surface gravity $\kappa^2$ at the event horizon $r_\mathrm{h} = 2 ( M - b )$ is reduced to
\begin{equation}
    \left. \kappa^2 \right|_{\{a = 0, \, \Lambda = 0\}} = \frac{1}{16 M^2}.
\end{equation}
     Assuming the vanishing angular momentum $( L = 0 )$, the conditions $r_0 > r_\mathrm{h}$ and $M > b$ ensure that $\kappa^2 - \lambda^2$ is strictly positive as
\begin{equation}
    \left. \kappa^2 - \lambda^2 \right|_{\{ a = 0, \, \Lambda = 0, \, L = 0 \}} = \frac{1}{16 M^2} - \frac{M^2}{\left( r_0 + 2 b \right)^4} > 0.
\end{equation}
    Notably, no radius $r_0$ satisfies the equilibrium condition $V'(r_0) = 0$ for a massless particle $(m = 0)$, implying the absence of such an unstable orbit in this case.

    \noindent \textbf{\textit{Asymptotic limit of cosmological constant}}---Regarding the opposite regime, we consider the asymptotic limit of the cosmological constant $(\Lambda \to -\infty)$, where the electric charge of the particle is determined as
\begin{equation}
    \left. q \right|_{\{ a = 0, \, \Lambda \to -\infty \}} = \frac{m^2 \left( r_0 + b \right) \left( r_0 + 2 b \right)^2}{\sqrt{3} Q \sqrt{L^2 + m^2 \left( r_0^2 + 2 b r_0 \right)}} \sqrt{-\Lambda} + \mathcal{O}(\Lambda)
    \label{q}
\end{equation}
    and the squared surface gravity at the horizon $r_\mathrm{h} = - 3 ( M - b) / ( 2 b^2 \Lambda ) + \mathcal{O}(\Lambda^{-2})$ is given by
\begin{equation}
    \left. \kappa^2 \right|_{\{ a = 0, \, \Lambda \to -\infty \}} = \frac{b^2}{9} \Lambda^2 + \mathcal{O}(\Lambda).
\end{equation}
    The Lyapunov exponent in this regime is
\begin{equation}
    \left. \lambda^2 \right|_{\{ a = 0, \, \Lambda \to -\infty \}} = - \frac{m^2 ( r_0^2 + 2 b r_0 ) \left[ 
L^2 ( 3 r_0^2 + 4 b r_0 ) + m^2 ( r_0^2 + 2 b r_0 ) \left( 2 r_0^2 + 2 b r_0 - b^2 \right) \right]}{9 \left[ L^2 + m^2 (r_0^2 + 2 b r_0) \right]^2} \Lambda^2 + \mathcal{O}(\Lambda).
\end{equation}
    Thus, $\kappa^2 - \lambda^2$ becomes
\begin{eqnarray}
    \left. \kappa^2 - \lambda^2 \right|_{\{ a = 0, \, \Lambda \to -\infty \}} &=& \frac{L^4 \Xi^4 b^2 + 2 m^4 r_0^3 \left( r_0 + b \right) ( r_0 + 2 b )^2 + L^2 m^2 ( r_0^2 + 2 b r_0 ) \left( 3 r_0^2 + 4 b r_0 + 2 b^2 \right)}{9 \left[ L^2 \Xi^2 + m^2 (r_0^2 + 2 b r_0) \right]^2} \Lambda^2 \nonumber
    \\
    && \quad + \mathcal{O}(\Lambda) > 0,
    \label{Lambda_infinity}
\end{eqnarray}
    which remains positive. The bound is satisfied only for sufficiently large values of $q$ because the electric charge \eqref{q} scales as $q \propto \sqrt{-\Lambda}$.

    \noindent \textbf{\textit{Near-horizon limit}}---We consider the near-horizon limit as $r_0 = r_\mathrm{h} + \epsilon$, where $\epsilon$ represents a small deviation. In this regime, the Lyapunov exponent is given by
\begin{equation}
    \left. \lambda^2 \right|_{\{ a = 0, \, r_0 = r_\mathrm{h} + \epsilon \}} = \frac{\left[ 1 - \frac{\Lambda}{3} \left( r_\mathrm{h} + 2 b \right) \left( 3 r_\mathrm{h} + 2 b \right) \right]^2}{4 \left( r_\mathrm{h} + 2 b \right)^2} + \mathcal{O}(\epsilon)
\end{equation}
    and $\kappa^2 - \lambda^2$ is expressed as
\begin{equation}
    \left. \kappa^2 - \lambda^2 \right|_{\{ a = 0, \, r_0 = r_\mathrm{h} + \epsilon \}} = -\frac{L^2 \left( r_\mathrm{h} + b \right) \left[ 1 - \frac{\Lambda}{3} \left( r_\mathrm{h} + 2 b \right) \left( 3 r_+ + 2 b \right) \right]^2}{r_\mathrm{h} \left( r_\mathrm{h} + 2 b \right)^3 \left[ L^2 + m^2 \left( r_\mathrm{h}^2 + 2 b r_\mathrm{h} \right) \right]} \epsilon + \mathcal{O}(\epsilon^2) < 0,
\end{equation}
    which is manifestly negative. The violation of the bound on chaos occurs in the vicinity of the horizon.
    
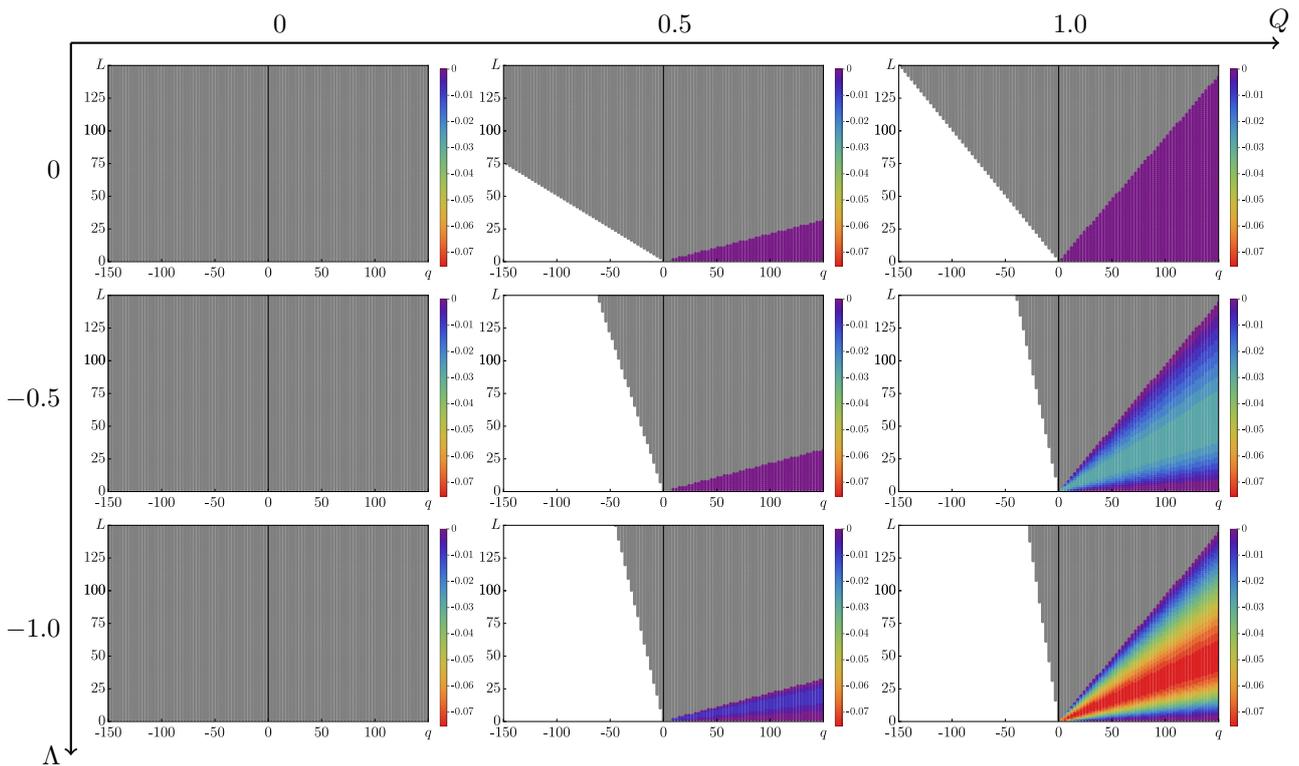
\begin{figure}[H]
    \centering
    \begin{tikzpicture}
        \matrix (M) [matrix of nodes, nodes={inner sep=0cm, outer sep=0.15cm, anchor=center}, row sep=0cm, column sep=0cm]
        {
            \includegraphics[width=5.2cm]{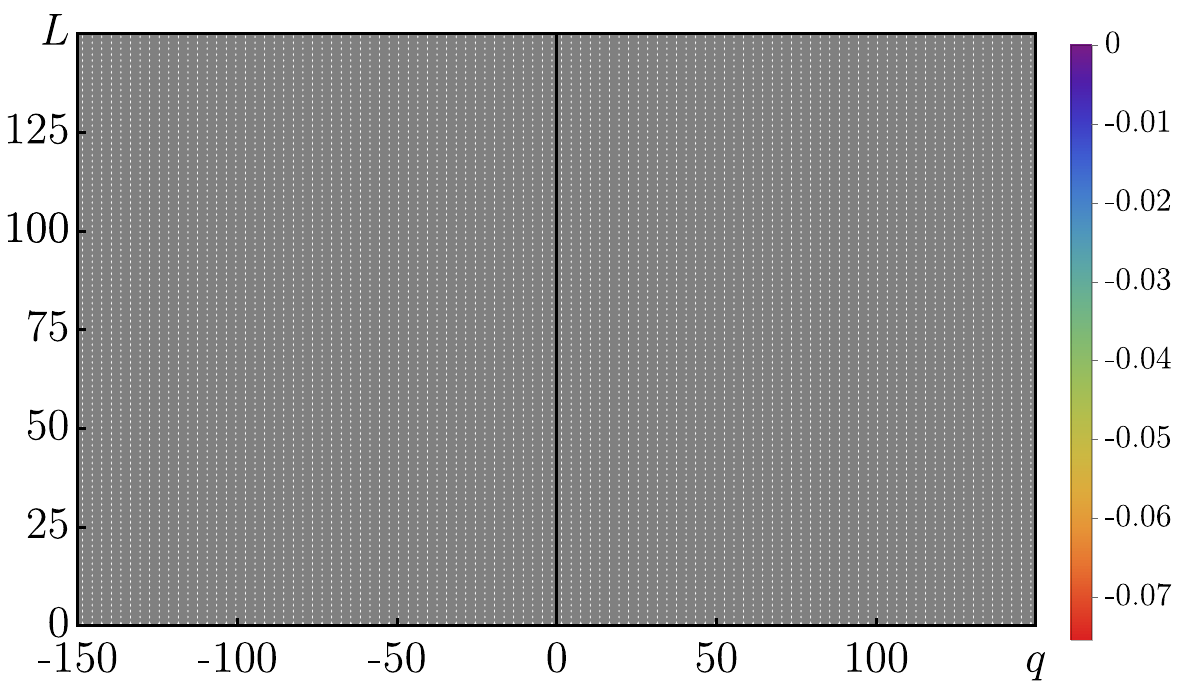} &
            \includegraphics[width=5.2cm]{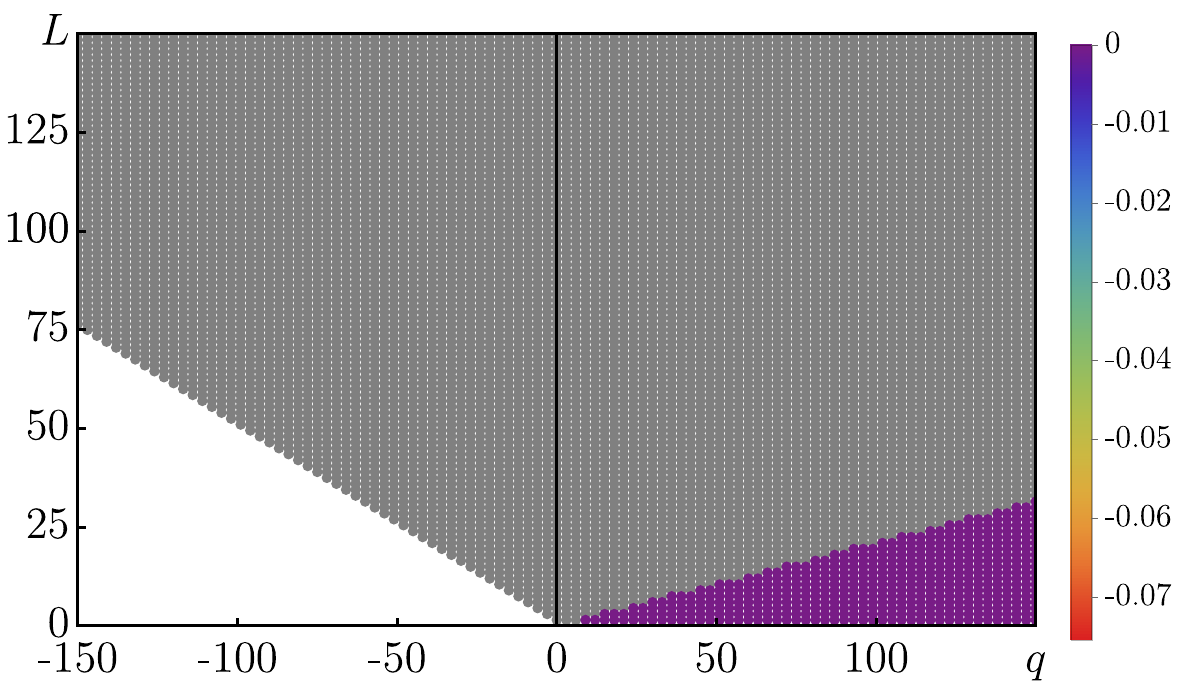} &
            \includegraphics[width=5.2cm]{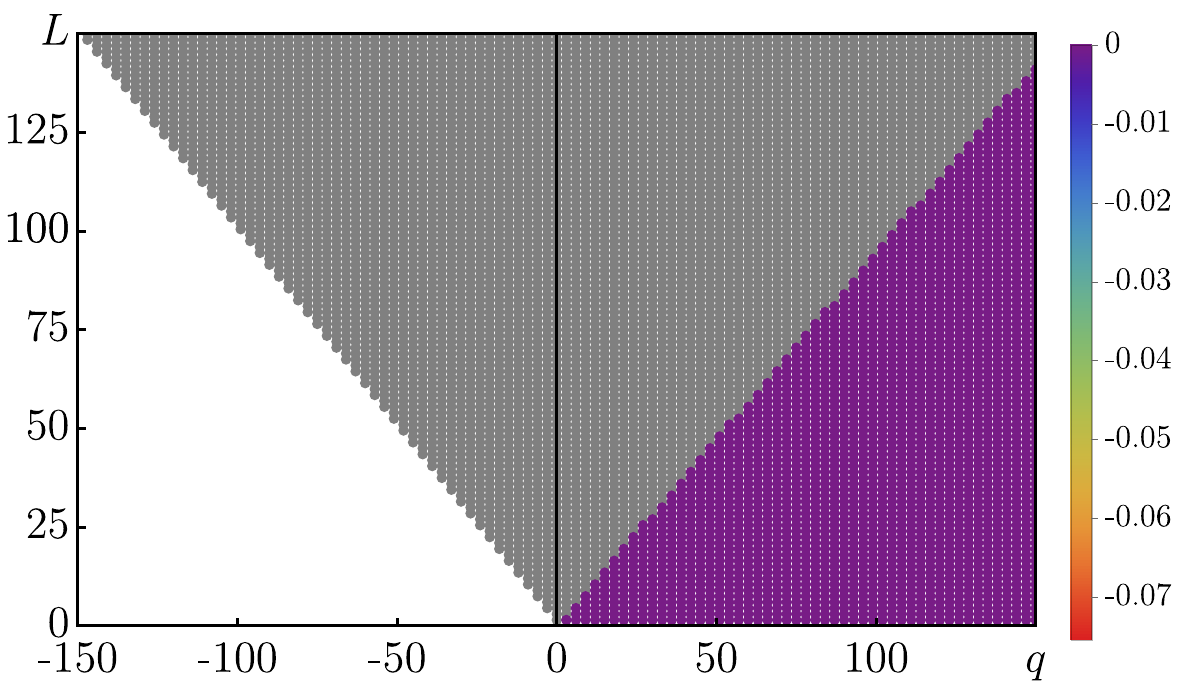} \\
            \includegraphics[width=5.2cm]{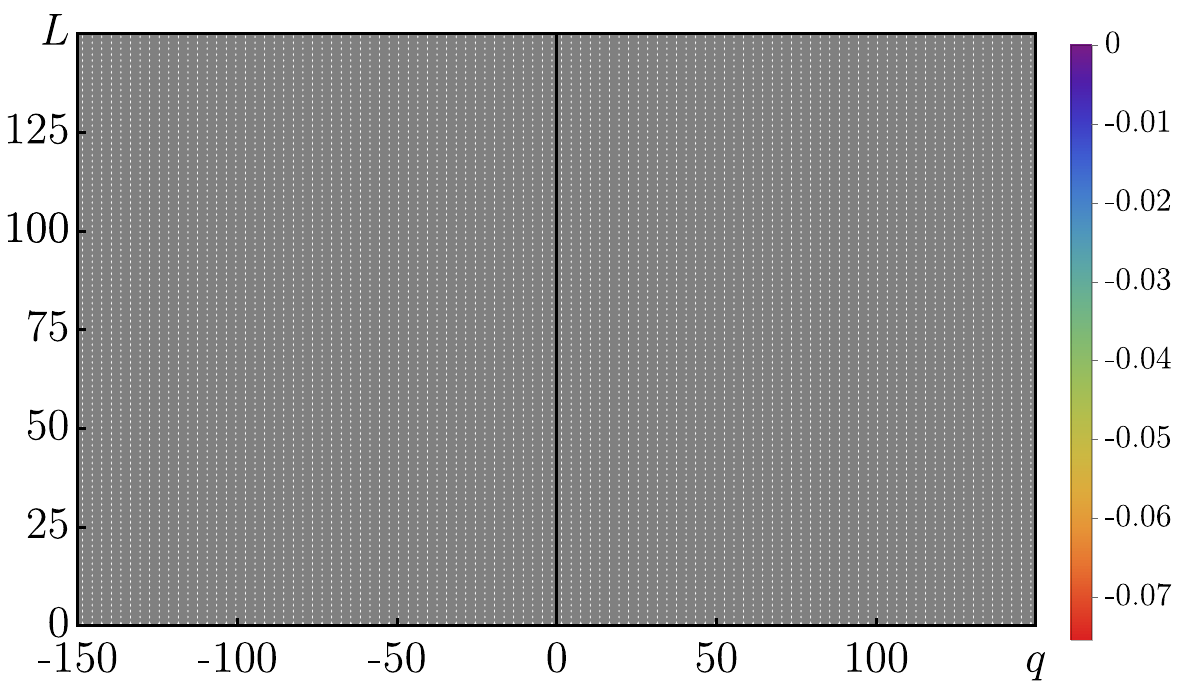} &
            \includegraphics[width=5.2cm]{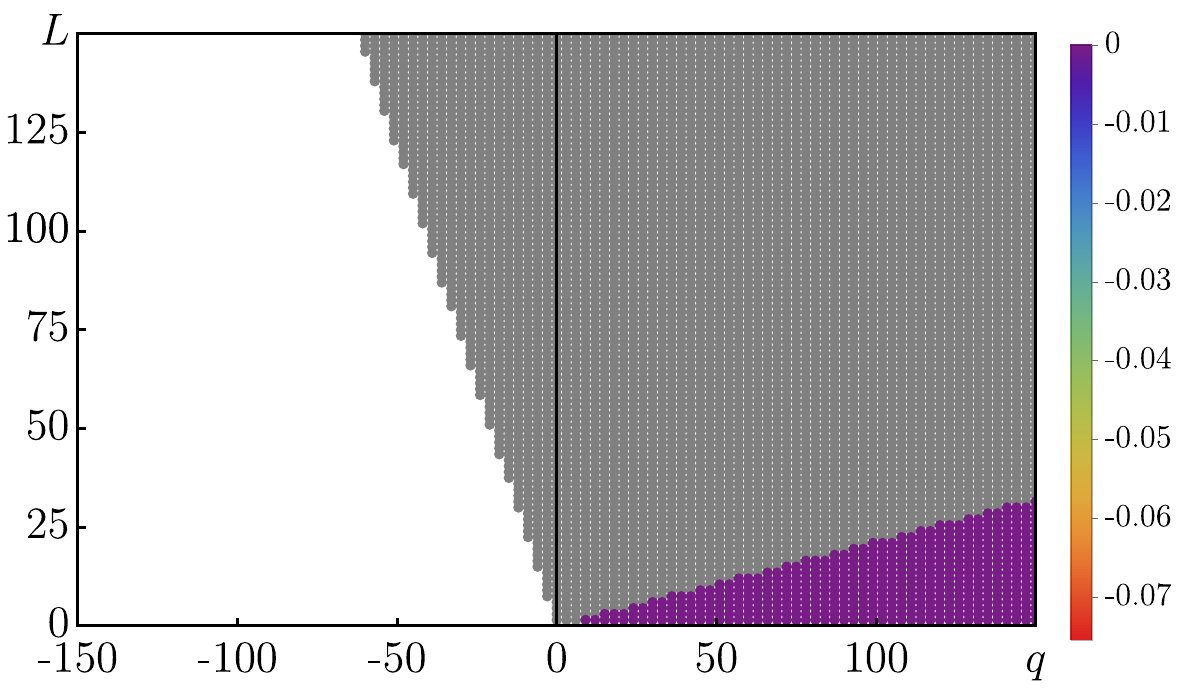} &
            \includegraphics[width=5.2cm]{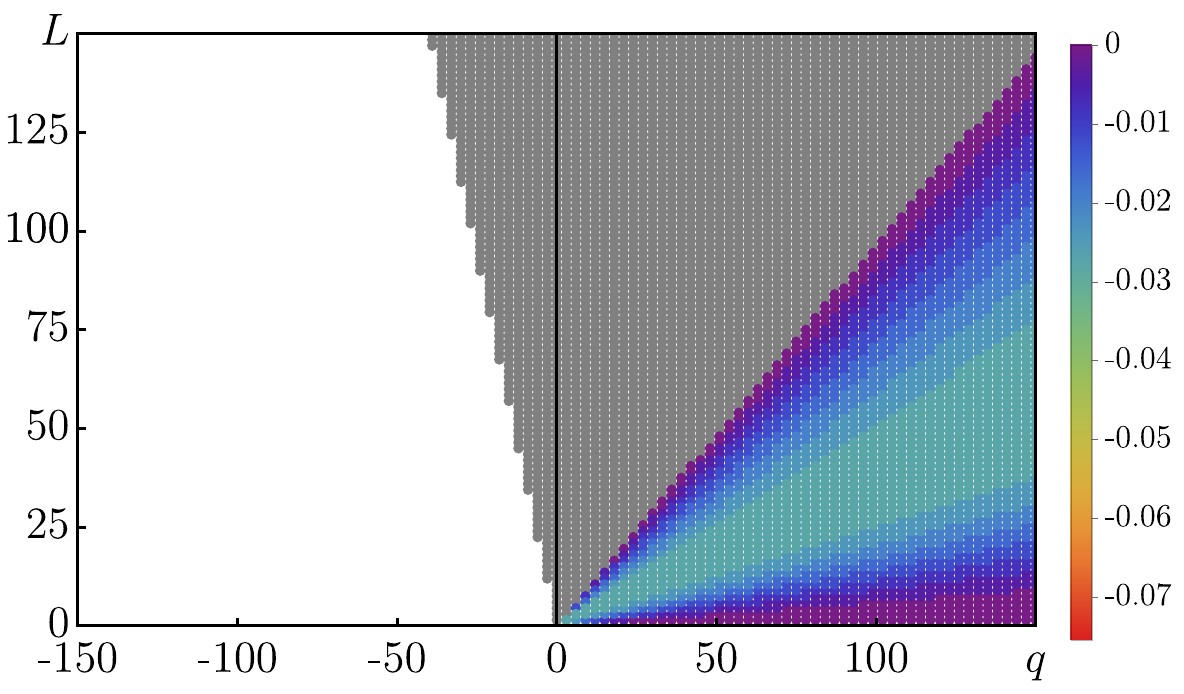} \\
            \includegraphics[width=5.2cm]{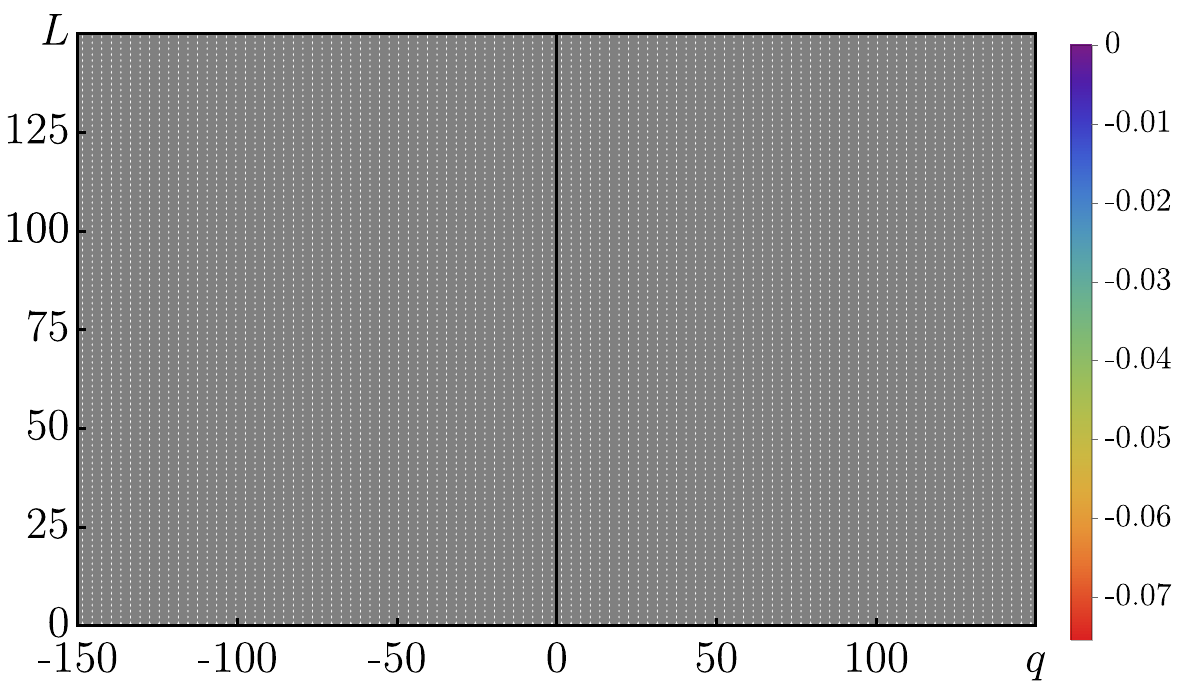} &
            \includegraphics[width=5.2cm]{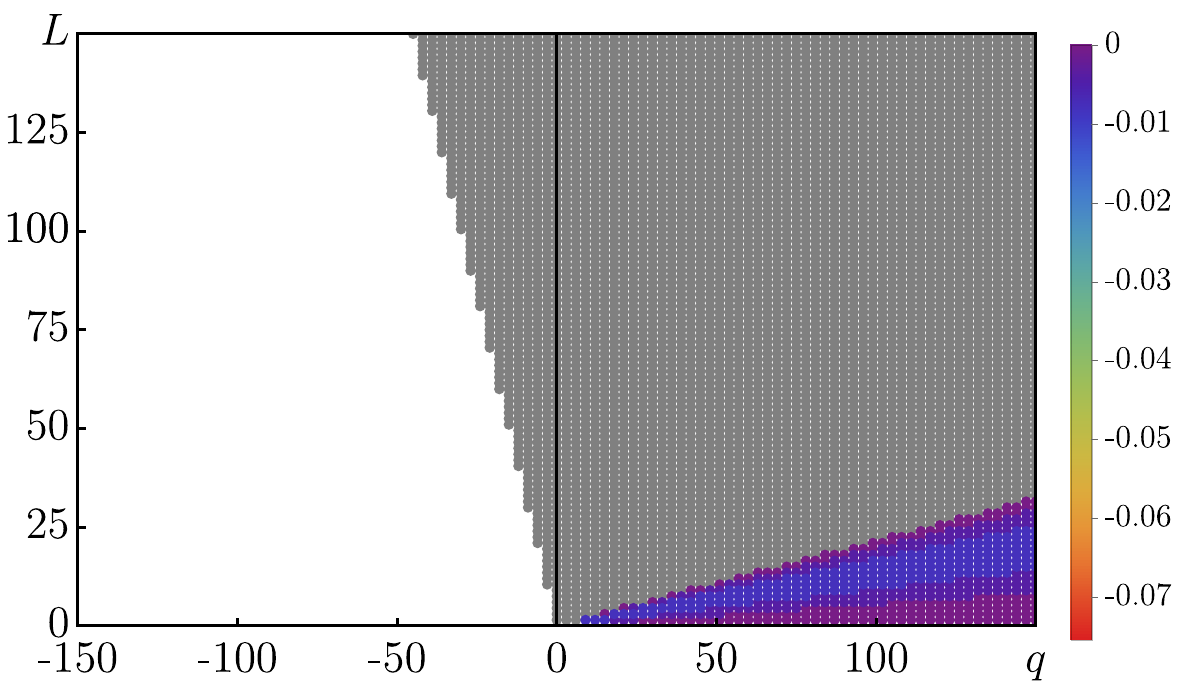} &
            \includegraphics[width=5.2cm]{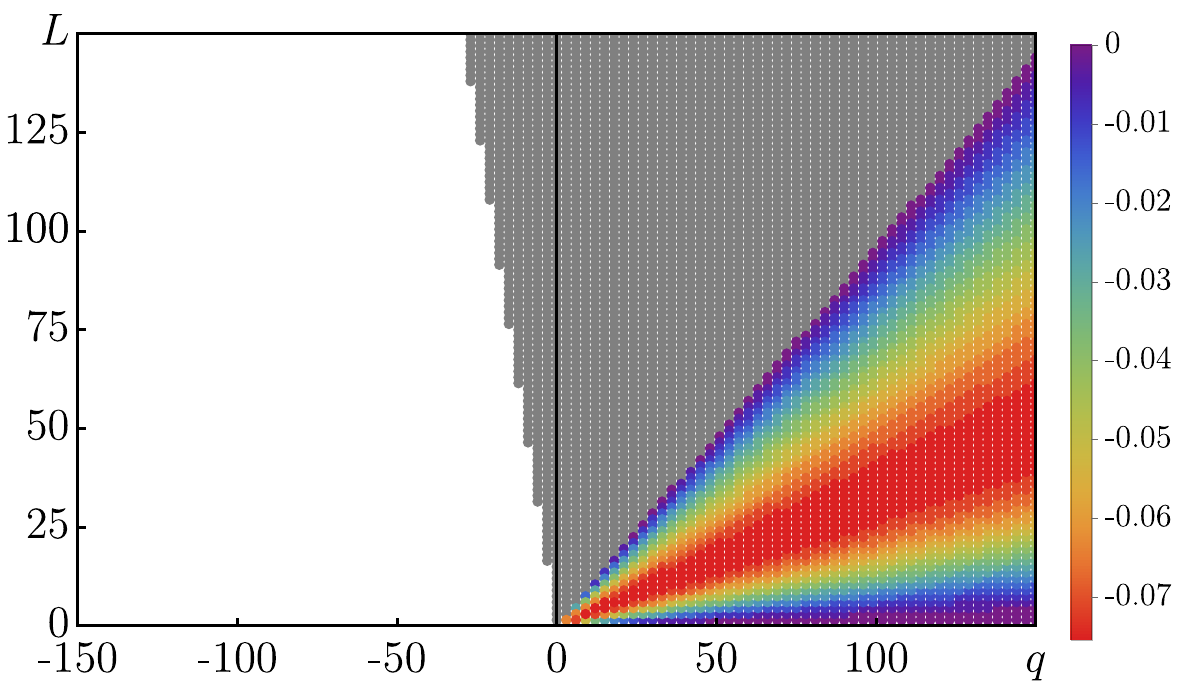} \\
        };

        \coordinate (O) at (M-1-1.north west);
        \coordinate (R) at (M-1-3.north east);
        \coordinate (B) at (M-3-1.south west);

        \draw[->, thick] (O) -- ($(R)$);
        \foreach \col/\xlab in {1/0, 2/0.5, 3/1.0} {\node[above=0cm, font=\fontsize{10pt}{10pt}\selectfont] at (M-1-\col.north) {\(\xlab\)};}
        \node[above=0cm, font=\fontsize{10pt}{10pt}\selectfont] at ($(R)$) {\(Q\)};

        \draw[->, thick] (O) -- ($(B)$);
        \foreach \row/\ylab in {1/0, 2/\mathrm{-}0.5, 3/\mathrm{-}1.0} {\node[left=0cm, font=\fontsize{10pt}{10pt}\selectfont] at (M-\row-1.west) {\(\ylab\)};}
        \node[left=0cm, font=\fontsize{10pt}{10pt}\selectfont] at ($(B)$) {\(\Lambda\)};
    \end{tikzpicture}
    \vspace{-1cm}
    \caption{Massless particle $( m = 0 )$ in the GMGHS black hole $( a = 0 )$ background.}
    \label{GMGHSm0}
\end{figure}

\begin{figure}[H]
    \centering
    \begin{tikzpicture}
        \matrix (M) [matrix of nodes, nodes={inner sep=0cm, outer sep=0.15cm, anchor=center}, row sep=0cm, column sep=0cm]
        {
            \includegraphics[width=5.2cm]{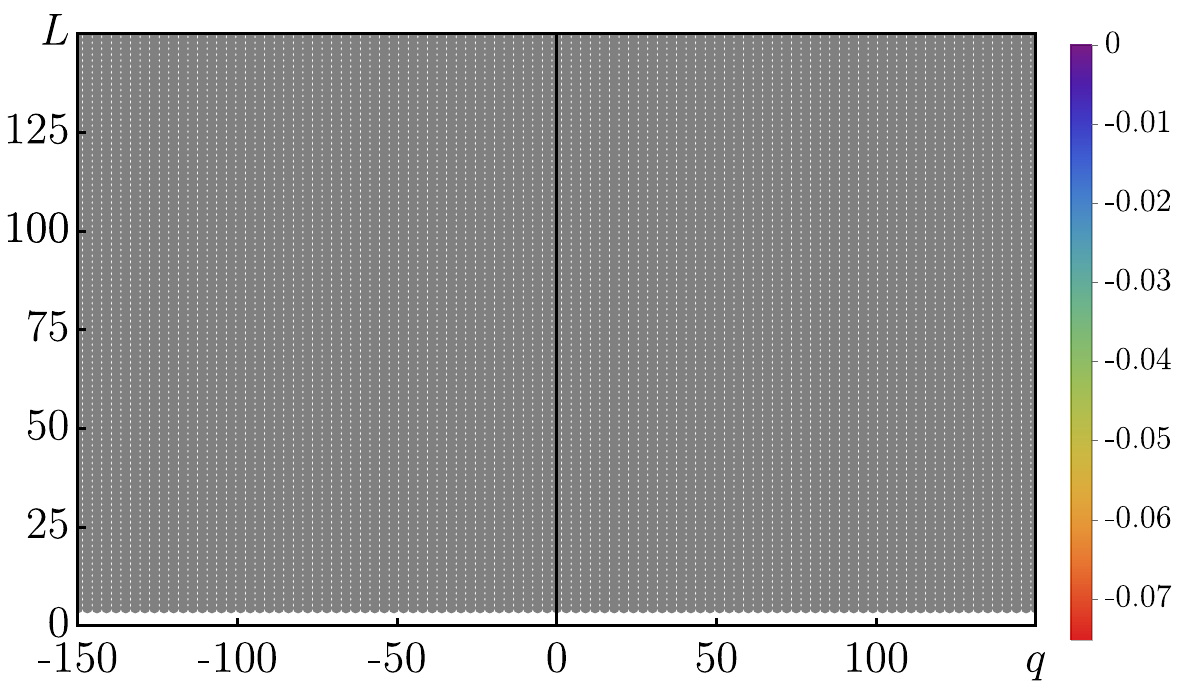} &
            \includegraphics[width=5.2cm]{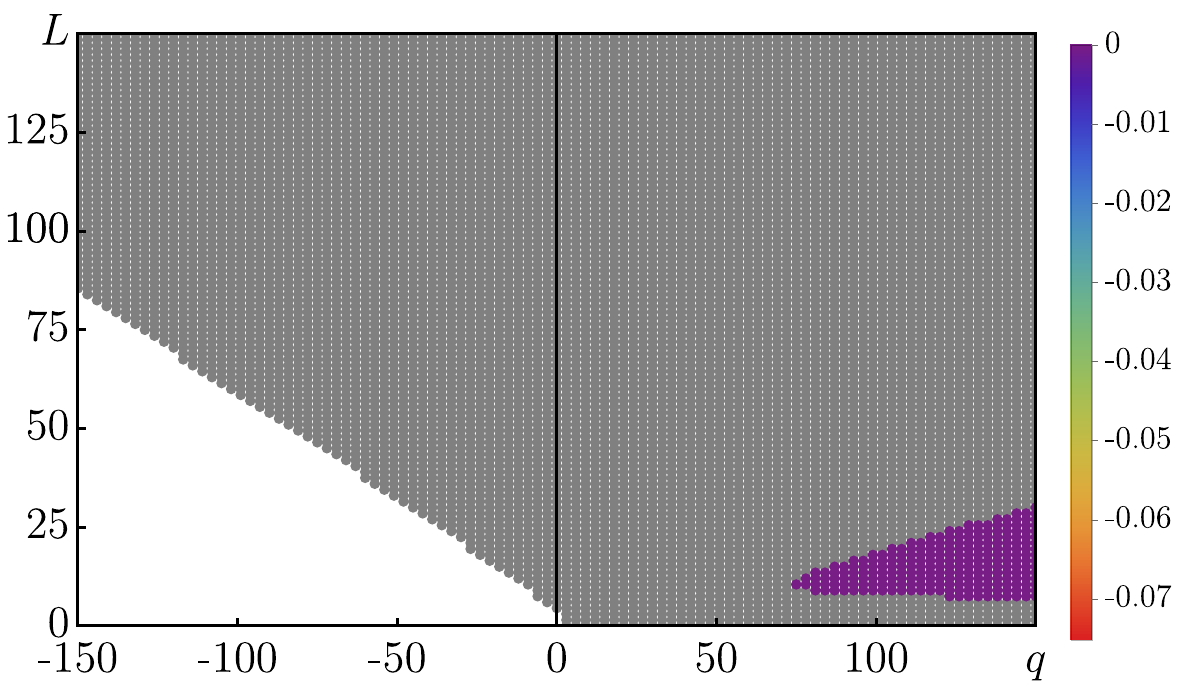} &
            \includegraphics[width=5.2cm]{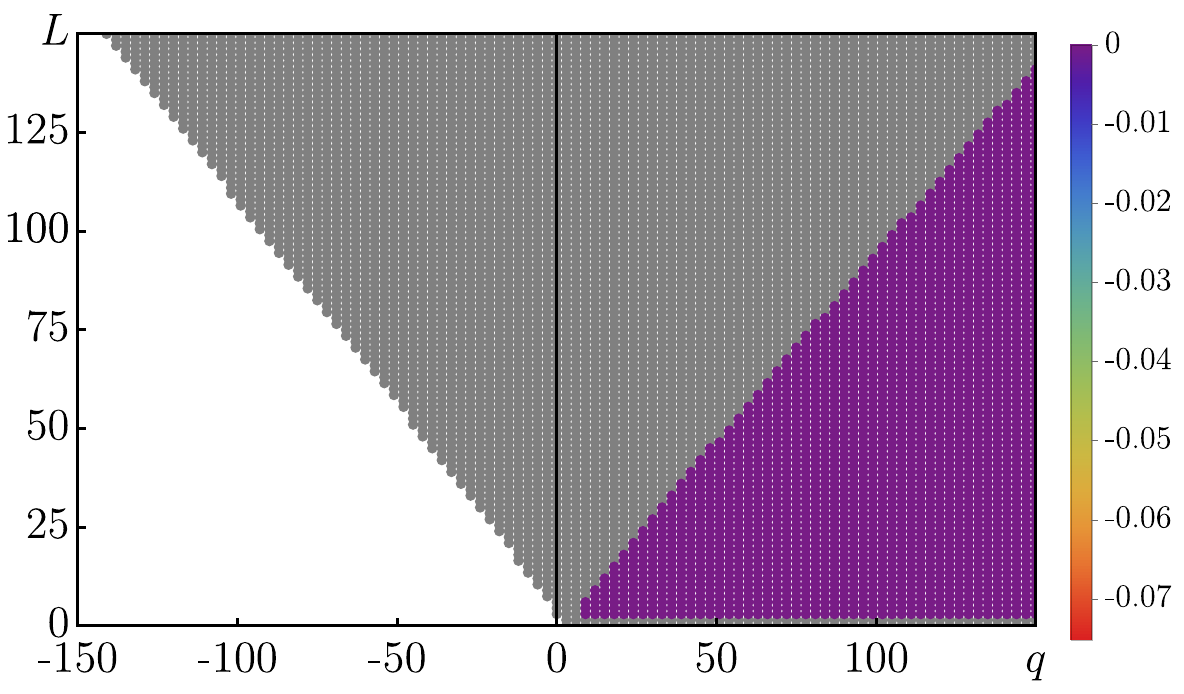} \\
            \includegraphics[width=5.2cm]{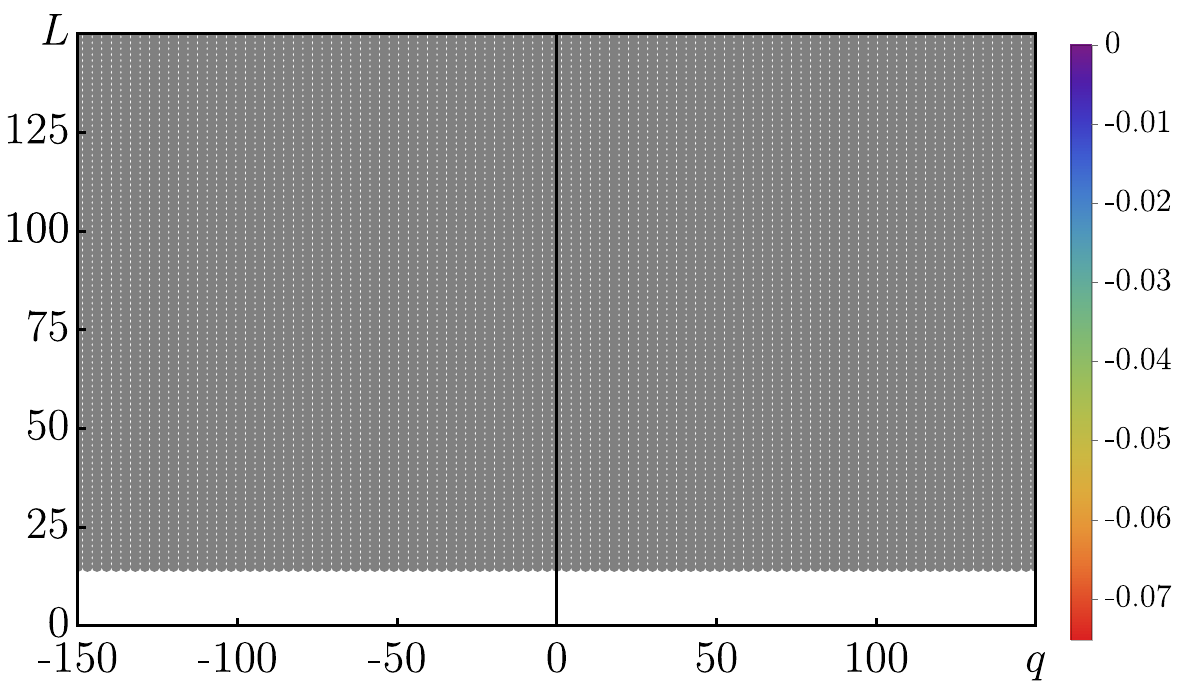} &
            \includegraphics[width=5.2cm]{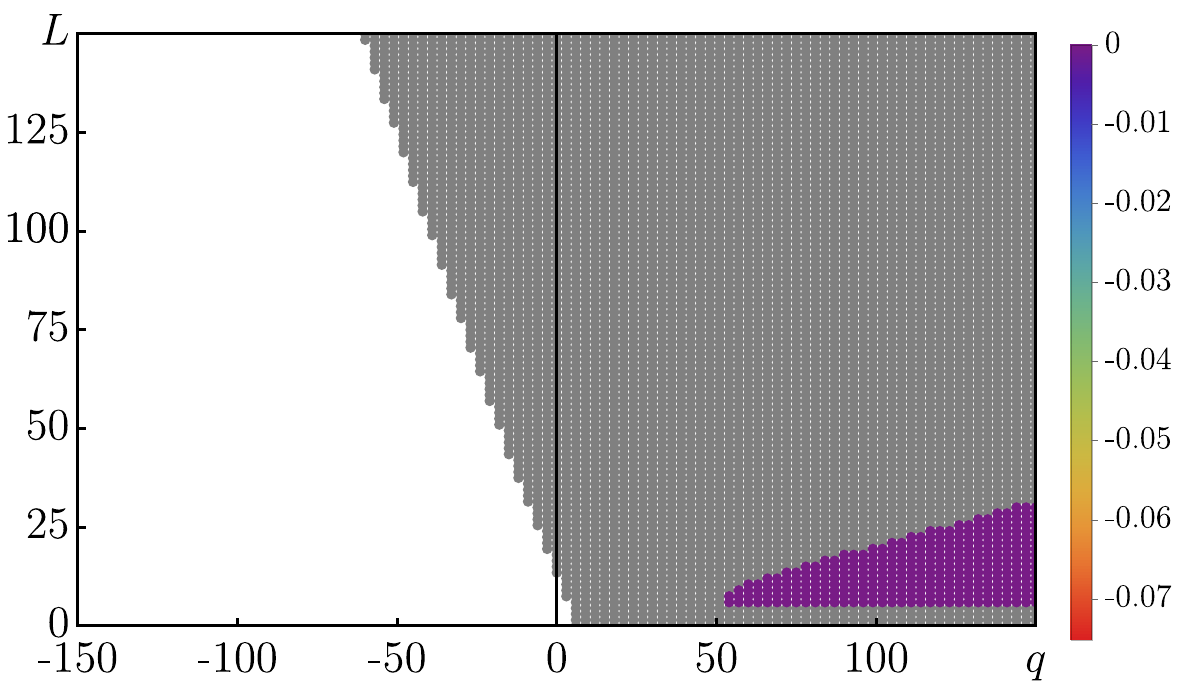} &
            \includegraphics[width=5.2cm]{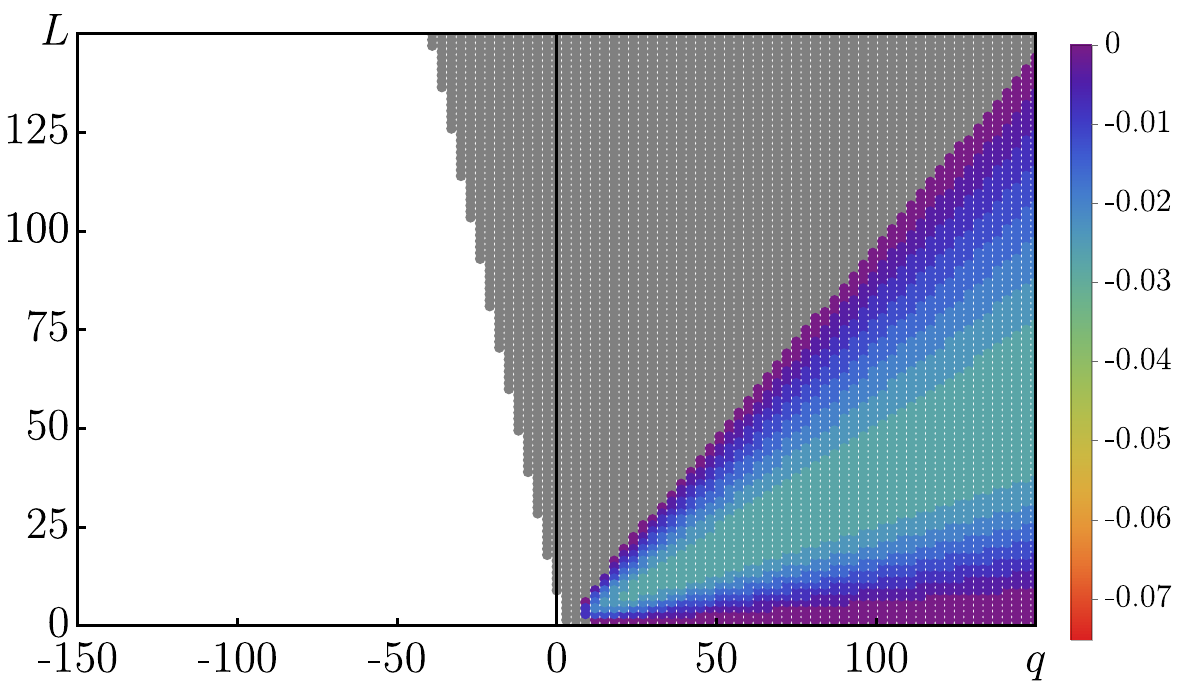} \\
            \includegraphics[width=5.2cm]{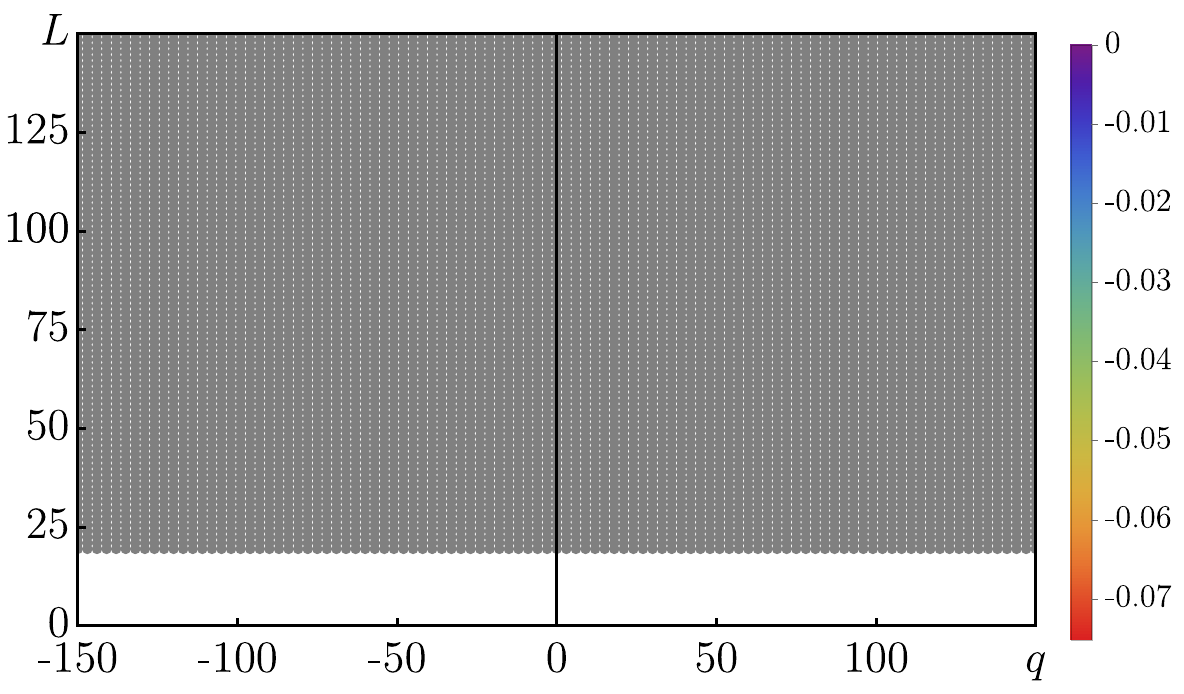} &
            \includegraphics[width=5.2cm]{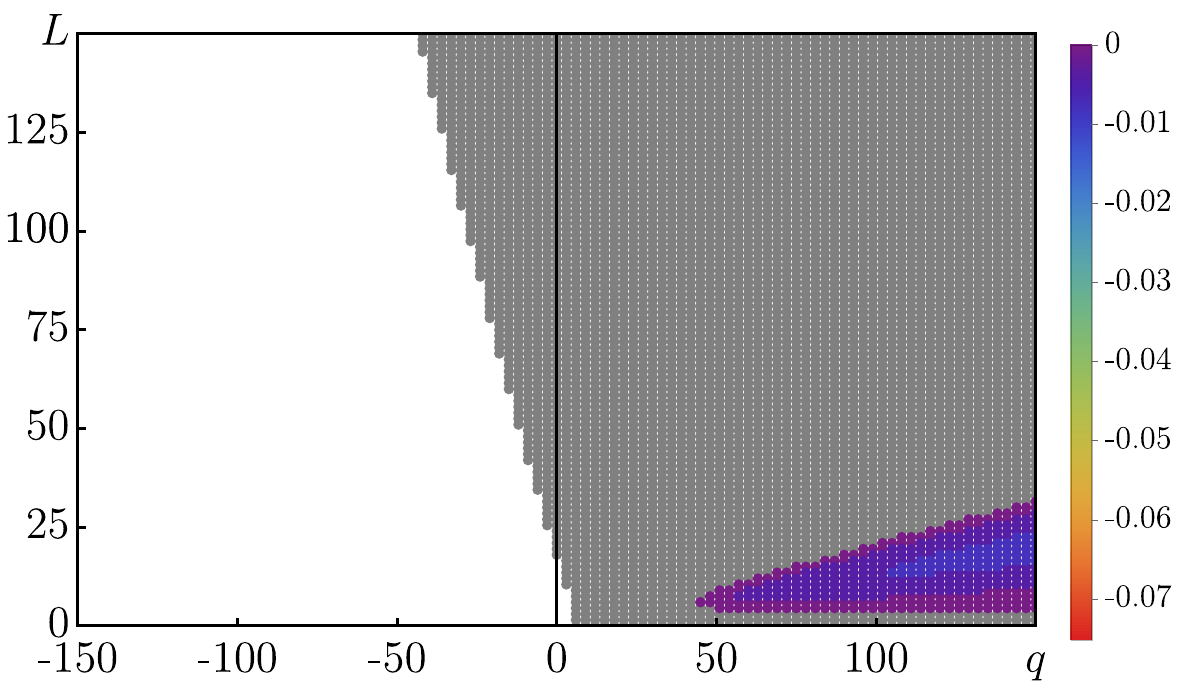} &
            \includegraphics[width=5.2cm]{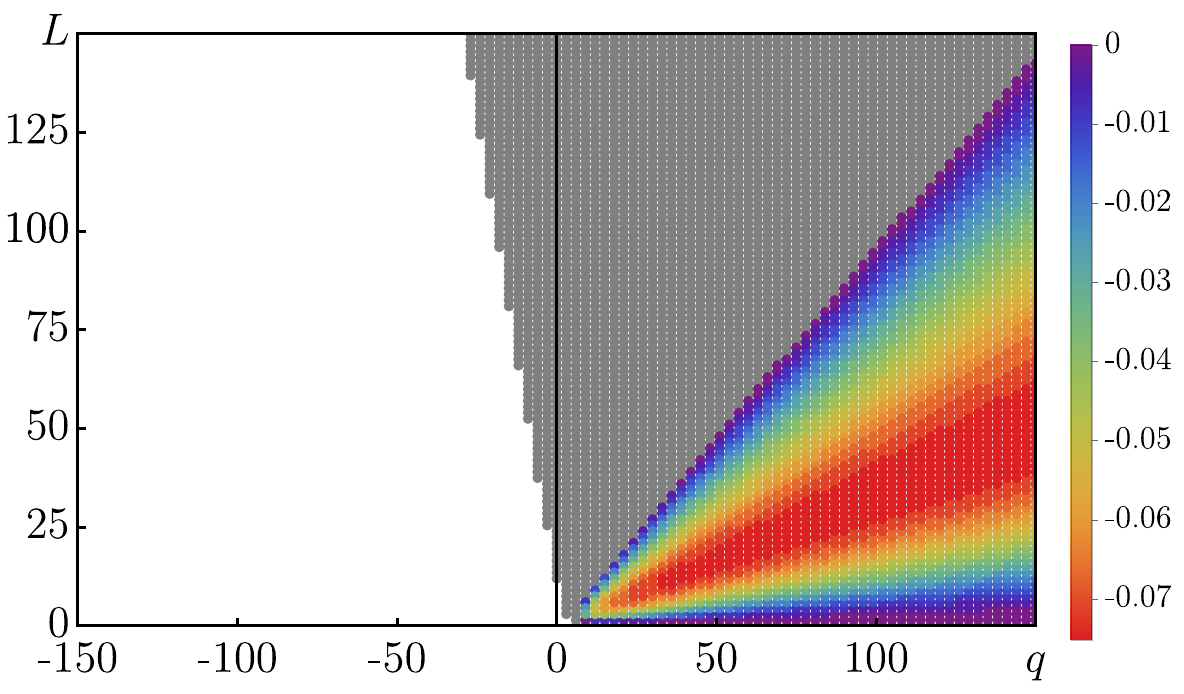} \\
        };

        \coordinate (O) at (M-1-1.north west);
        \coordinate (R) at (M-1-3.north east);
        \coordinate (B) at (M-3-1.south west);

        \draw[->, thick] (O) -- ($(R)$);
        \foreach \col/\xlab in {1/0, 2/0.5, 3/1.0} {\node[above=0cm, font=\fontsize{10pt}{10pt}\selectfont] at (M-1-\col.north) {\(\xlab\)};}
        \node[above=0cm, font=\fontsize{10pt}{10pt}\selectfont] at ($(R)$) {\(Q\)};

        \draw[->, thick] (O) -- ($(B)$);
        \foreach \row/\ylab in {1/0, 2/\mathrm{-}0.5, 3/\mathrm{-}1.0} {\node[left=0cm, font=\fontsize{10pt}{10pt}\selectfont] at (M-\row-1.west) {\(\ylab\)};}
        \node[left=0cm, font=\fontsize{10pt}{10pt}\selectfont] at ($(B)$) {\(\Lambda\)};
    \end{tikzpicture}
    \vspace{-1cm}
    \caption{Massive particle $(m = 1)$ in the GMGHS black hole $( a = 0 )$ background.}
    \label{GMGHSm1}
\end{figure}
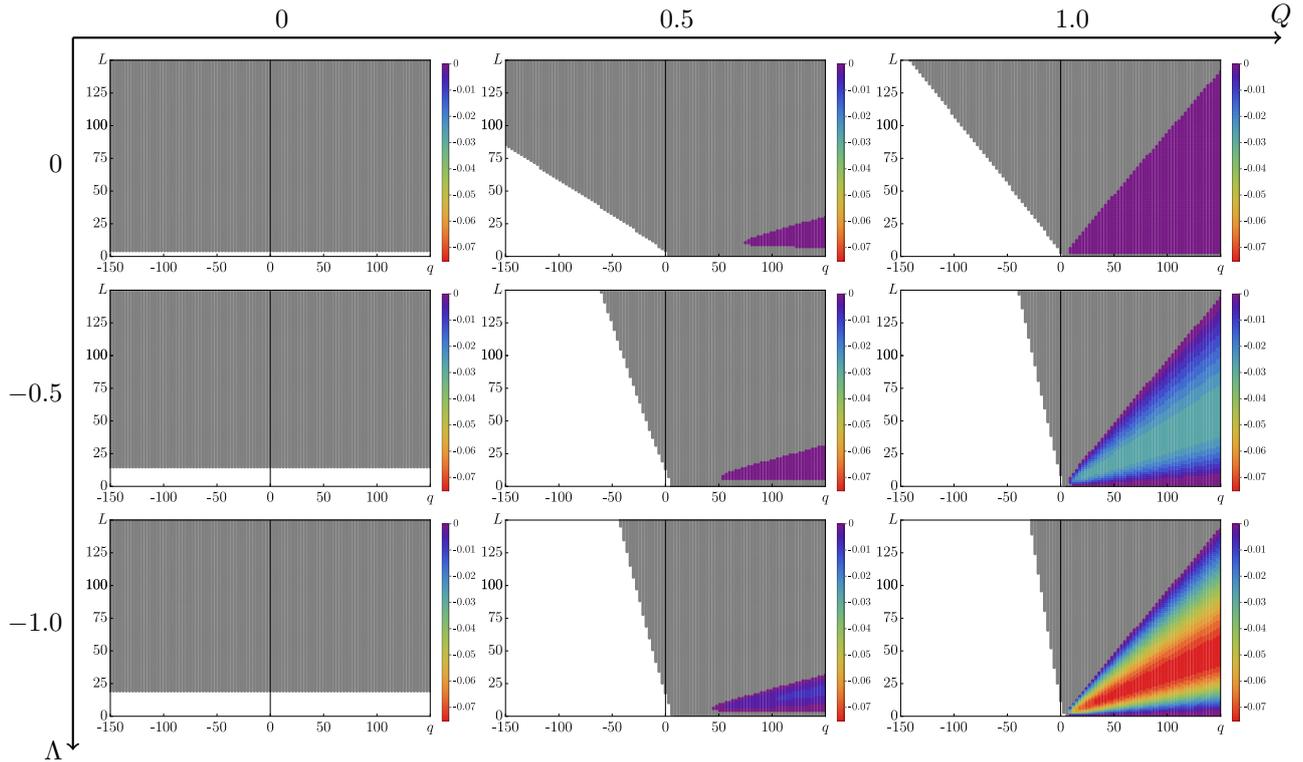

    Furthermore, we used numerical analysis to extend the analysis beyond analytically tractable limits. Figs.~\ref{GMGHSm0} and \ref{GMGHSm1} depict the violation of the bound on chaos based on analyses of the Lyapunov exponent for particle motion in the GMGHS black hole spacetime. We set the black hole mass at $M = 1$ in the numerical analysis presented in the figures for easy comparison. This is equivalent to normalizing all physical quantities by the black hole mass, thereby yielding dimensionless variables,
\begin{equation}
    \tilde{t} \equiv \frac{t}{M}, \quad \tilde{r} \equiv \frac{r}{M}, \quad \tilde{\Lambda} \equiv M^2 \Lambda, \quad \tilde{Q} \equiv \frac{Q}{M}, \quad \tilde{a} \equiv \frac{a}{M}, \quad \tilde{m} \equiv \frac{m}{M}, \quad \tilde{q} \equiv \frac{q}{M}, \quad \tilde{L} \equiv M^2 L
\end{equation}    
    where the tilde notation denotes the dimensionless variables. Each figure comprises nine plots arranged in a $3 \times 3$ grid to explore the influences of the positive electric charge $Q$ and cosmological constant $\Lambda$ on the violation of the bound on chaos. In each plot, the horizontal axis represents the electric charge of the particle $q$, while the vertical axis represents its angular momentum $L$, both ranging from $-150$ to $150$. Plots in the same row correspond to black holes with the same cosmological constant $\Lambda$, while varying the electric charge $Q$ of the black hole. Conversely, plots in the same column share the same $Q$, with varying $\Lambda$ values. The top row represents asymptotically flat black holes ({\it i.e.}, $\Lambda = 0$), while the bottom row represents AdS black holes with $\Lambda = -1$, the most negative cosmological constant among the plots. The middle row corresponds to an intermediate case with $\Lambda = -1/2$. The colored regions indicate $\kappa^2 - \lambda^2 < 0$, signaling a violation of the bound on chaos. The gray regions depict $\kappa^2 - \lambda^2 \geq 0$, signifying that the bound on chaos is satisfied, and the white regions denote the absence of chaotic behavior.
    
    The effective Lagrangian \eqref{effective_lagrangian} is invariant under the reversal of the particle's angular momentum $(L \to -L)$; therefore, we can consider only positive values of $L$ in the GMGHS black hole cases. The top-left plot depicts a pure Schwarzschild black hole, while the top-right plot illustrates a GMGHS black hole with an electric charge $Q = 1$. The bottom-left corner represents a Schwarzschild-AdS black hole, and the bottom-right corner corresponds to a GMGHS--AdS black hole with a cosmological constant $\Lambda = -1$. In each column, from left to right, the electric charge takes values $Q = 0$, $Q = -0.5$, and $Q = -1$, respectively. Fig. \ref{GMGHSm0} presents the results of the massless particle $( m = 0 )$, while Fig. \ref{GMGHSm1} presents those of the massive particle $( m = 1 )$. The plots in the leftmost column correspond to the Schwarzschild(-AdS) black hole.

    The middle column in Fig. \ref{GMGHSm1} examines the influence of the cosmological constant. In the plot for the asymptotically flat black hole (top plot), the bound is satisfied in nearly the entire parameter region, with a small region in the bottom-right quadrant depicting a mild violation. Introducing the cosmological constant reduces the size of the chaotic region (gray region), while the area of the violation region (colored region) increases. This indicates that the cosmological constant reduces the parameter space where chaos is observed, and, as in Eq. \eqref{Lambda_infinity}, the gray region emerges at large $q$ values when the limit of the cosmological constant tends to infinity.
    
    Furthermore, a large electric charge $Q$ on the black hole leads to a gradual expansion of the region where the Lyapunov bound is violated. As Fig. \ref{GMGHSm1} illustrates, the parameter space satisfying $\kappa^2 \geq \lambda^2$ shrinks with increasing $Q$; moreover, the area of the violation region and the magnitude of its deviation from the bound increase. The bound holds across nearly the entire domain for neutral black holes; however, prominent and increasingly widespread regions of violation appear with increasing electric charge, particularly where the black hole and the particle charges share the same sign, {\it i.e.}, $Qq > 0$. Fig. \ref{GMGHSm0} illustrates a similar trend, indicating the consistency of this behavior under variations in the cosmological constant and black hole charge.

\subsection{Kerr--Sen--AdS Black Hole}
    We investigated the effects of black hole rotation on the bound on chaos, extending the previous analysis to the Kerr--Sen--AdS black hole spacetime. In this rotating case, the geometry introduces frame-dragging and axial asymmetry, considerably altering the structure of the effective potential that governs particle dynamics.

    \noindent \textbf{\textit{Asymptotic limit of cosmological constant}}---In the asymptotic limit of the cosmological constant $(\Lambda \to -\infty)$, the effective potential is given by
\begin{equation}
    \left. V(r) \right|_{\Lambda \to -\infty} = - \sqrt{\frac{L^2 \left( r^2 + 2 b r \right)}{3 \left( r^2 + 2 b r + a^2 \right)}} \sqrt{-\Lambda} + \mathcal{O}\left( \frac{1}{\sqrt{-\Lambda}} \right).
\end{equation}
    When the cosmological constant becomes highly negative, the effective potential monotonically decreases without developing a local maximum. Consequently, no unstable orbit forms, precluding the possibility of chaotic motion in this regime.
    
    Beyond this analytic insight, we conducted a numerical analysis to further investigate the behavior of the Lyapunov exponent. The results are presented in the following figures.
    
\begin{figure}[H]
    \centering
    \begin{tikzpicture}
        \matrix (M) [matrix of nodes, nodes={inner sep=0cm, outer sep=0.15cm, anchor=center}, row sep=0cm, column sep=0cm]
        {
            \includegraphics[width=5.2cm]{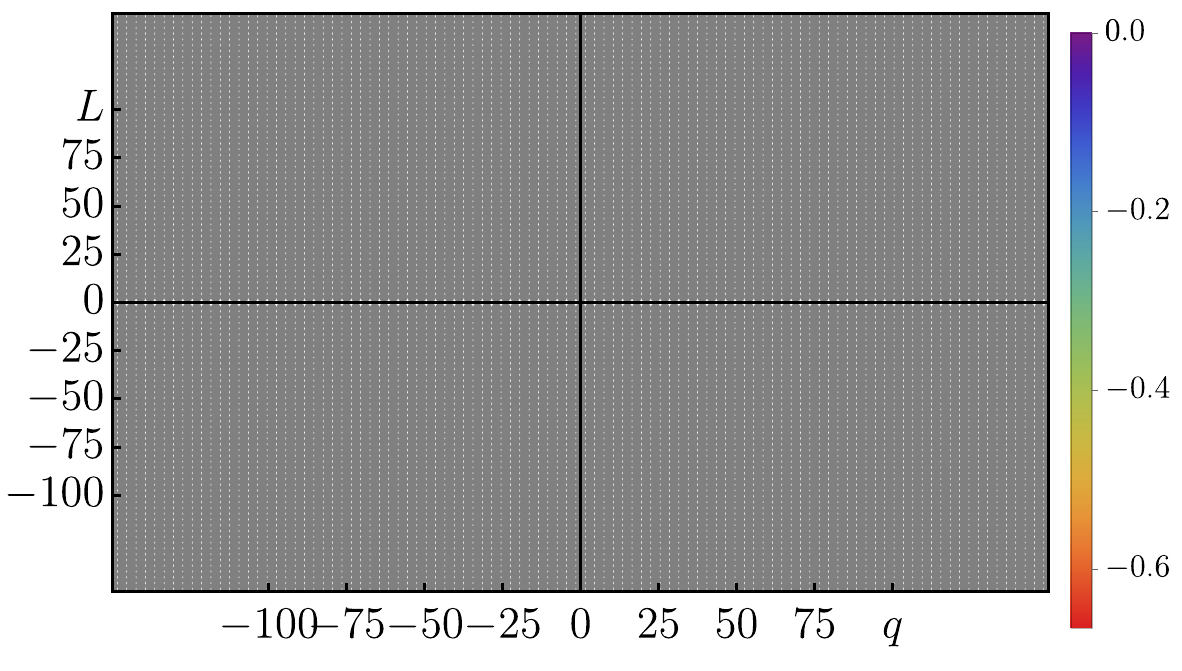} &
            \includegraphics[width=5.2cm]{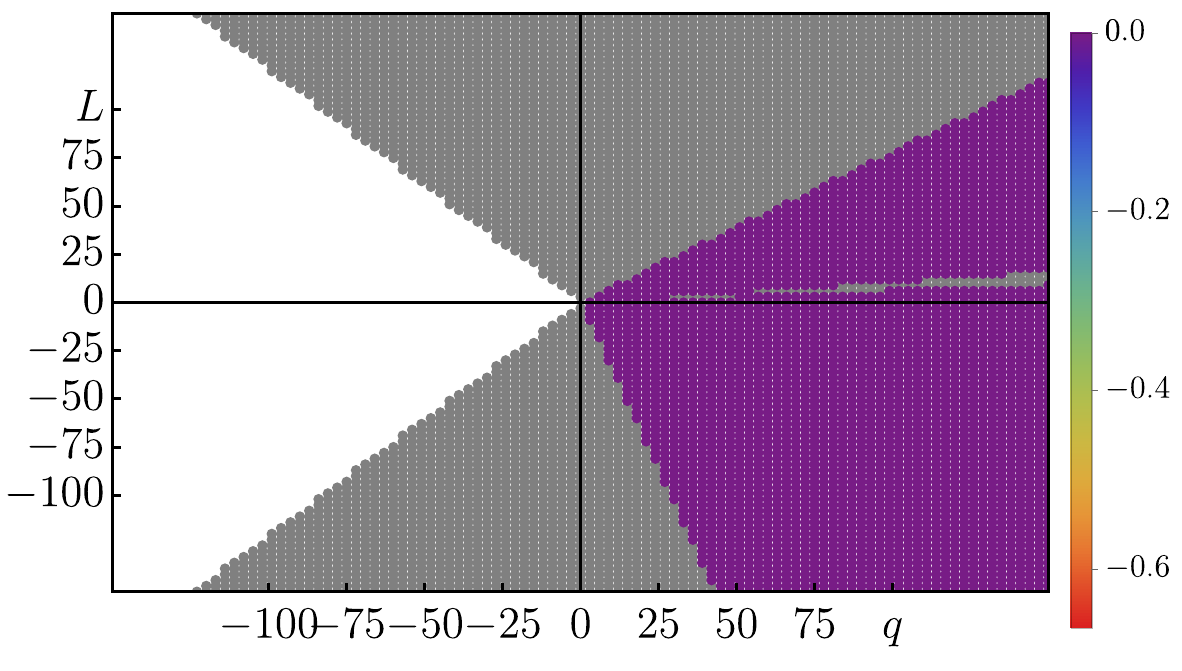} &
            \includegraphics[width=5.2cm]{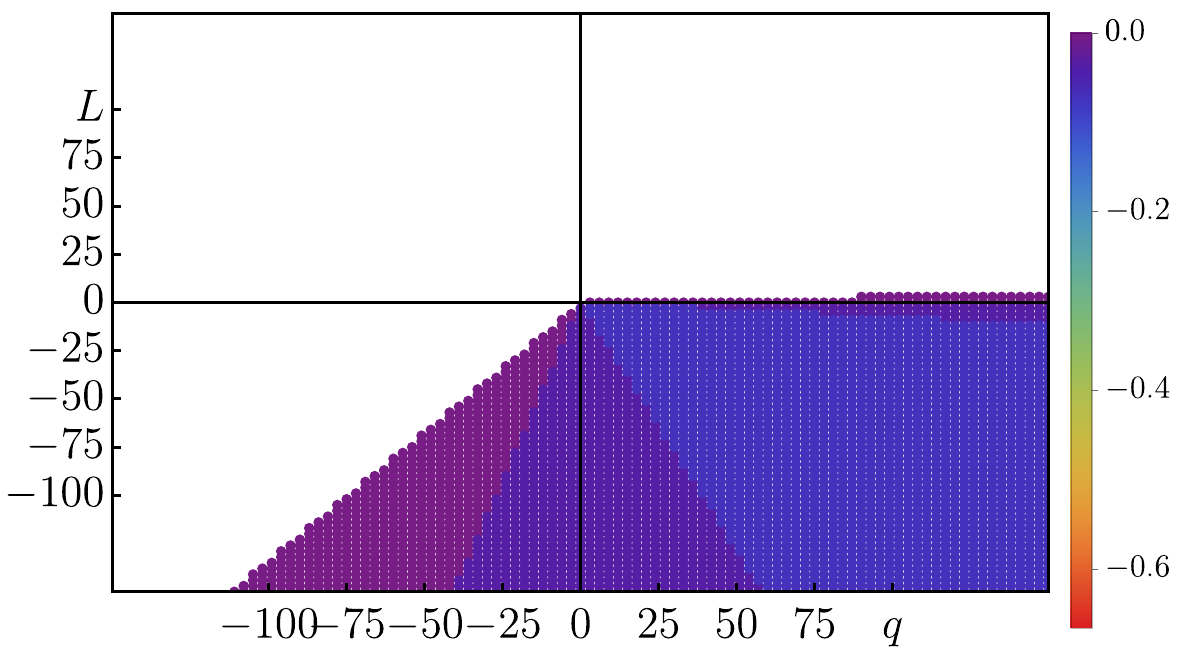} \\
            \includegraphics[width=5.2cm]{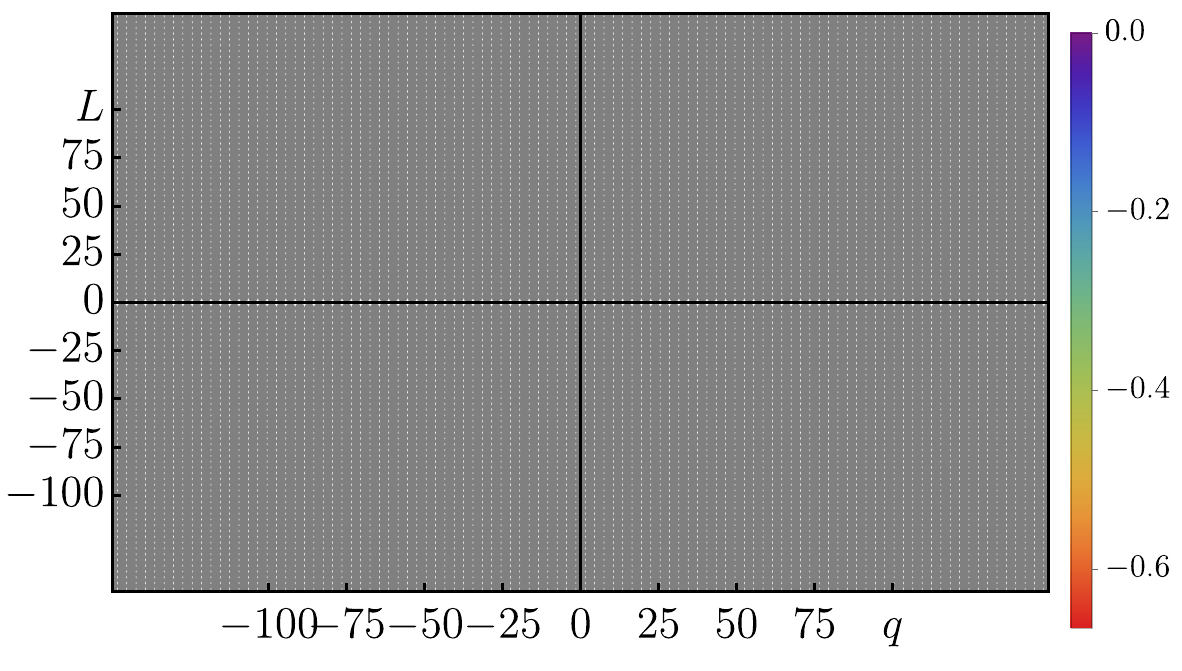} &
            \includegraphics[width=5.2cm]{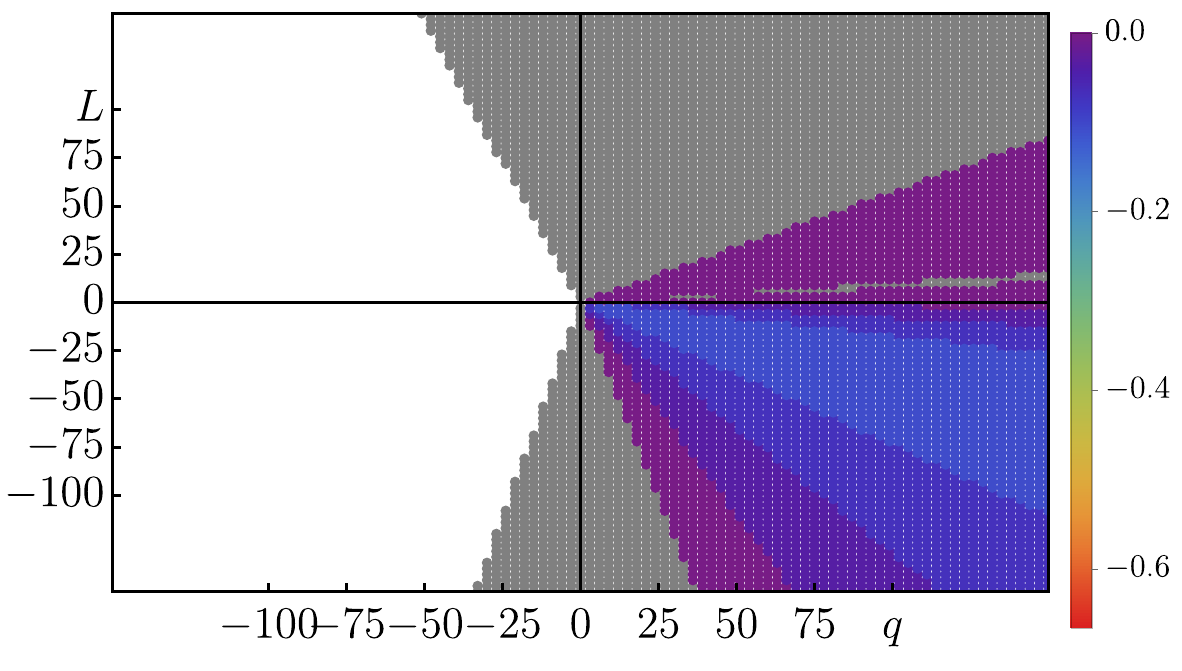} &
            \includegraphics[width=5.2cm]{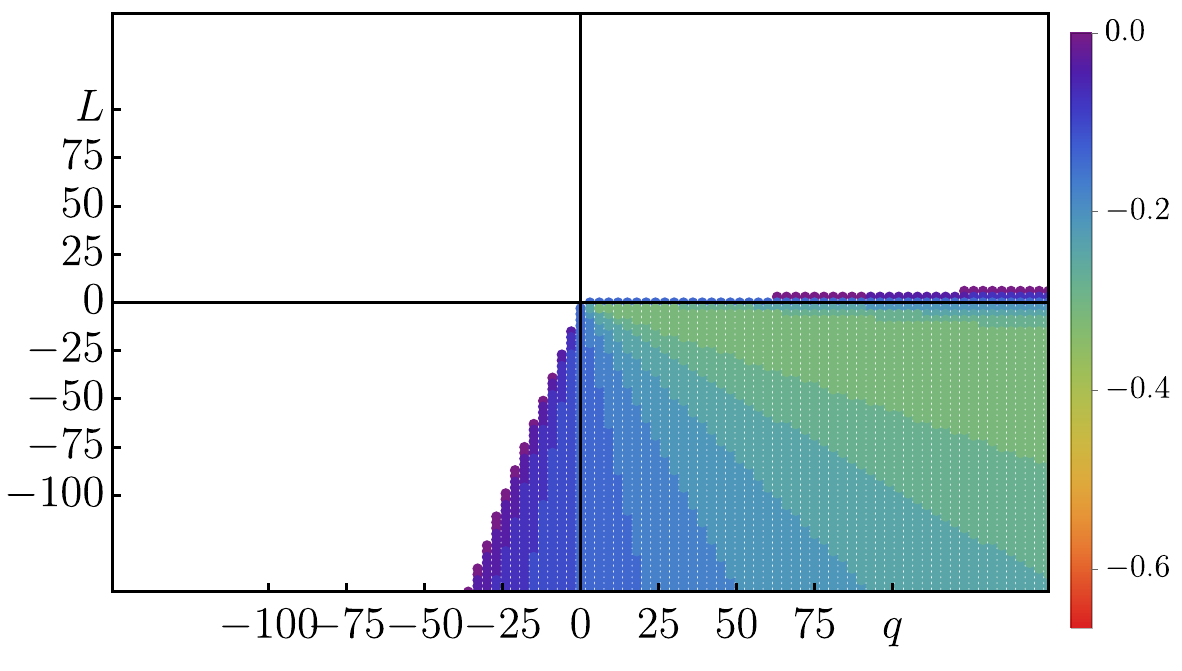} \\
            \includegraphics[width=5.2cm]{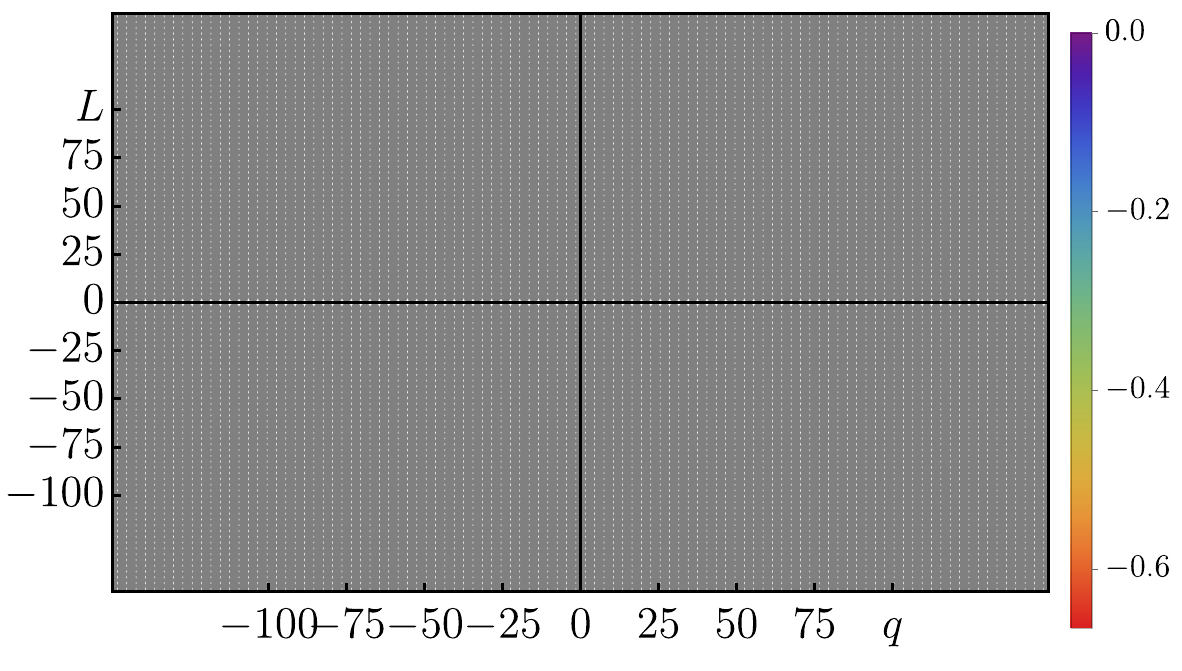} &
            \includegraphics[width=5.2cm]{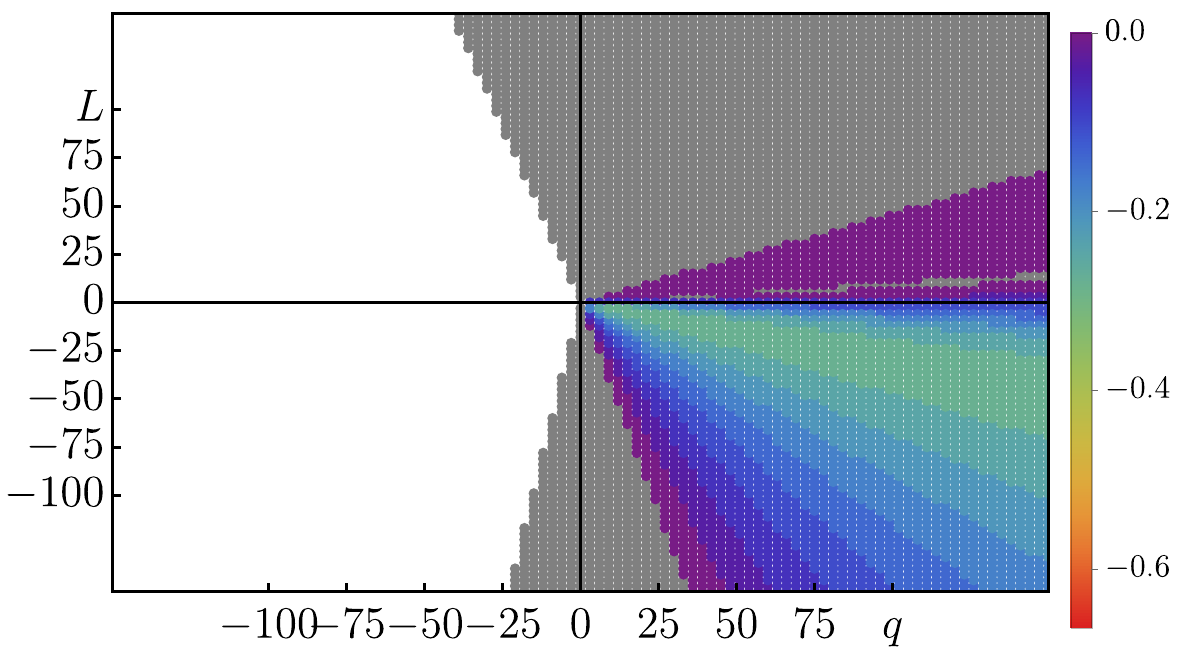} &
            \includegraphics[width=5.2cm]{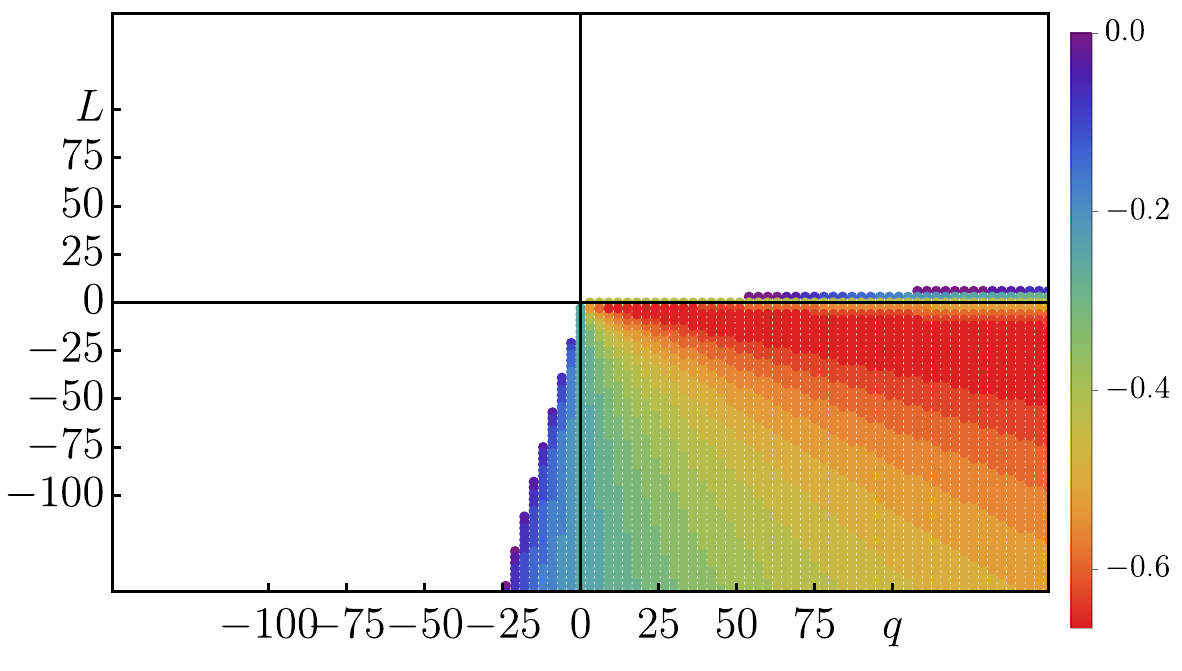} \\
        };

        \coordinate (O) at (M-1-1.north west);
        \coordinate (R) at (M-1-3.north east);
        \coordinate (B) at (M-3-1.south west);

        \draw[->, thick] (O) -- ($(R)$);
        \foreach \col/\xlab in {1/0, 2/0.9 Q_\mathrm{Max}, 3/Q_\mathrm{Max}} {\node[above=0cm, font=\fontsize{10pt}{10pt}\selectfont] at (M-1-\col.north) {\(\xlab\)};}
        \node[above=0cm, font=\fontsize{10pt}{10pt}\selectfont] at ($(R)$) {\(Q\)};

        \draw[->, thick] (O) -- ($(B)$);
        \foreach \row/\ylab in {1/0, 2/\mathrm{-}0.5, 3/\mathrm{-}1.0} {\node[left=0cm, font=\fontsize{10pt}{10pt}\selectfont] at (M-\row-1.west) {\(\ylab\)};}
        \node[left=0cm, font=\fontsize{10pt}{10pt}\selectfont] at ($(B)$) {\(\Lambda\)};
    \end{tikzpicture}
    \vspace{-1cm}
    \caption{Massless particle ($m = 0$) in the Kerr-Sen-AdS black hole ($a = 0.1$) background.}
    \label{KSAdSa0.1m0}
\end{figure}
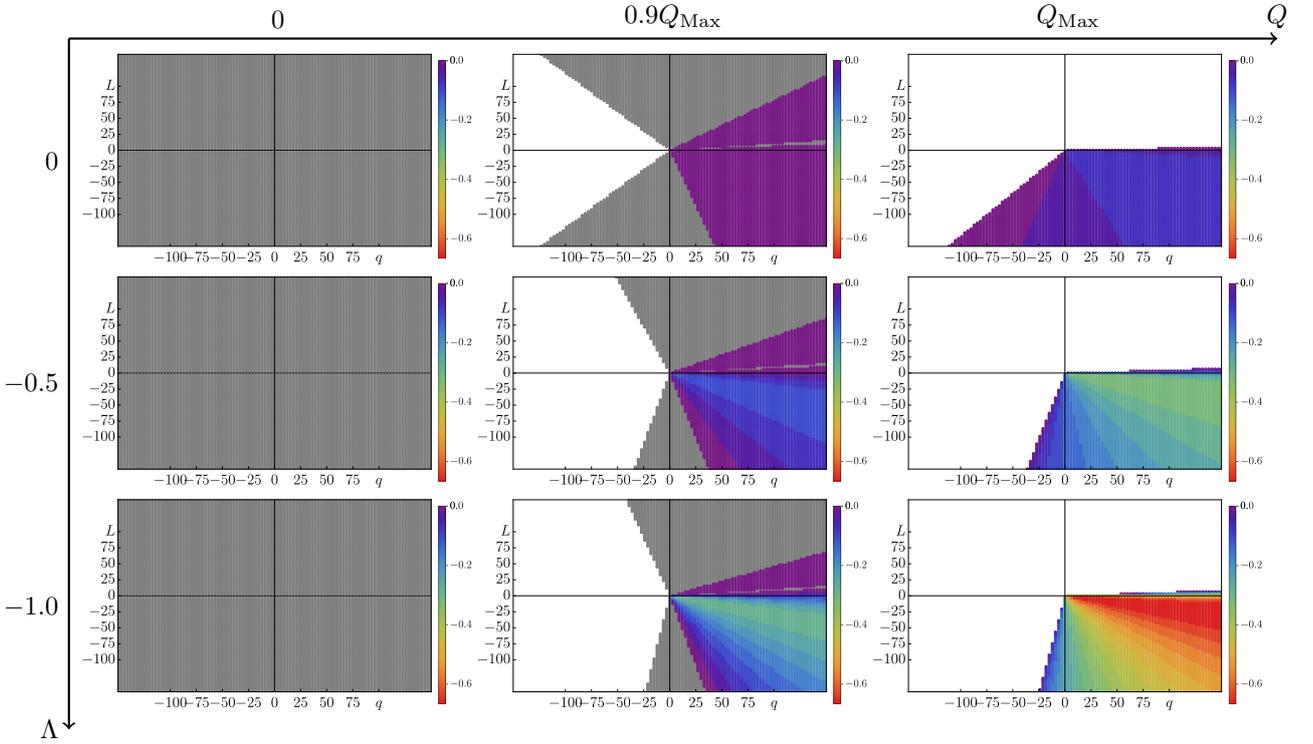
\begin{figure}[H]
    \centering
    \begin{tikzpicture}
        \matrix (M) [matrix of nodes, nodes={inner sep=0cm, outer sep=0.15cm, anchor=center}, row sep=0cm, column sep=0cm]
        {
            \includegraphics[width=5.2cm]{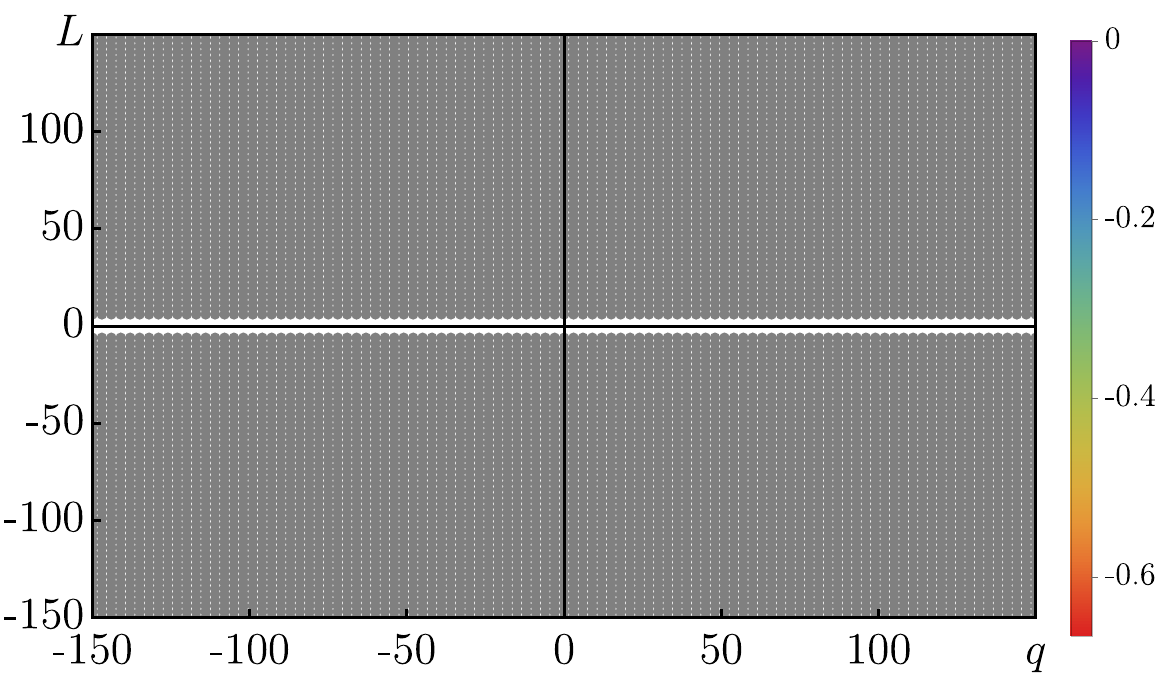} &
            \includegraphics[width=5.2cm]{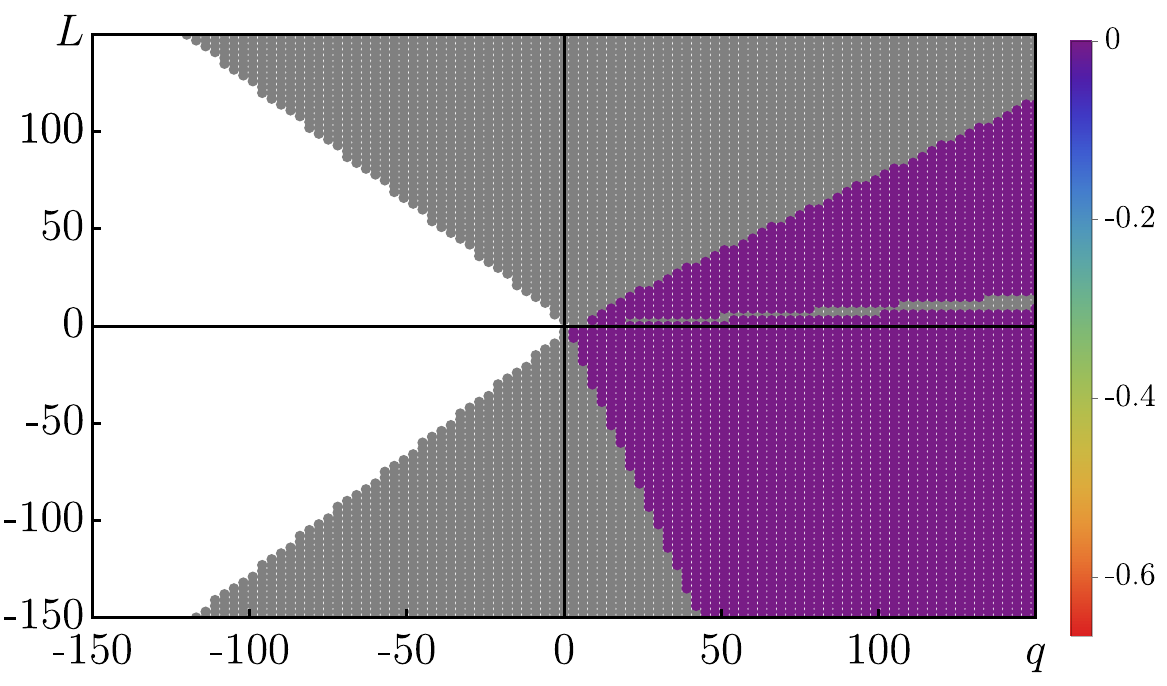} &
            \includegraphics[width=5.2cm]{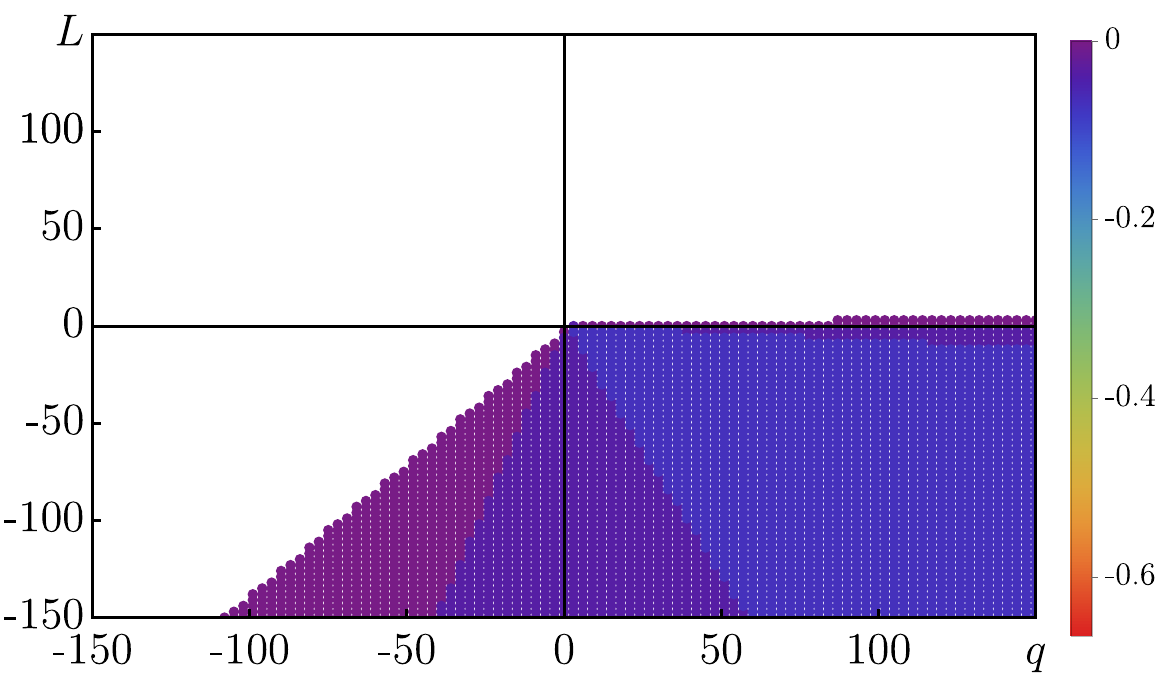} \\
            \includegraphics[width=5.2cm]{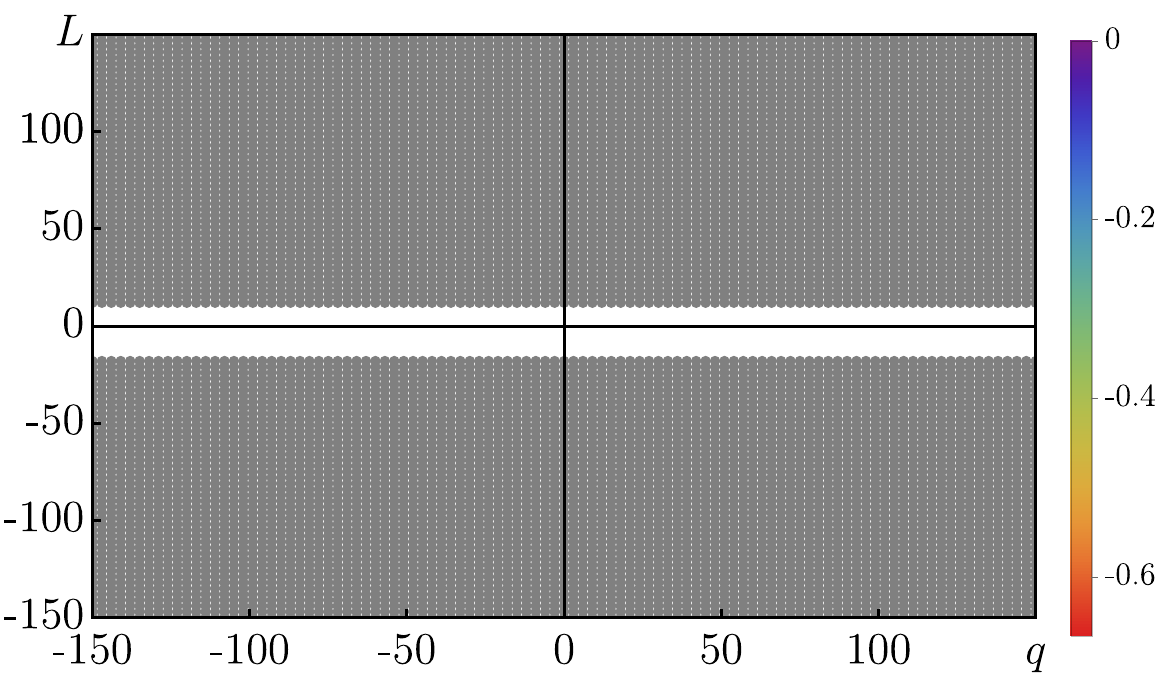} &
            \includegraphics[width=5.2cm]{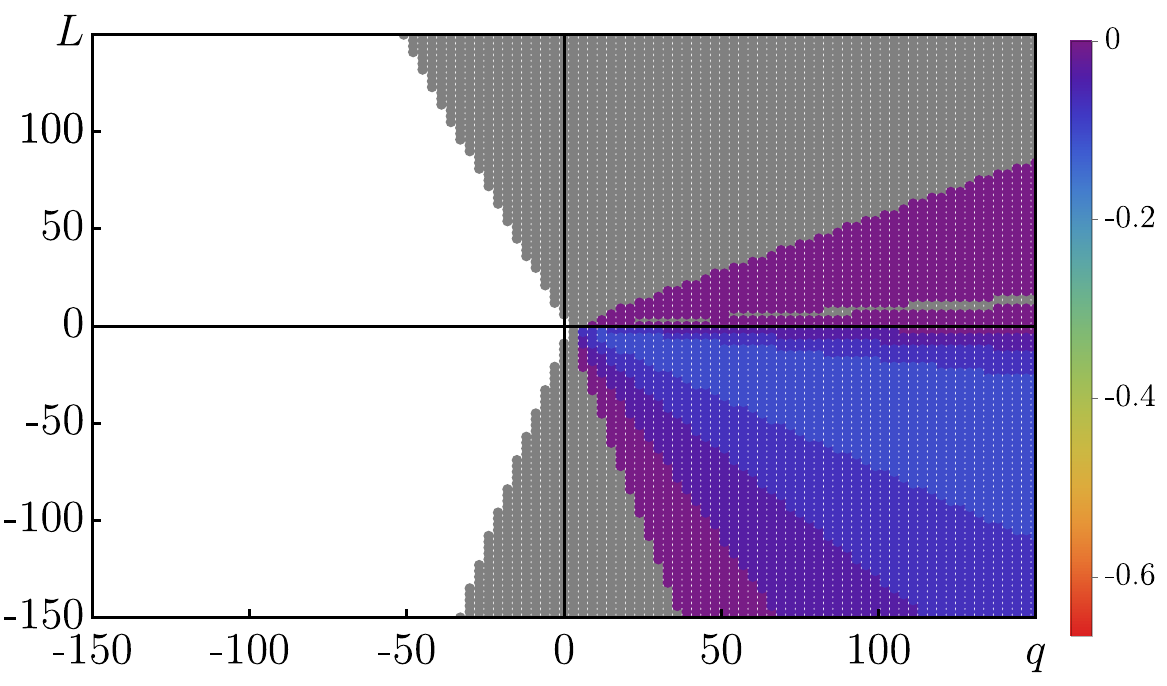} &
            \includegraphics[width=5.2cm]{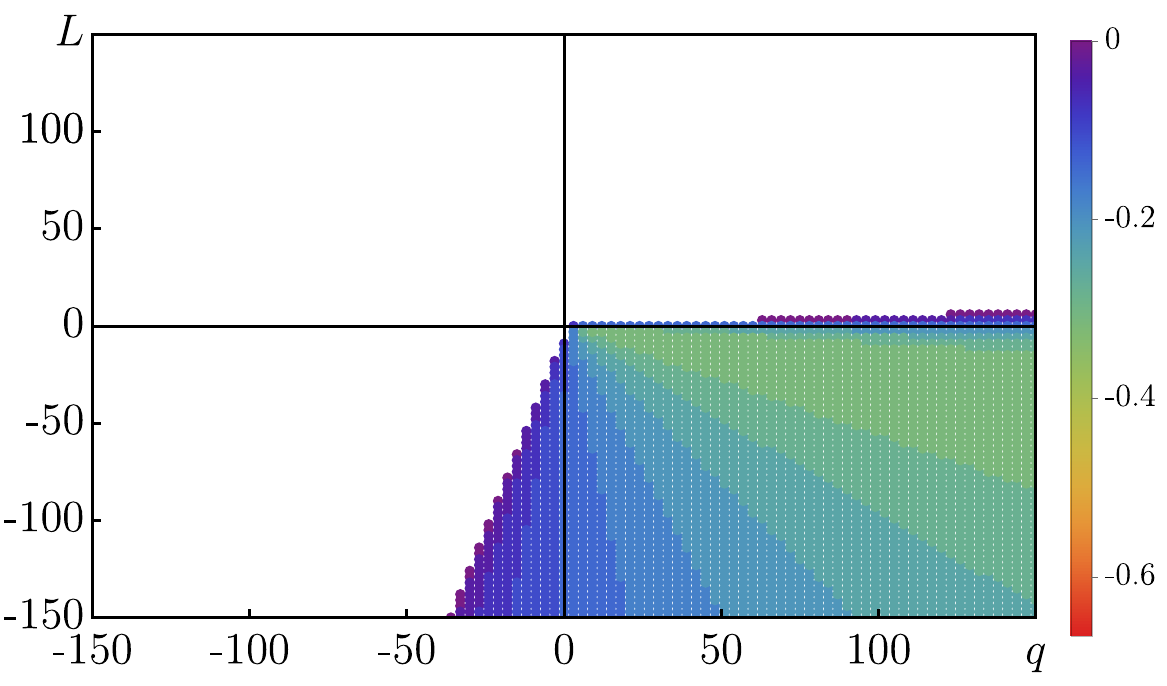} \\
            \includegraphics[width=5.2cm]{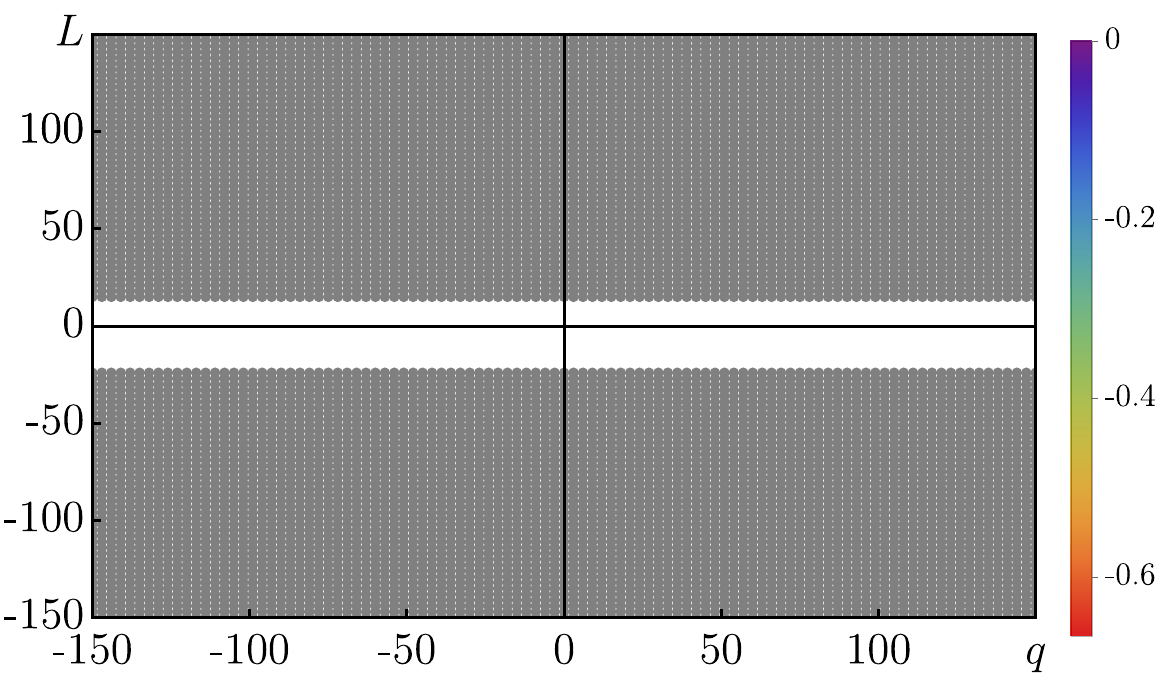} &
            \includegraphics[width=5.2cm]{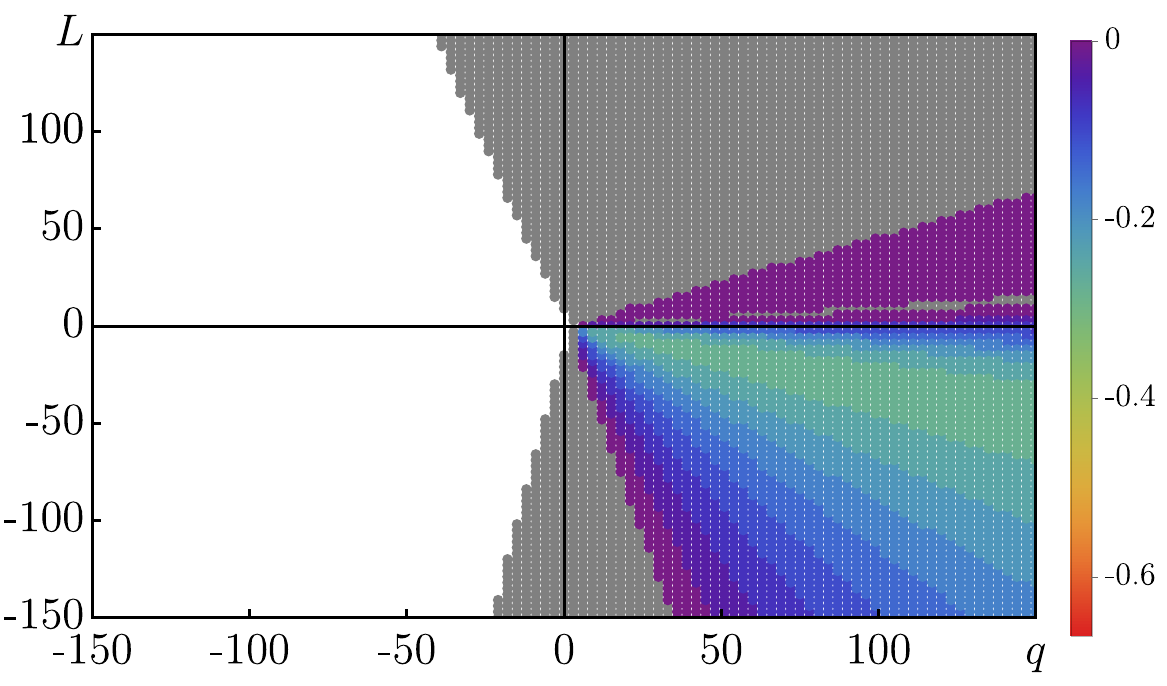} &
            \includegraphics[width=5.2cm]{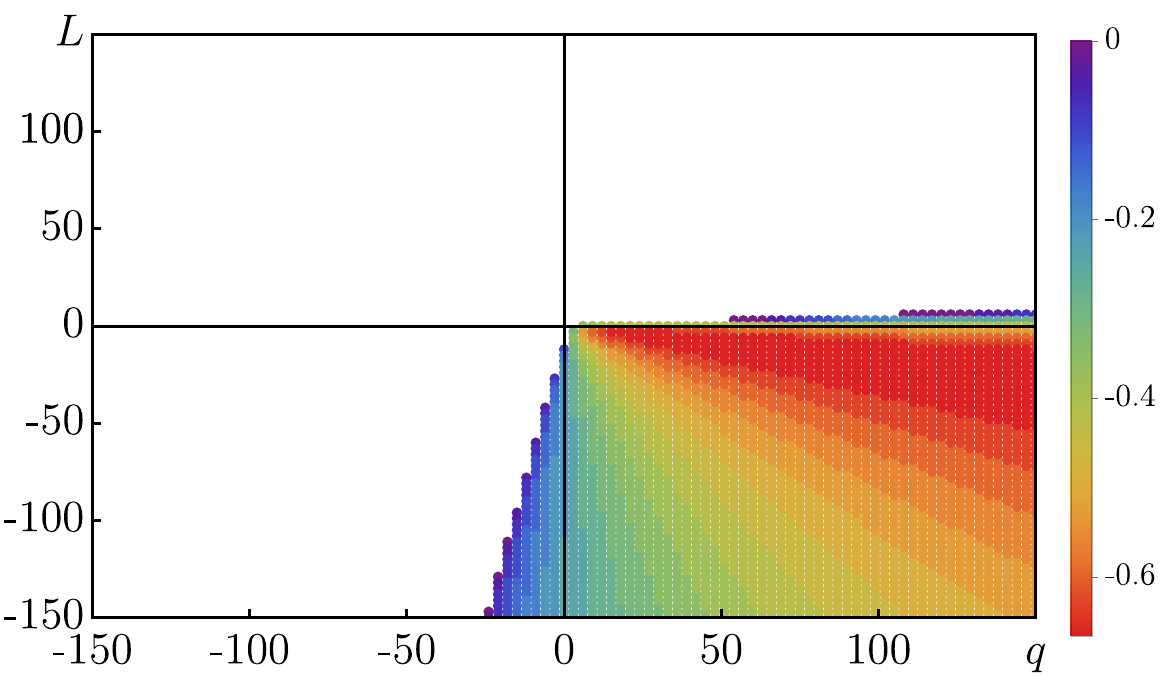} \\
        };

        \coordinate (O) at (M-1-1.north west);
        \coordinate (R) at (M-1-3.north east);
        \coordinate (B) at (M-3-1.south west);

        \draw[->, thick] (O) -- ($(R)$);
        \foreach \col/\xlab in {1/0, 2/0.9 Q_\mathrm{Max}, 3/Q_\mathrm{Max}} {\node[above=0cm, font=\fontsize{10pt}{10pt}\selectfont] at (M-1-\col.north) {\(\xlab\)};}
        \node[above=0cm, font=\fontsize{10pt}{10pt}\selectfont] at ($(R)$) {\(Q\)};

        \draw[->, thick] (O) -- ($(B)$);
        \foreach \row/\ylab in {1/0, 2/\mathrm{-}0.5, 3/\mathrm{-}1.0} {\node[left=0cm, font=\fontsize{10pt}{10pt}\selectfont] at (M-\row-1.west) {\(\ylab\)};}
        \node[left=0cm, font=\fontsize{10pt}{10pt}\selectfont] at ($(B)$) {\(\Lambda\)};
    \end{tikzpicture}
    \vspace{-1cm}
    \caption{Massive particle ($m = 1$) in the Kerr-Sen-AdS black hole ($a = 0.1$) background.}
    \label{KSAdSa0.1m1}
\end{figure}
\begin{figure}[H]
    \centering
    \begin{tikzpicture}
        \matrix (M) [matrix of nodes, nodes={inner sep=0cm, outer sep=0.15cm, anchor=center}, row sep=0cm, column sep=0cm]
        {
            \includegraphics[width=5.2cm]{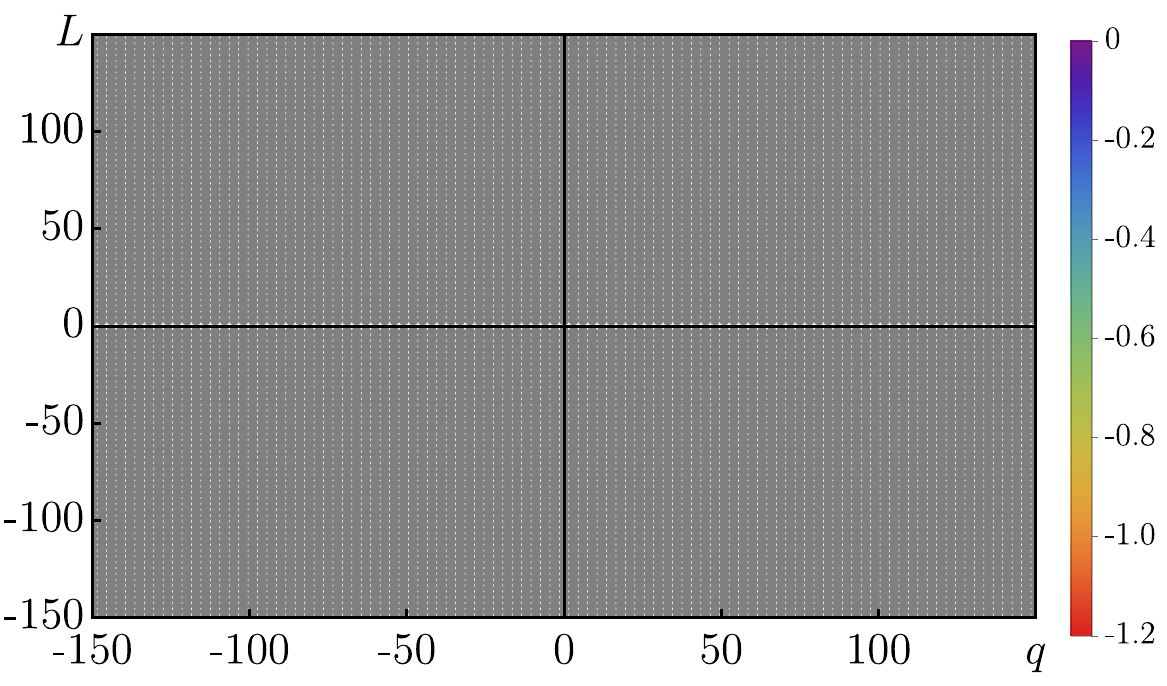} &
            \includegraphics[width=5.2cm]{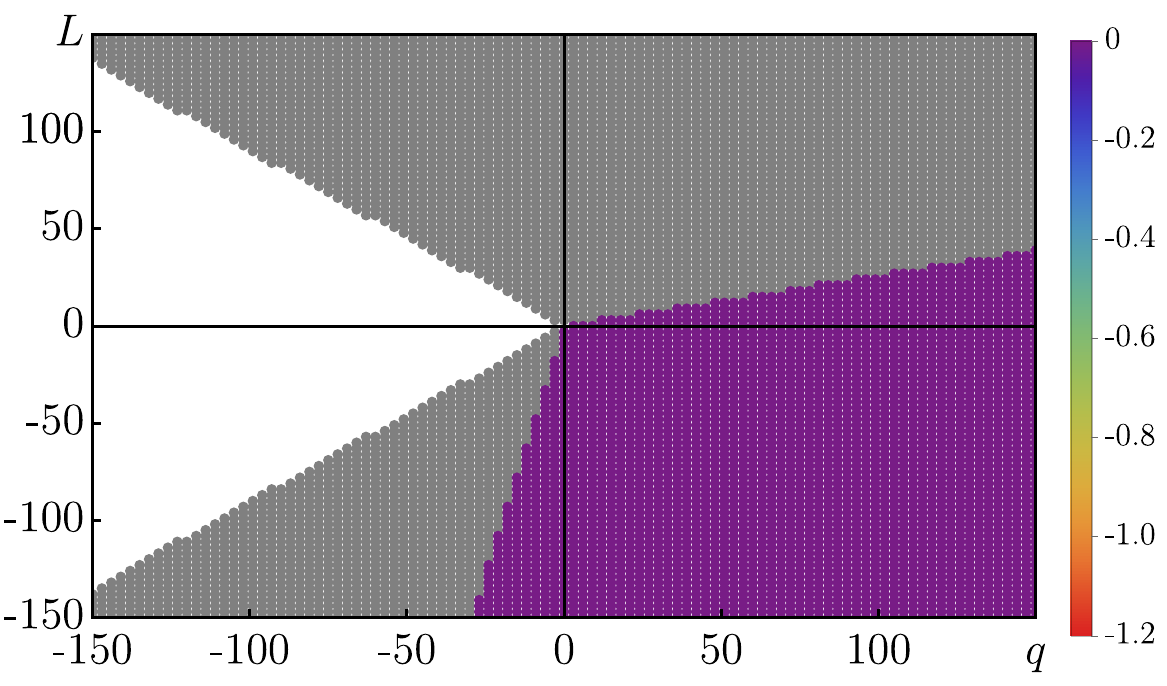} &
            \includegraphics[width=5.2cm]{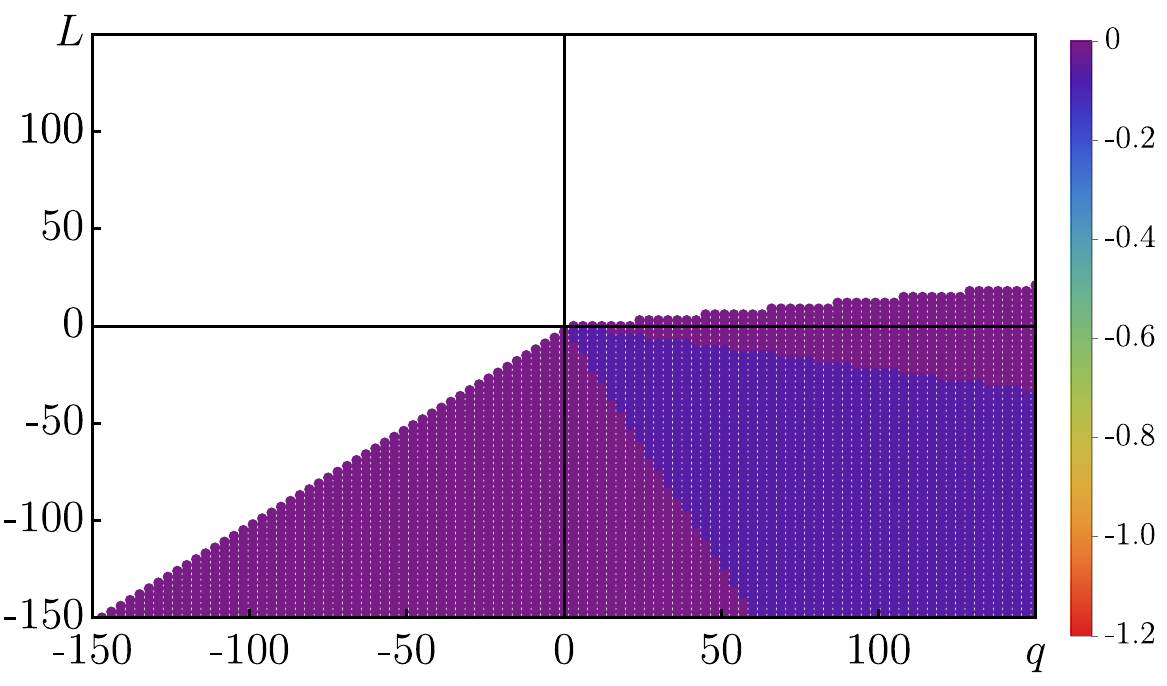} \\
            \includegraphics[width=5.2cm]{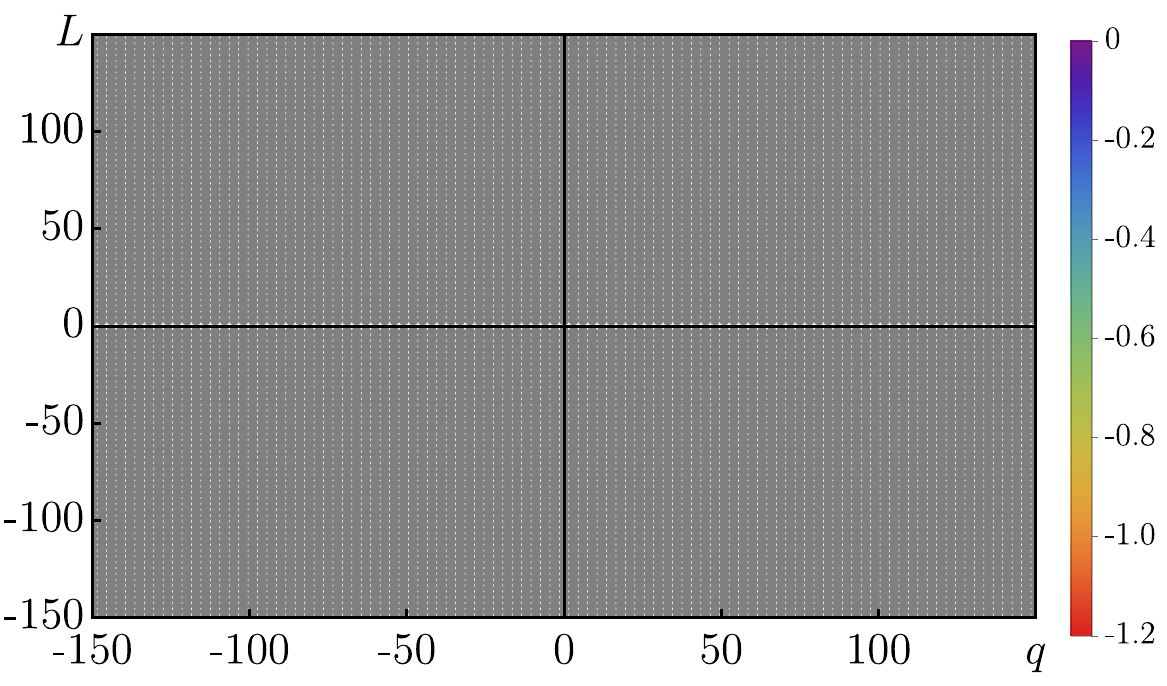} &
            \includegraphics[width=5.2cm]{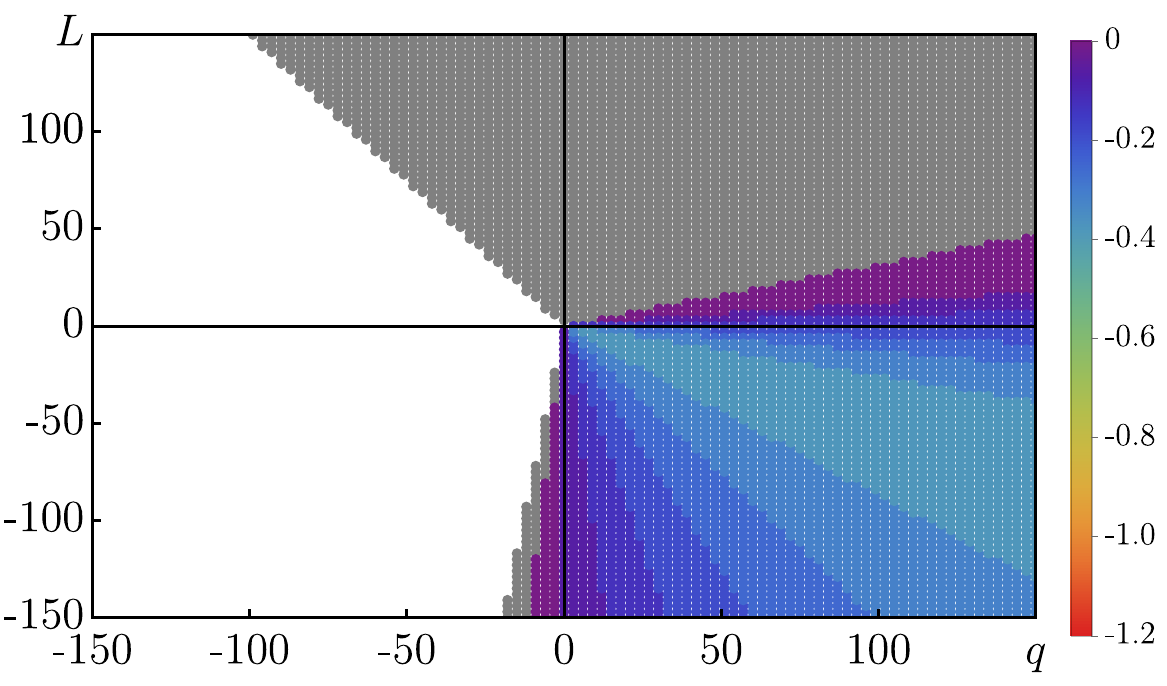} &
            \includegraphics[width=5.2cm]{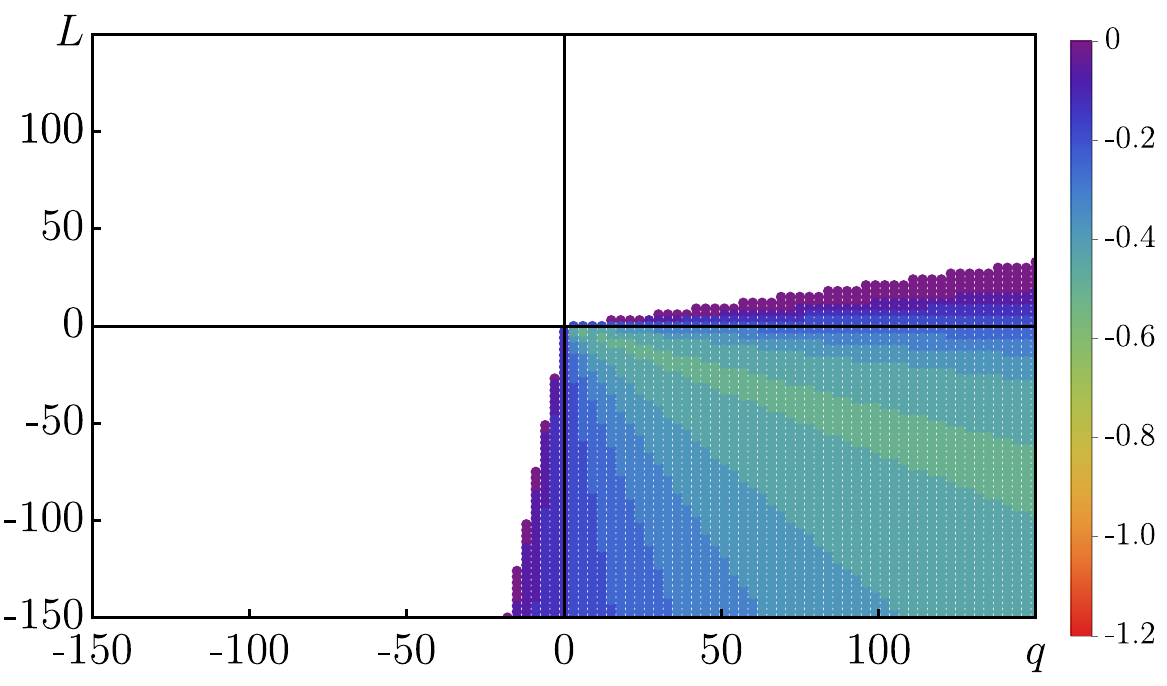} \\
            \includegraphics[width=5.2cm]{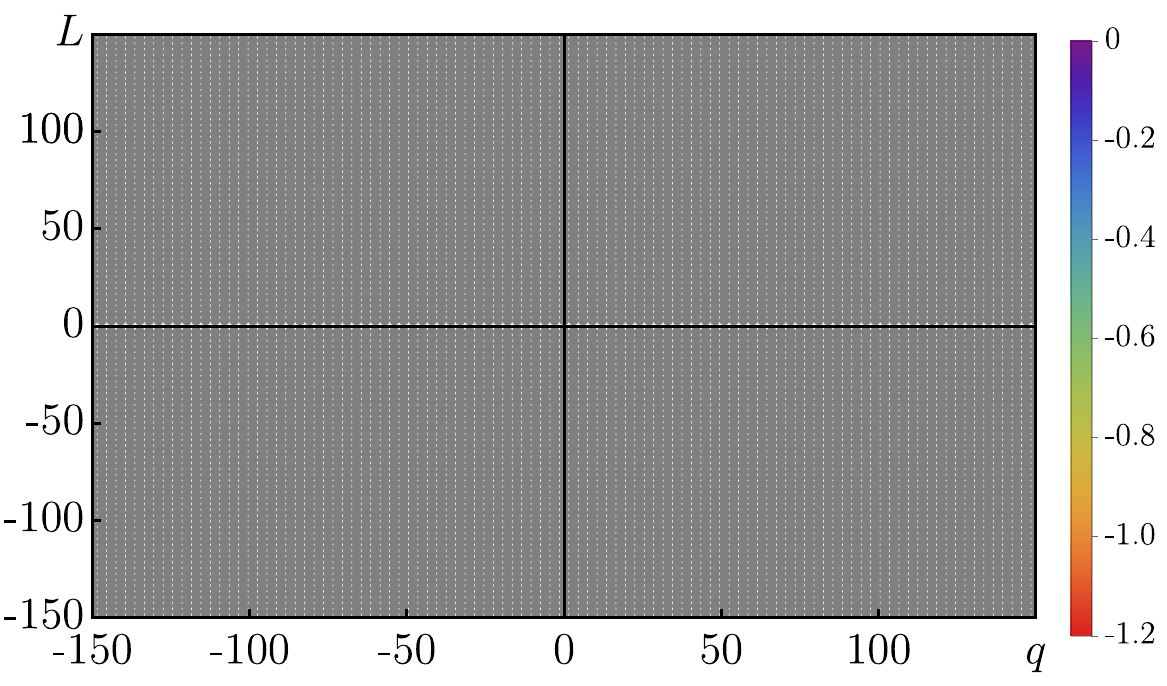} &
            \includegraphics[width=5.2cm]{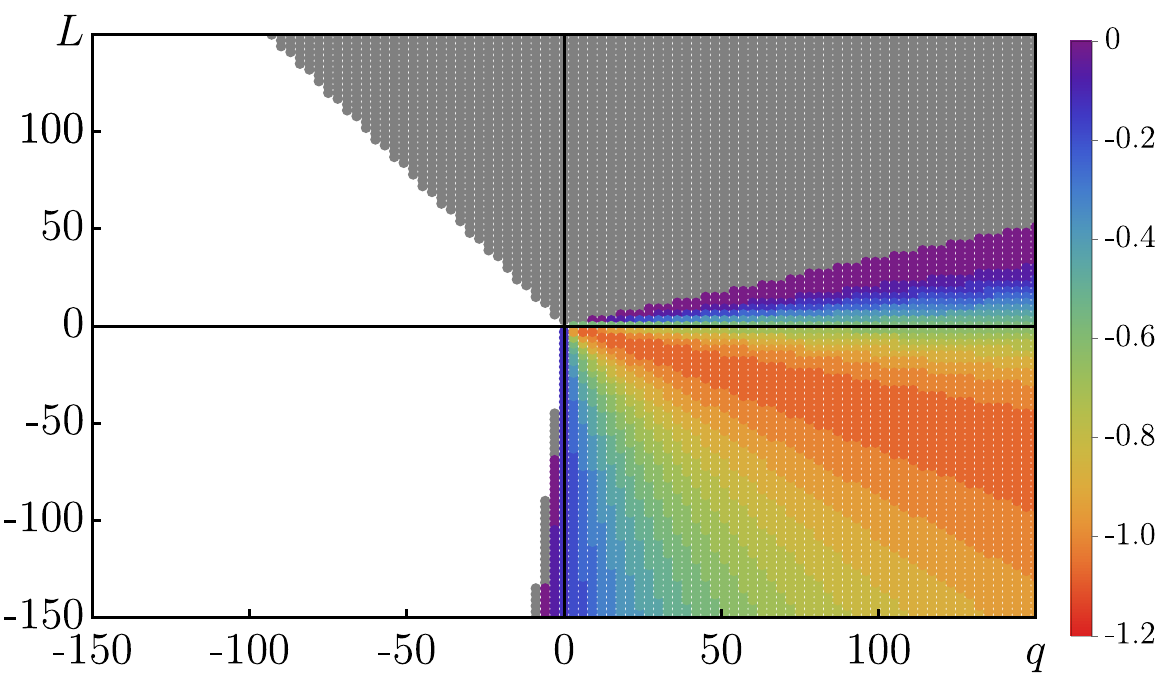} &
            \includegraphics[width=5.2cm]{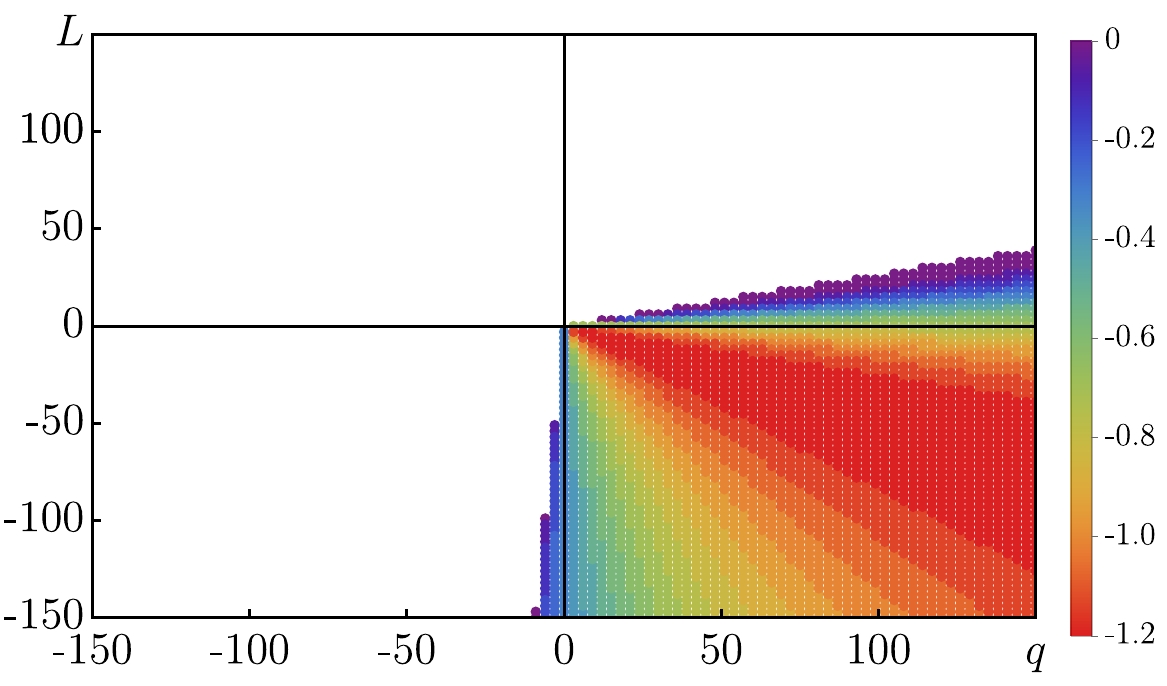} \\
        };

        \coordinate (O) at (M-1-1.north west);
        \coordinate (R) at (M-1-3.north east);
        \coordinate (B) at (M-3-1.south west);

        \draw[->, thick] (O) -- ($(R)$);
        \foreach \col/\xlab in {1/0, 2/0.9 Q_\mathrm{Max}, 3/Q_\mathrm{Max}} {\node[above=0cm, font=\fontsize{10pt}{10pt}\selectfont] at (M-1-\col.north) {\(\xlab\)};}
        \node[above=0cm, font=\fontsize{10pt}{10pt}\selectfont] at ($(R)$) {\(Q\)};

        \draw[->, thick] (O) -- ($(B)$);
        \foreach \row/\ylab in {1/0, 2/\mathrm{-}0.5, 3/\mathrm{-}1.0} {\node[left=0cm, font=\fontsize{10pt}{10pt}\selectfont] at (M-\row-1.west) {\(\ylab\)};}
        \node[left=0cm, font=\fontsize{10pt}{10pt}\selectfont] at ($(B)$) {\(\Lambda\)};
    \end{tikzpicture}
    \vspace{-1cm}
    \caption{Massless particle ($m = 0$) in the Kerr-Sen-AdS black hole ($a = 0.5$) background.}
    \label{KSAdSa0.5m0}
\end{figure}
\begin{figure}[H]
    \centering
    \begin{tikzpicture}
        \matrix (M) [matrix of nodes, nodes={inner sep=0cm, outer sep=0.15cm, anchor=center}, row sep=0cm, column sep=0cm]
        {
            \includegraphics[width=5.2cm]{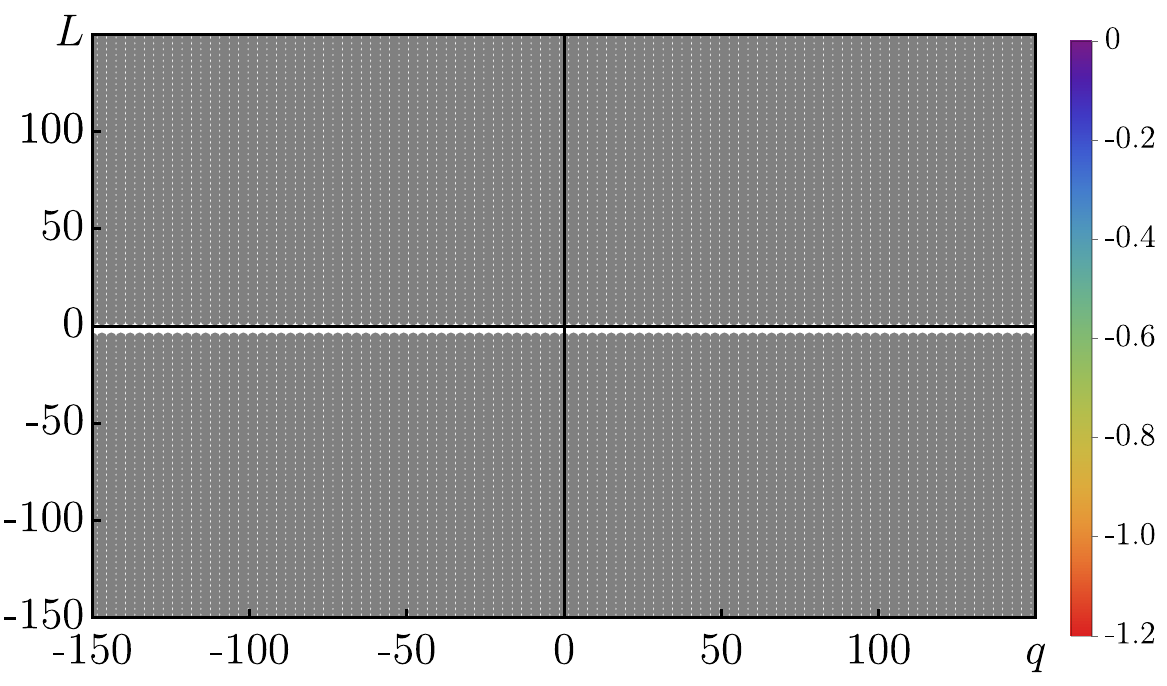} &
            \includegraphics[width=5.2cm]{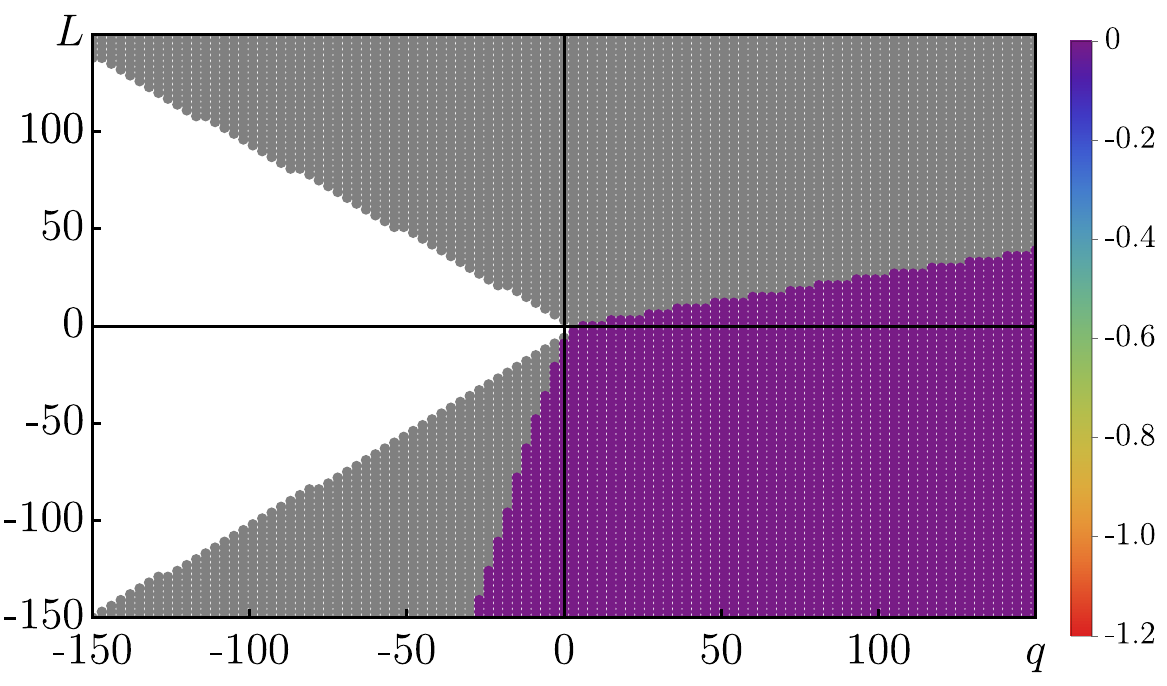} &
            \includegraphics[width=5.2cm]{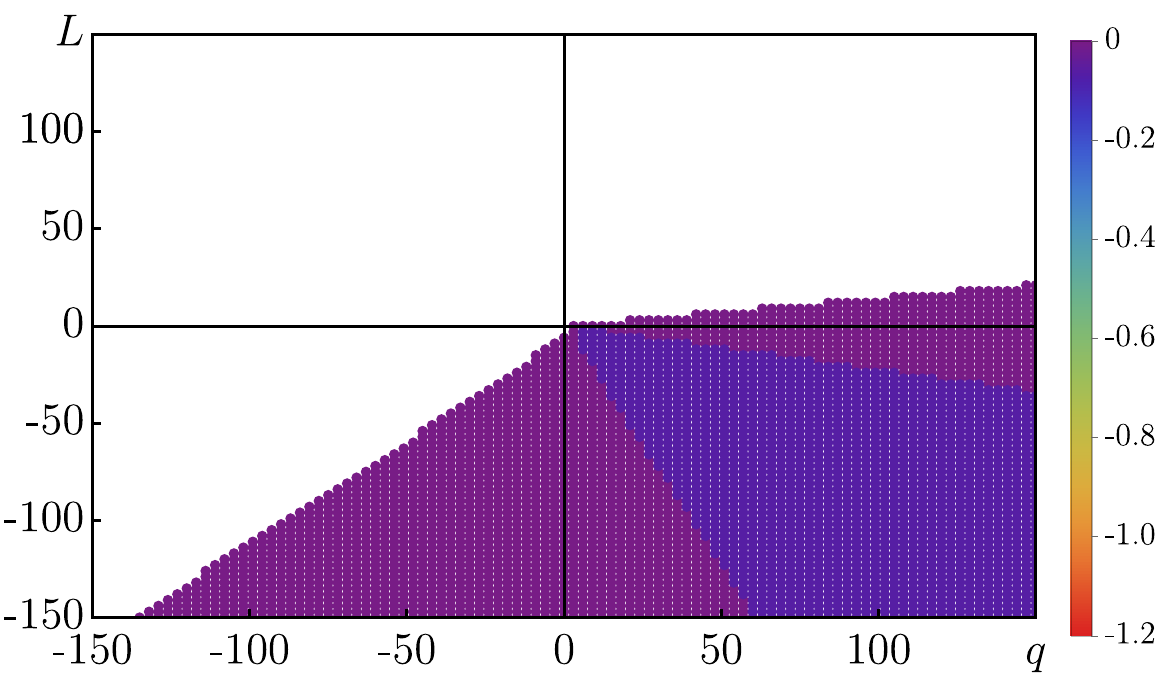} \\
            \includegraphics[width=5.2cm]{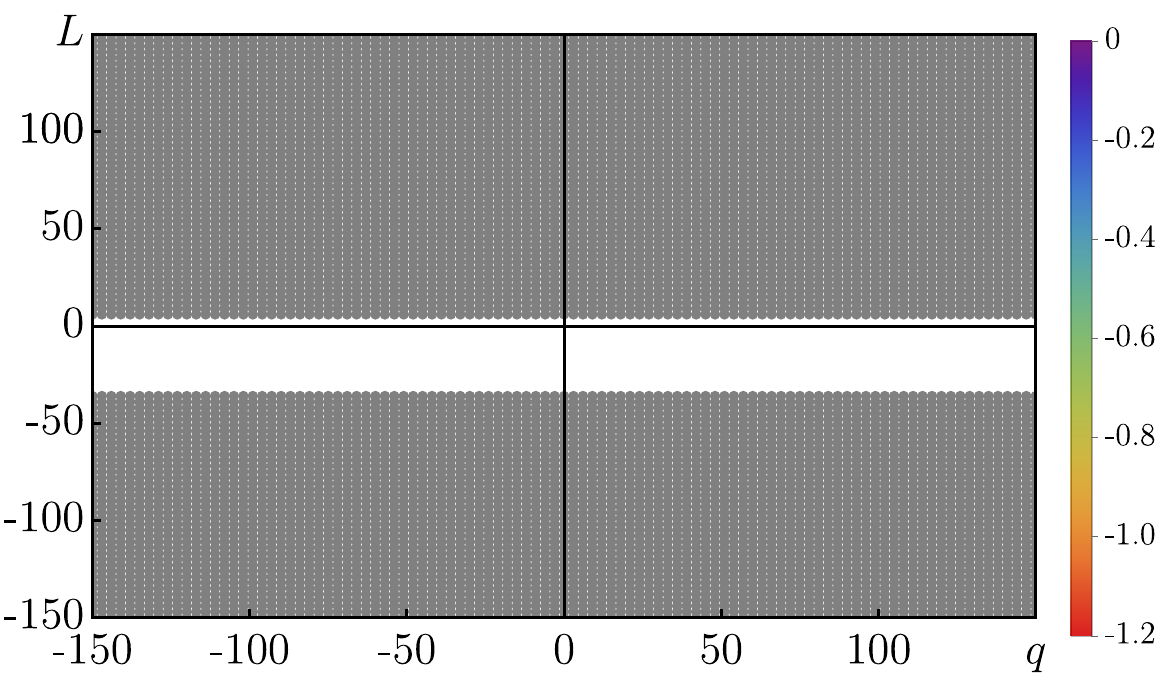} &
            \includegraphics[width=5.2cm]{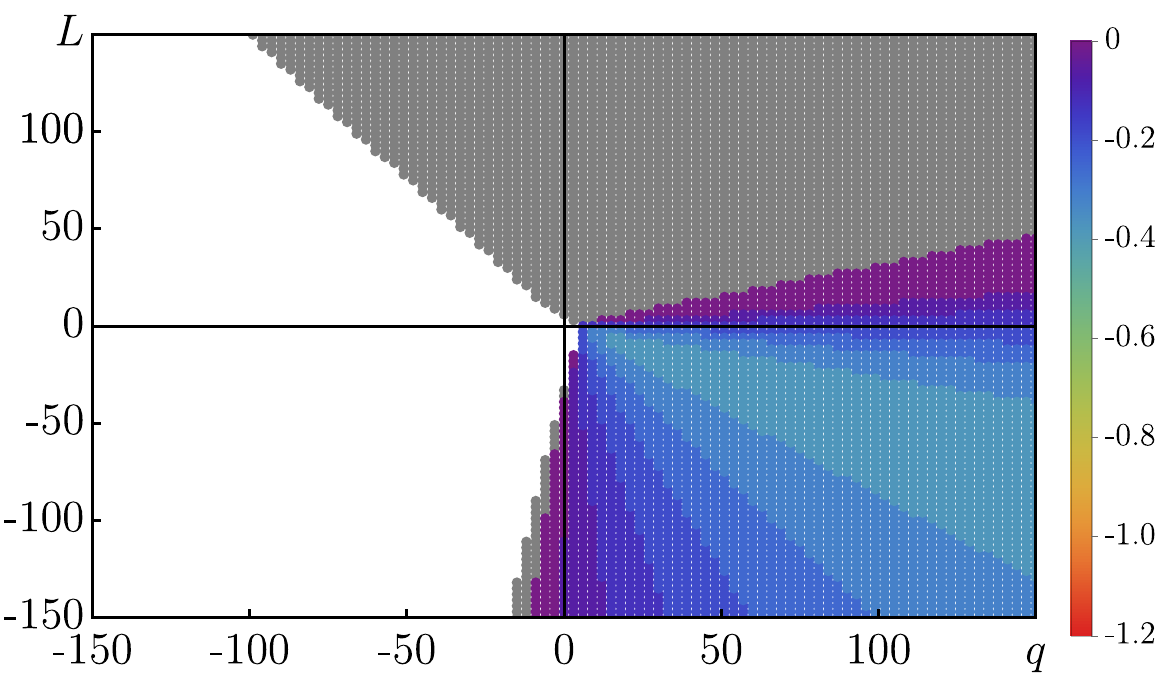} &
            \includegraphics[width=5.2cm]{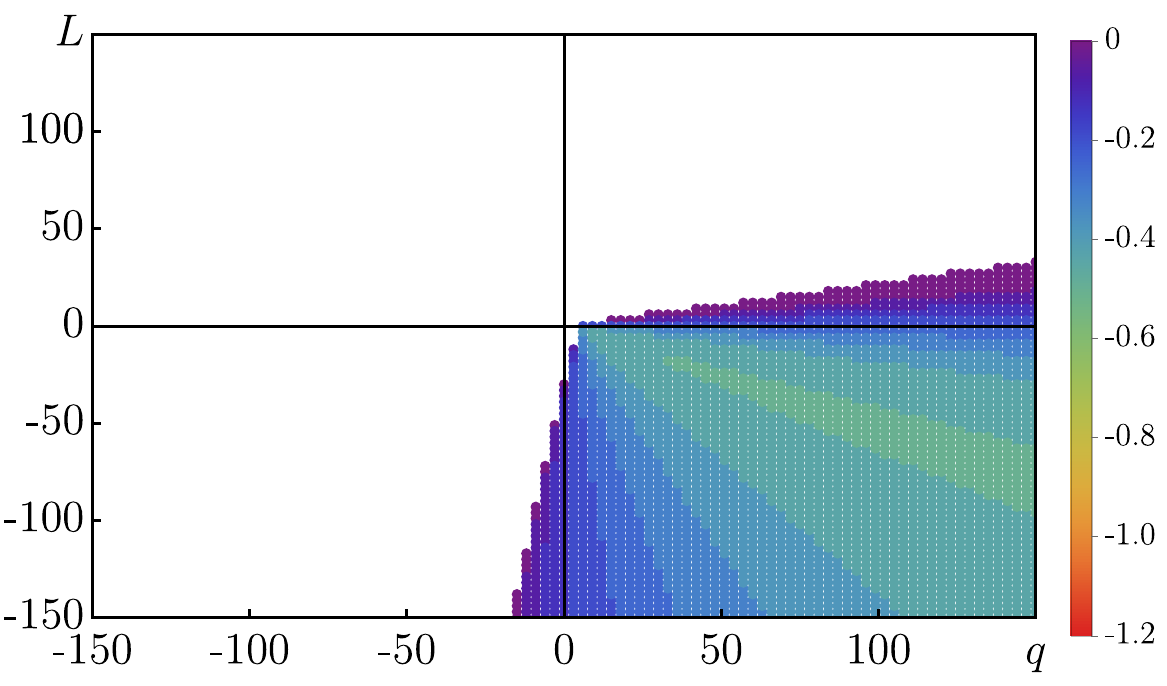} \\
            \includegraphics[width=5.2cm]{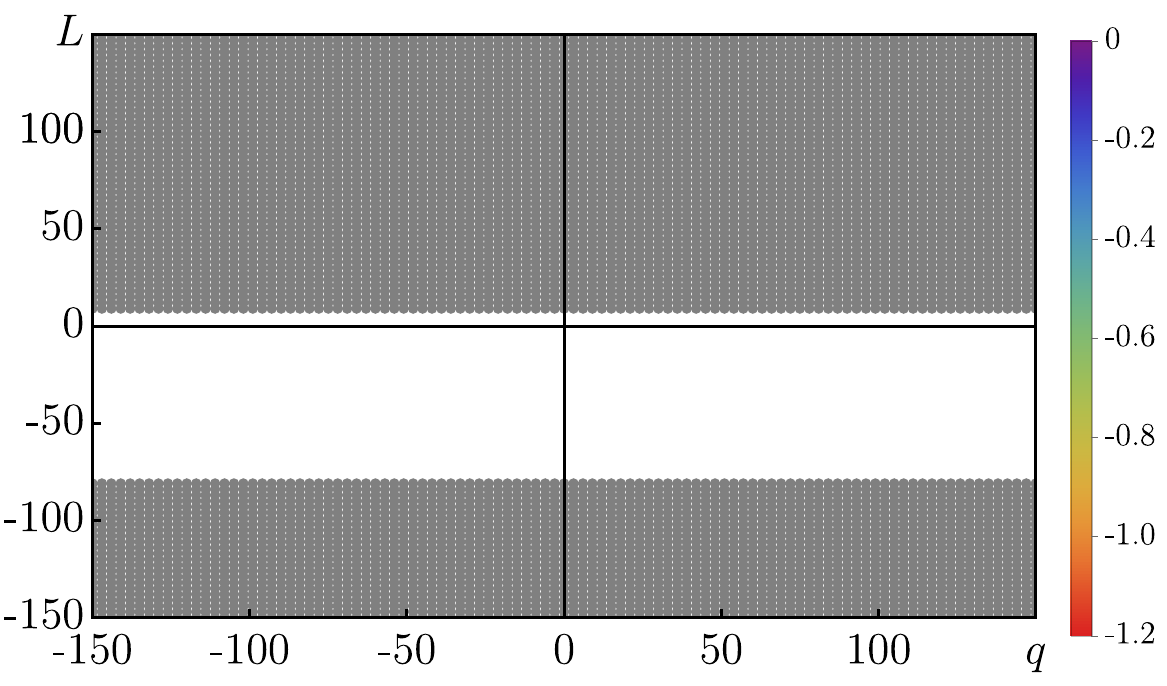} &
            \includegraphics[width=5.2cm]{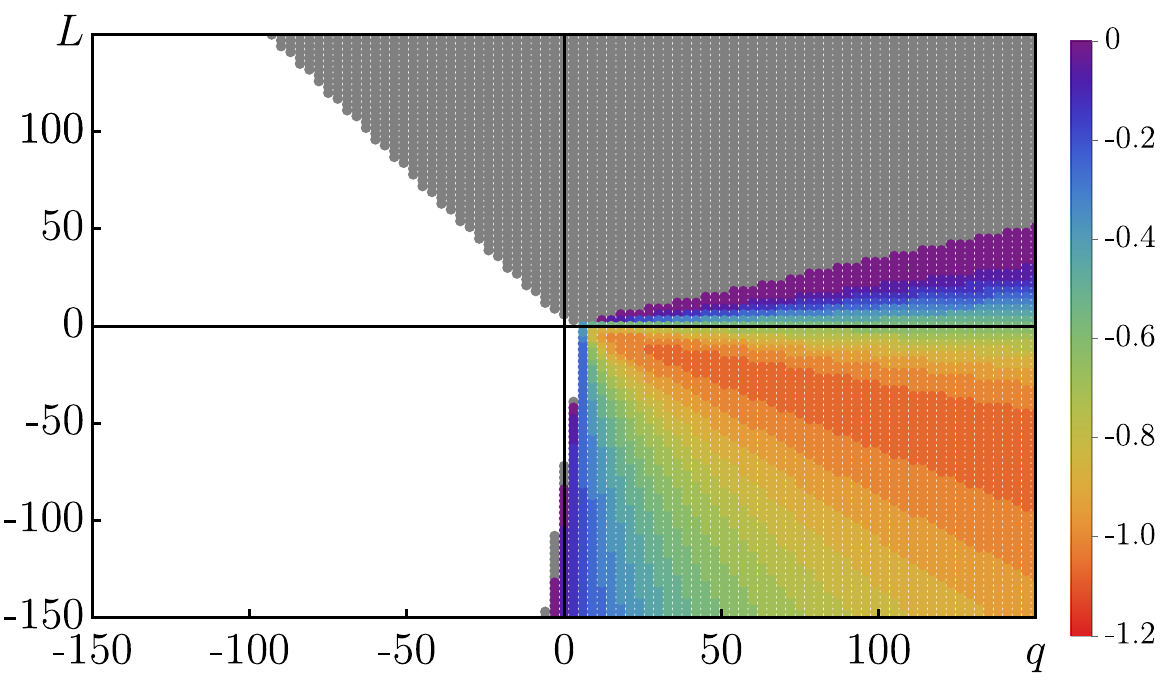} &
            \includegraphics[width=5.2cm]{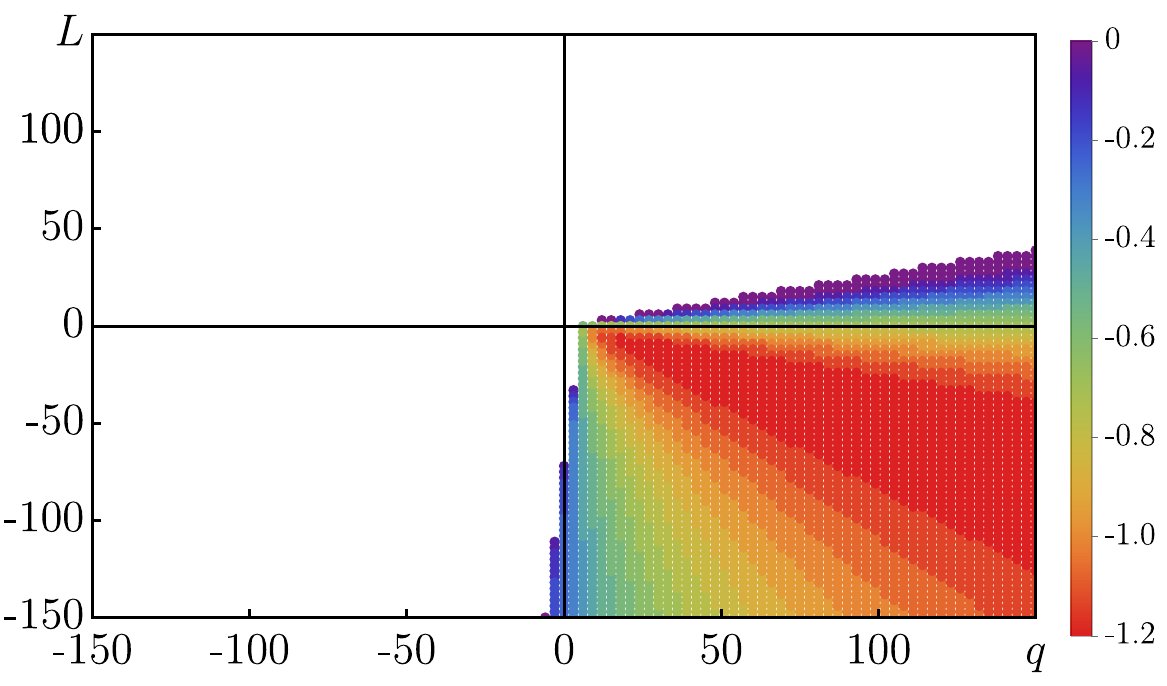} \\
        };

        \coordinate (O) at (M-1-1.north west);
        \coordinate (R) at (M-1-3.north east);
        \coordinate (B) at (M-3-1.south west);

        \draw[->, thick] (O) -- ($(R)$);
        \foreach \col/\xlab in {1/0, 2/0.9 Q_\mathrm{Max}, 3/Q_\mathrm{Max}} {\node[above=0cm, font=\fontsize{10pt}{10pt}\selectfont] at (M-1-\col.north) {\(\xlab\)};}
        \node[above=0cm, font=\fontsize{10pt}{10pt}\selectfont] at ($(R)$) {\(Q\)};

        \draw[->, thick] (O) -- ($(B)$);
        \foreach \row/\ylab in {1/0, 2/\mathrm{-}0.5, 3/\mathrm{-}1.0} {\node[left=0cm, font=\fontsize{10pt}{10pt}\selectfont] at (M-\row-1.west) {\(\ylab\)};}
        \node[left=0cm, font=\fontsize{10pt}{10pt}\selectfont] at ($(B)$) {\(\Lambda\)};
    \end{tikzpicture}
    \vspace{-1cm}
    \caption{Massive particle ($m = 1$) in the Kerr-Sen-AdS black hole ($a = 0.5$) background.}
    \label{KSAdSa0.5m1}
\end{figure}
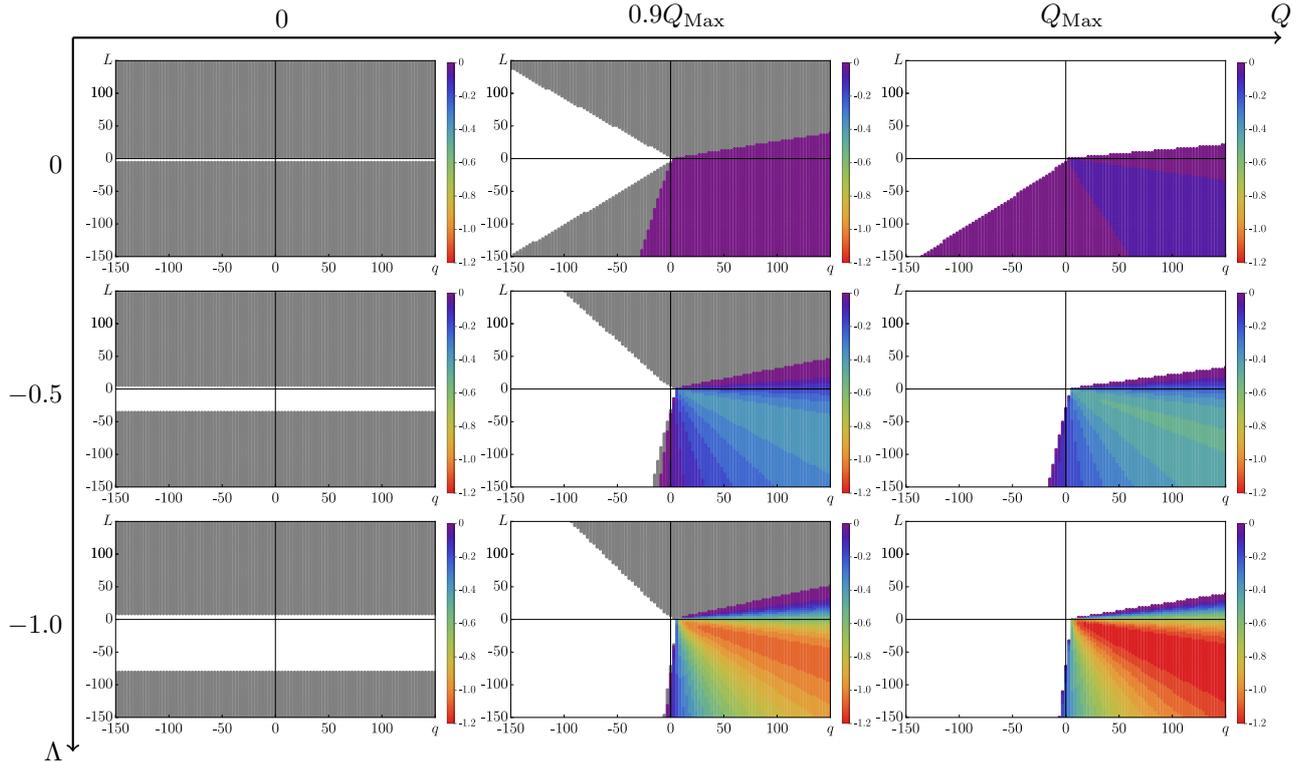
    Figs. \ref{KSAdSa0.1m0}, \ref{KSAdSa0.1m1}, \ref{KSAdSa0.5m0}, and \ref{KSAdSa0.5m1} present the violation of the bound on chaos based on analyses of the behavior of the Lyapunov exponent for particle motion in the Kerr--Sen--AdS black hole spacetime. The plots on the extreme left of each row represent an uncharged black hole, while those on the far left denote an extremally charged black hole. The middle plots in each row present black holes with charges set to $90\%$ of the extremal value. The plot in the upper left corner corresponds to a pure Schwarzschild black hole, the upper right corner corresponds to an extremal Kerr--Sen black hole, the lower left corner corresponds to a Schwarzschild--AdS black hole, and the lower right corner corresponds to an extremal Kerr--Sen--AdS black hole. The surface gravity of the extremal black hole vanishes; therefore, the plots in the rightmost column include only the violation of the bound on chaos, indicated by the colored region. The plots in the leftmost column, corresponding to the Kerr(--AdS) black hole, were previously studied \cite{Kan:2021blg, Gwak:2022xje}, and their consistency with our results reinforces the validity of our analysis.

    Figs. \ref{KSAdSa0.1m0} and \ref{KSAdSa0.1m1} correspond to a spin parameter of $a = 0.1$. The extremal black holes have cosmological constants  $\Lambda = 0$, $-0.5$, and $-1$ with maximum electric charges $Q_\mathrm{Max} = 1.342$, $1.322$, and $1.307$, respectively. Figs. \ref{KSAdSa0.5m0} and \ref{KSAdSa0.5m1} correspond to a spin parameter of $a = 0.5$. The extremal black holes have $\Lambda = 0$, $-0.5$, and $-1$ with maximum electric charges $Q_\mathrm{Max} = 1$, $0.908$, and $0.847$, respectively. In these figures, the violation and satisfaction of the bound on chaos predominantly occur in the region where the product of the particle charge $q$ and the black hole charge $Q$ is positive $(qQ > 0)$. Notably, the bound violation is more pronounced for negative angular momentum and when the spin parameter and angular momentum of the particle have opposite signs $(a L < 0)$, reflecting the enhanced instability in the Kerr--Sen--AdS spacetime.
    
    Moreover, in the middle column of Figs. \ref{KSAdSa0.1m0} and \ref{KSAdSa0.1m1}, which correspond to black holes with $90\%$ of the extremal charge, small gray regions satisfying the bound on chaos appear within the broader violation regions. This structure suggests that, even in parameter spaces dominated by bound violation, narrow windows exist where the bound on chaos remains intact. As the spin parameter $a$ increases, the gray regions gradually shrink. In Figs. \ref{KSAdSa0.5m0} and \ref{KSAdSa0.5m1}, the gray regions become increasingly suppressed and eventually vanish. This behavior suggests that a higher black hole spin amplifies orbital instability, reducing the likelihood of satisfying the bound on chaos. Consequently, at sufficiently high $a$, the violation regions become dominant, leaving no region where the bound holds.
    
\begin{figure}[H]
    \centering
    \includegraphics[width=8.3cm]{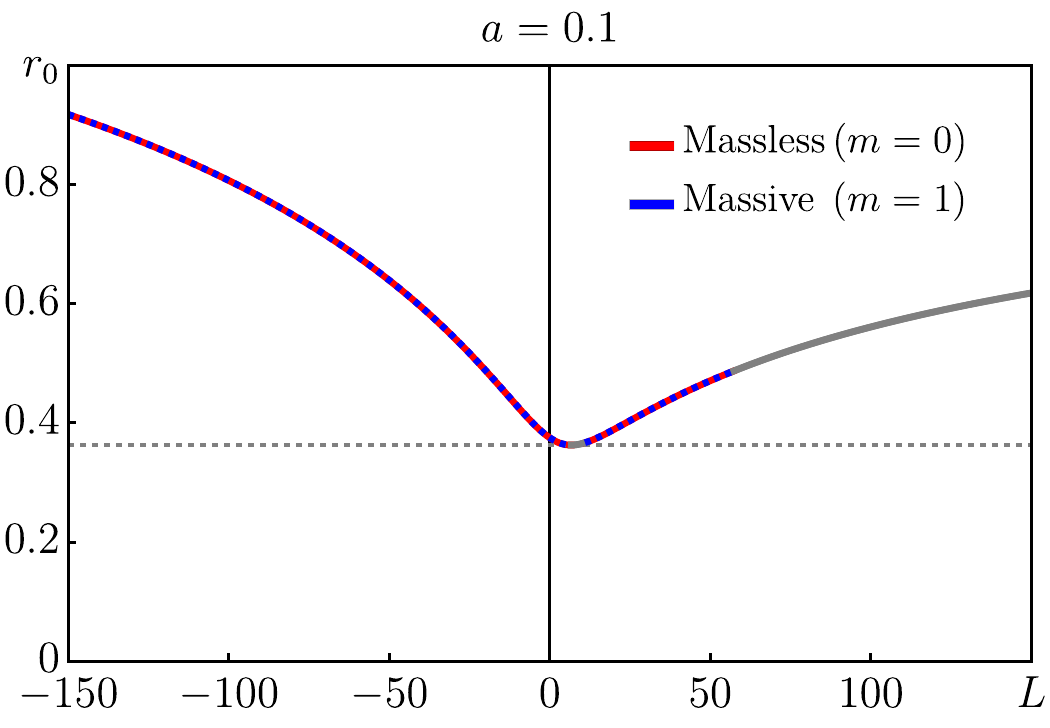}
    \includegraphics[width=8.3cm]{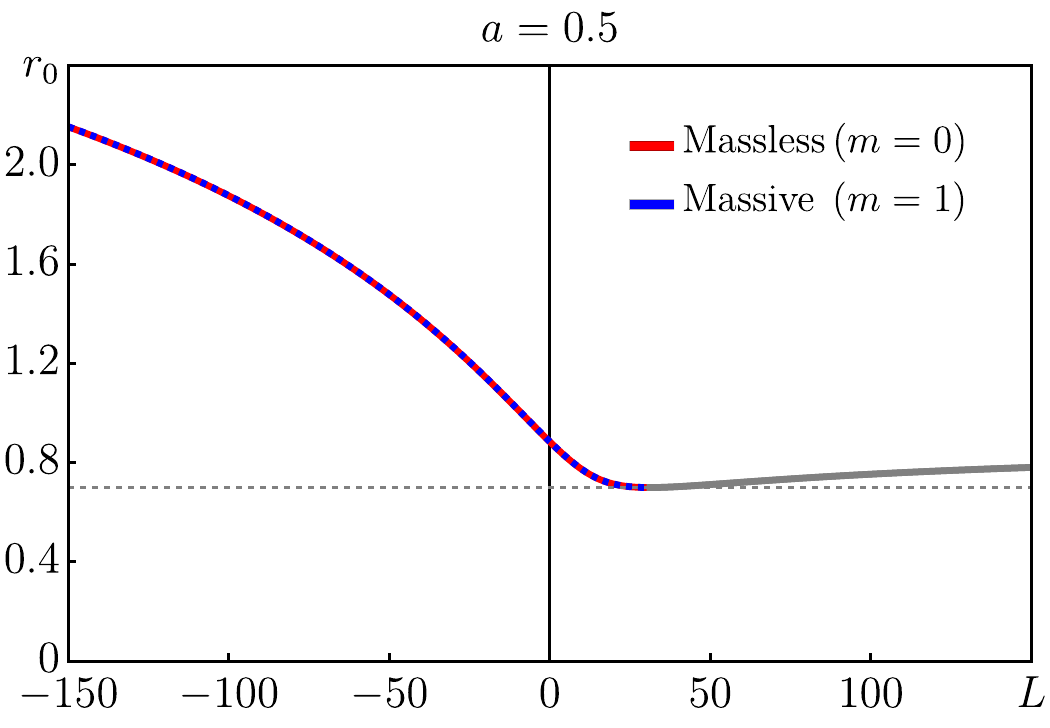}
    \caption{Location of the unstable radial equilibrium point $r_0$ as a function of the particle's angular momentum $L$ for the charged probe particle $(q = 100)$ in the Kerr-Sen-AdS black hole background.}
    \label{Lvsr0}
\end{figure}
    Fig. \ref{Lvsr0} depicts the dependence of the unstable radial equilibrium point $r_0$ on the angular momentum $L$ for massless and massive charged probe particles, with a fixed electric charge $q = 100$, indicated in red and blue, respectively. The dotted horizontal line denotes the location of the outer horizon, while the gray line indicates the regions where the bound on chaos is satisfied, consistent with the previous figures.

    The left plot corresponds to the middle plot of Figures \ref{KSAdSa0.5m0} and \ref{KSAdSa0.5m1}, representing Kerr--Sen--AdS black holes with a cosmological constant $(\Lambda = -0.5)$. As in those figures, a narrow gray region appears in the range $7 \lesssim L \lesssim 10$, indicating a brief interval where the bound on chaos is satisfied. This interval is bound on both sides by regions of violation, suggesting that the bound on chaos depends on angular momentum in a non-monotonic manner. Specifically, as $L$ increases from zero, the system transitions from violating the bound on chaos to satisfying it and then reenters a regime of violation as $L$ increases further. At larger values of $L$, specifically $L \approx 57$, the system enters a regime where the bound on chaos is satisfied.

   A similar pattern is observed in the right plot, where the gray region becomes visible for $L \gtrsim 32$. In both plots, this transition is closely associated with the proximity of $r_0$ to the outer horizon, particularly when $a L > 0$. These results indicate the high sensitivity of the bound on chaos to the interplay between the angular momentum of the particle and the near-horizon geometry. This underscores the nontrivial influence of angular momentum on the dynamical stability and chaotic behavior of probe particle orbits in Kerr--Sen--AdS spacetimes.

\section{Conclusion} \label{section_conclusion}
    In this study, we investigated the bound on the Lyapunov exponent for a charged probe particle moving in the background of GMGHS--AdS and Kerr--Sen--AdS black holes. We analyzed the dynamics near unstable orbits, evaluated the local Lyapunov exponent, and compared the exponent to the surface gravity of the black hole, which sets the upper bound proposed by Maldacena, Shenker, and Stanford \cite{Maldacena:2015waa}. This comparative analysis revealed several notable features of the chaotic behavior in string-inspired AdS spacetimes.

    We derived the effective Lagrangian for a charged particle confined to the equatorial plane of the Kerr--Sen--AdS geometry, incorporating contributions from the dilaton and axion fields characteristic to the heterotic string theory. The Lyapunov exponent was computed analytically near the potential extremum, and its squared exponent was compared to the squared surface gravity to assess whether the bound on chaos was satisfied or violated.

    Our analysis revealed that the bound on chaos can be violated under specific conditions, particularly in the near-horizon limit of extremal or near-extremal black holes. Regarding the GMGHS--AdS case, which corresponds to the non-rotating limit, the violation of the bound on chaos occurs generically near the horizon. It is amplified by a negative cosmological constant, which considerably enlarges the violation region in parameter space.

    In the rotating Kerr--Sen--AdS spacetime, the rotation amplifies the instability of particle orbits and broadens the parameter space where the bound on chaos is violated. The violation is more pronounced for massless and massive particles when the product of the particle and black hole charges is positive $(qQ > 0)$. Furthermore, the violation is prominent when the angular momentum of the particle is anti-aligned with the black hole spin $(aL < 0)$ because frame-dragging effects increase the effective instability.

    Our numerical analysis confirmed these results, highlighting the complex interplay among the cosmological constant, electric charge, and angular momentum in determining the emergence of chaotic behavior. Narrow parameter windows where the bound remains satisfied exist; however, these regions shrink rapidly with increasing spin and vanish in the high-spin regime. In the extremal limit of Kerr--Sen--AdS black holes, where the surface gravity vanishes, the Lyapunov bound is systematically violated, as expected.

    Thus, our analysis reveals that violating the bound on chaos is a robust feature in Kerr-Sen-AdS black holes, particularly in regimes characterized by strong coupling, rotation, and a negative cosmological constant. This suggests that string-inspired theories may considerably impact the universality of the bound on chaos. These recurring violations could reflect new dynamical features rooted in the stringy structure of spacetime or indicate limitations in the current understanding of holographic dualities.

    Future research may explore these issues further by incorporating higher-order stringy corrections, accounting for backreaction effects, or analyzing genuinely stringy probes such as D-branes and fundamental strings. These directions could help clarify the scope and limitations of the bound on chaos and provide deeper insights into the interplay between string theory, gravitational dynamics, and quantum mechanics.

\section*{Acknowledgement}
    This research was supported by Basic Science Research Program through the National Research Foundation of Korea (NRF) funded by the Ministry of Education (NRF-2022R1I1A2063176) and the Dongguk University Research Fund of 2025.

\bibliographystyle{unsrt}
\bibliography{Refs}

\begin{thebibliography}{10}

\bibitem{EventHorizonTelescope:2019dse}
Kazunori Akiyama et~al.
\newblock {First M87 Event Horizon Telescope Results. I. The Shadow of the Supermassive Black Hole}.
\newblock {\em Astrophys. J. Lett.}, 875:L1, 2019.

\bibitem{LIGOScientific:2016aoc}
B.~P. Abbott et~al.
\newblock {Observation of Gravitational Waves from a Binary Black Hole Merger}.
\newblock {\em Phys. Rev. Lett.}, 116(6):061102, 2016.

\bibitem{Hawking:1975vcx}
S.~W. Hawking.
\newblock {Particle Creation by Black Holes}.
\newblock {\em Commun. Math. Phys.}, 43:199--220, 1975.
\newblock [Erratum: Commun.Math.Phys. 46, 206 (1976)].

\bibitem{Bekenstein:1973ur}
Jacob~D. Bekenstein.
\newblock {Black holes and entropy}.
\newblock {\em Phys. Rev. D}, 7:2333--2346, 1973.

\bibitem{Maldacena:1997re}
Juan~Martin Maldacena.
\newblock {The Large $N$ limit of superconformal field theories and supergravity}.
\newblock {\em Adv. Theor. Math. Phys.}, 2:231--252, 1998.

\bibitem{Strominger:2001pn}
Andrew Strominger.
\newblock {The dS / CFT correspondence}.
\newblock {\em JHEP}, 10:034, 2001.

\bibitem{Strominger:2001gp}
Andrew Strominger.
\newblock {Inflation and the dS / CFT correspondence}.
\newblock {\em JHEP}, 11:049, 2001.

\bibitem{McInnes:2001zw}
Brett McInnes.
\newblock {The dS / CFT correspondence and the big smash}.
\newblock {\em JHEP}, 08:029, 2002.

\bibitem{Anninos:2011ui}
Dionysios Anninos, Thomas Hartman, and Andrew Strominger.
\newblock {Higher Spin Realization of the dS/CFT Correspondence}.
\newblock {\em Class. Quant. Grav.}, 34(1):015009, 2017.

\bibitem{Guica:2008mu}
Monica Guica, Thomas Hartman, Wei Song, and Andrew Strominger.
\newblock {The Kerr/CFT Correspondence}.
\newblock {\em Phys. Rev. D}, 80:124008, 2009.

\bibitem{Matsuo:2009sj}
Yoshinori Matsuo, Takuya Tsukioka, and Chul-Moon Yoo.
\newblock {Another Realization of Kerr/CFT Correspondence}.
\newblock {\em Nucl. Phys. B}, 825:231--241, 2010.

\bibitem{Castro:2010fd}
Alejandra Castro, Alexander Maloney, and Andrew Strominger.
\newblock {Hidden Conformal Symmetry of the Kerr Black Hole}.
\newblock {\em Phys. Rev. D}, 82:024008, 2010.

\bibitem{Compere:2012jk}
Geoffrey Comp\`ere.
\newblock {The Kerr/CFT Correspondence and its Extensions}.
\newblock {\em Living Rev. Rel.}, 15(1):11--81, 2012.

\bibitem{Garousi:2009zx}
Mohammad~R. Garousi and Ahmad Ghodsi.
\newblock {The RN/CFT Correspondence}.
\newblock {\em Phys. Lett. B}, 687:79--83, 2010.

\bibitem{Sakai:2004cn}
Tadakatsu Sakai and Shigeki Sugimoto.
\newblock {Low energy hadron physics in holographic QCD}.
\newblock {\em Prog. Theor. Phys.}, 113:843--882, 2005.

\bibitem{Erlich:2005qh}
Joshua Erlich, Emanuel Katz, Dam~T. Son, and Mikhail~A. Stephanov.
\newblock {QCD and a holographic model of hadrons}.
\newblock {\em Phys. Rev. Lett.}, 95:261602, 2005.

\bibitem{Karch:2006pv}
Andreas Karch, Emanuel Katz, Dam~T. Son, and Mikhail~A. Stephanov.
\newblock {Linear confinement and AdS/QCD}.
\newblock {\em Phys. Rev. D}, 74:015005, 2006.

\bibitem{Gubser:2008px}
Steven~S. Gubser.
\newblock {Breaking an Abelian gauge symmetry near a black hole horizon}.
\newblock {\em Phys. Rev. D}, 78:065034, 2008.

\bibitem{Hartnoll:2008vx}
Sean~A. Hartnoll, Christopher~P. Herzog, and Gary~T. Horowitz.
\newblock {Building a Holographic Superconductor}.
\newblock {\em Phys. Rev. Lett.}, 101:031601, 2008.

\bibitem{Hartnoll:2008kx}
Sean~A. Hartnoll, Christopher~P. Herzog, and Gary~T. Horowitz.
\newblock {Holographic Superconductors}.
\newblock {\em JHEP}, 12:015, 2008.

\bibitem{Ryu:2006bv}
Shinsei Ryu and Tadashi Takayanagi.
\newblock {Holographic derivation of entanglement entropy from AdS/CFT}.
\newblock {\em Phys. Rev. Lett.}, 96:181602, 2006.

\bibitem{Hubeny:2007xt}
Veronika~E. Hubeny, Mukund Rangamani, and Tadashi Takayanagi.
\newblock {A Covariant holographic entanglement entropy proposal}.
\newblock {\em JHEP}, 07:062, 2007.

\bibitem{Sekino:2008he}
Yasuhiro Sekino and Leonard Susskind.
\newblock {Fast Scramblers}.
\newblock {\em JHEP}, 10:065, 2008.

\bibitem{Maldacena:2016hyu}
Juan Maldacena and Douglas Stanford.
\newblock {Remarks on the Sachdev-Ye-Kitaev model}.
\newblock {\em Phys. Rev. D}, 94(10):106002, 2016.

\bibitem{Maldacena:2015waa}
Juan Maldacena, Stephen~H. Shenker, and Douglas Stanford.
\newblock {A bound on chaos}.
\newblock {\em JHEP}, 08:106, 2016.

\bibitem{Hashimoto:2016dfz}
Koji Hashimoto and Norihiro Tanahashi.
\newblock {Universality in Chaos of Particle Motion near Black Hole Horizon}.
\newblock {\em Phys. Rev. D}, 95(2):024007, 2017.

\bibitem{Hashimoto:2022kfv}
Koji Hashimoto and Kakeru Sugiura.
\newblock {Causality bounds chaos in geodesic motion}.
\newblock {\em Phys. Rev. D}, 107(6):066005, 2023.

\bibitem{Gao:2022ybw}
Chuanhong Gao, Deyou Chen, Chengye Yu, and Peng Wang.
\newblock {Chaos bound and its violation in charged Kiselev black hole}.
\newblock {\em Phys. Lett. B}, 833:137343, 2022.

\bibitem{Guo:2022kio}
Xiaobo Guo, Yuhang Lu, Benrong Mu, and Peng Wang.
\newblock {Probing phase structure of black holes with Lyapunov exponents}.
\newblock {\em JHEP}, 08:153, 2022.

\bibitem{Jeong:2023hom}
Soyeon Jeong, Bum-Hoon Lee, Hocheol Lee, and Wonwoo Lee.
\newblock {Homoclinic orbit and the violation of the chaos bound around a black hole with anisotropic matter fields}.
\newblock {\em Phys. Rev. D}, 107(10):104037, 2023.

\bibitem{Das:2024iuf}
Surajit Das, Surojit Dalui, and Rickmoy Samanta.
\newblock {Near-horizon chaos beyond Einstein gravity}.
\newblock {\em Phys. Rev. D}, 110(12):124037, 2024.

\bibitem{Gallo:2024wju}
Emanuel Gallo and Thomas M\"adler.
\newblock {Bounds for Lyapunov exponent of circular light orbits in black holes}.
\newblock {\em Eur. Phys. J. C}, 85(3):299, 2025.

\bibitem{Gogoi:2024akv}
Naba~Jyoti Gogoi, Saumen Acharjee, and Prabwal Phukon.
\newblock {Lyapunov exponents and phase transition of Hayward AdS black hole}.
\newblock {\em Eur. Phys. J. C}, 84(11):1144, 2024.

\bibitem{Lei:2024qpu}
Yu-Qi Lei, Xian-Hui Ge, and Surojit Dalui.
\newblock {Thermodynamic stability versus chaos bound violation in D-dimensional RN black holes: Angular momentum effects and phase transitions}.
\newblock {\em Phys. Lett. B}, 856:138929, 2024.

\bibitem{Shukla:2024tkw}
Bhaskar Shukla, Pranaya~Pratik Das, David Dudal, and Subhash Mahapatra.
\newblock {Interplay between the Lyapunov exponents and phase transitions of charged AdS black holes}.
\newblock {\em Phys. Rev. D}, 110(2):024068, 2024.

\bibitem{Kan:2021blg}
Naoto Kan and Bogeun Gwak.
\newblock {Bound on the Lyapunov exponent in Kerr-Newman black holes via a charged particle}.
\newblock {\em Phys. Rev. D}, 105(2):026006, 2022.

\bibitem{Yu:2022tlr}
Chengye Yu, Deyou Chen, and Chuanhong Gao.
\newblock {Bound on Lyapunov exponent in Einstein-Maxwell-Dilaton-Axion black holes}.
\newblock {\em Chin. Phys. C}, 46(12):125106, 2022.

\bibitem{Prihadi:2023tvr}
Hadyan~Luthfan Prihadi, Freddy~Permana Zen, Donny Dwiputra, and Seramika Ariwahjoedi.
\newblock {Chaos and fast scrambling delays of a dyonic Kerr-Sen-AdS4 black hole and its ultraspinning version}.
\newblock {\em Phys. Rev. D}, 107(12):124053, 2023.

\bibitem{Prihadi:2023qmk}
Hadyan~Luthfan Prihadi, Freddy~Permana Zen, Donny Dwiputra, and Seramika Ariwahjoedi.
\newblock {Localized chaos due to rotating shock waves in Kerr\textendash{}AdS black holes and their ultraspinning version}.
\newblock {\em Gen. Rel. Grav.}, 56(8):90, 2024.

\bibitem{Giataganas:2024hil}
D.~Giataganas, A.~Kehagias, and A.~Riotto.
\newblock {Quasinormal modes and universality of the Penrose limit of black hole photon rings}.
\newblock {\em JHEP}, 09:168, 2024.

\bibitem{Yin:2022mjv}
Rui Yin, Jing Liang, and Benrong Mu.
\newblock {Chaos bound and its violation in the torus-like black hole}.
\newblock 10 2022.

\bibitem{Chen:2022tbb}
Deyou Chen and Chuanhong Gao.
\newblock {Angular momentum and chaos bound of charged particles around Einstein\textendash{}Euler\textendash{}Heisenberg AdS black holes}.
\newblock {\em New J. Phys.}, 24(12):123014, 2022.

\bibitem{Xie:2023tjc}
Jiayu Xie, Jie Wang, and Bing Tang.
\newblock {Circular motion and chaos bound of a charged particle near charged 4D Einstein\textendash{}Gauss\textendash{}Bonnet-AdS black holes}.
\newblock {\em Phys. Dark Univ.}, 42:101271, 2023.

\bibitem{Yu:2023spr}
Chengye Yu, Deyou Chen, Benrong Mu, and Yucheng He.
\newblock {Violating the chaos bound in five-dimensional, charged, rotating Einstein-Maxwell-Chern-Simons black holes}.
\newblock {\em Nucl. Phys. B}, 987:116093, 2023.

\bibitem{Kumara:2024obd}
A.~Naveena Kumara, Shreyas Punacha, and Md~Sabir Ali.
\newblock {Lyapunov exponents and phase structure of Lifshitz and hyperscaling violating black holes}.
\newblock {\em JCAP}, 07:061, 2024.

\bibitem{Chen:2023wph}
Deyou Chen and Chuanhong Gao.
\newblock {Chaos bound in Kerr-Newman-Taub-NUT black holes via circular motions}.
\newblock {\em Chin. Phys. C}, 47(1):015108, 2023.

\bibitem{Awal:2025irl}
Mozib~Bin Awal and Prabwal Phukon.
\newblock {Probing Thermodynamic Phase Transitions of 4D R-Charged Black Holes via Lyapunov Exponent}.
\newblock 5 2025.

\bibitem{Singh:2024qfw}
Balbeer Singh, Nibedita Padhi, and Rashmi~R. Nayak.
\newblock {Circular orbits and chaos bound in slow-rotating curved acoustic black holes}.
\newblock 5 2024.

\bibitem{Gwak:2022xje}
Bogeun Gwak, Naoto Kan, Bum-Hoon Lee, and Hocheol Lee.
\newblock {Violation of bound on chaos for charged probe in Kerr-Newman-AdS black hole}.
\newblock {\em JHEP}, 09:026, 2022.

\bibitem{Song:2022lhf}
Yue Song, Rui Yin, Yiqian He, and Benrong Mu.
\newblock {Chaos bound of charged particles around phantom AdS black hole}.
\newblock 11 2022.

\bibitem{Park:2023lfc}
Junsu Park and Bogeun Gwak.
\newblock {Bound on Lyapunov exponent in Kerr-Newman-de Sitter black holes by a charged particle}.
\newblock {\em JHEP}, 04:023, 2024.

\bibitem{R:2025gok}
Karthik R., Dillirajan D., K.~M. Ajith, Kartheek Hegde, Shreyas Punacha, and A.~Naveena Kumara.
\newblock {Euclidean Thermodynamics and Lyapunov Exponents of Einstein-Power-Yang-Mills AdS Black Holes}.
\newblock 4 2025.

\bibitem{Han:2023ckr}
Hyewon Han and Bogeun Gwak.
\newblock {Effects of fluctuations in higher-dimensional AdS black holes}.
\newblock {\em Phys. Rev. D}, 110(6):066013, 2024.

\bibitem{Lei:2023jqv}
Yu-Qi Lei and Xian-Hui Ge.
\newblock {Stationary equilibrium of test particles near charged black branes with the hyperscaling violating factor}.
\newblock {\em Phys. Rev. D}, 107(10):106002, 2023.

\bibitem{Dutta:2023yhx}
Pinaki Dutta, Kamal~L. Panigrahi, and Balbeer Singh.
\newblock {Circular string in a black p-brane leading to chaos}.
\newblock {\em JHEP}, 10:189, 2023.

\bibitem{Dutta:2024rta}
Pinaki Dutta, Kamal~L. Panigrahi, and Balbeer Singh.
\newblock {Chaos bound and its violation in black p-brane}.
\newblock {\em JHEP}, 02:043, 2025.

\bibitem{Gibbons:1987ps}
G.~W. Gibbons and Kei-ichi Maeda.
\newblock {Black Holes and Membranes in Higher Dimensional Theories with Dilaton Fields}.
\newblock {\em Nucl. Phys. B}, 298:741--775, 1988.

\bibitem{Garfinkle:1990qj}
David Garfinkle, Gary~T. Horowitz, and Andrew Strominger.
\newblock {Charged black holes in string theory}.
\newblock {\em Phys. Rev. D}, 43:3140, 1991.
\newblock [Erratum: Phys.Rev.D 45, 3888 (1992)].

\bibitem{Hassan:1991mq}
S.~F. Hassan and Ashoke Sen.
\newblock {Twisting classical solutions in heterotic string theory}.
\newblock {\em Nucl. Phys. B}, 375:103--118, 1992.

\bibitem{Sen:1992ua}
Ashoke Sen.
\newblock {Rotating charged black hole solution in heterotic string theory}.
\newblock {\em Phys. Rev. Lett.}, 69:1006--1009, 1992.

\bibitem{Chong:2004na}
Z.~W. Chong, Mirjam Cvetic, H.~Lu, and C.~N. Pope.
\newblock {Charged rotating black holes in four-dimensional gauged and ungauged supergravities}.
\newblock {\em Nucl. Phys. B}, 717:246--271, 2005.

\bibitem{Chow:2013gba}
David D.~K. Chow and Geoffrey Comp\`ere.
\newblock {Dyonic AdS black holes in maximal gauged supergravity}.
\newblock {\em Phys. Rev. D}, 89(6):065003, 2014.

\bibitem{Wu:2020cgf}
Di~Wu, Puxun Wu, Hongwei Yu, and Shuang-Qing Wu.
\newblock {Are ultraspinning Kerr-Sen- AdS$_4$ black holes always superentropic?}
\newblock {\em Phys. Rev. D}, 102(4):044007, 2020.

\bibitem{Ali:2023ppg}
Md~Sabir Ali, Sushant~G. Ghosh, and Anzhong Wang.
\newblock {Thermodynamics of Kerr-Sen-AdS black holes in the restricted phase space}.
\newblock {\em Phys. Rev. D}, 108(4):044045, 2023.

\bibitem{Hawking:1998kw}
S.~W. Hawking, C.~J. Hunter, and Marika Taylor.
\newblock {Rotation and the AdS / CFT correspondence}.
\newblock {\em Phys. Rev. D}, 59:064005, 1999.

\bibitem{Gwak:2018akg}
Bogeun Gwak.
\newblock {Weak Cosmic Censorship Conjecture in Kerr-(Anti-)de Sitter Black Hole with Scalar Field}.
\newblock {\em JHEP}, 09:081, 2018.

\bibitem{Gwak:2021tcl}
Bogeun Gwak.
\newblock {Weak cosmic censorship conjecture in Kerr-Newman-(anti-)de Sitter black hole with charged scalar field}.
\newblock {\em JCAP}, 10:012, 2021.

\bibitem{Cardoso:2008bp}
Vitor Cardoso, Alex~S. Miranda, Emanuele Berti, Helvi Witek, and Vilson~T. Zanchin.
\newblock {Geodesic stability, Lyapunov exponents and quasinormal modes}.
\newblock {\em Phys. Rev. D}, 79(6):064016, 2009.

\bibitem{Pradhan:2012rkk}
Parthapratim Pradhan.
\newblock {Stability analysis and quasinormal modes of Reissner\textendash{}Nordstr\o{}m space-time via Lyapunov exponent}.
\newblock {\em Pramana}, 87(1):5, 2016.

\bibitem{Pradhan:2013bli}
Partha~Pratim Pradhan.
\newblock {Lyapunov Exponent and Charged Myers Perry Spacetimes}.
\newblock {\em Eur. Phys. J. C}, 73(6):2477, 2013.

\bibitem{Pradhan:2014tva}
Parthapratim Pradhan.
\newblock {Circular Geodesics in Tidal Charged Black Hole}.
\newblock {\em Int. J. Geom. Meth. Mod. Phys.}, 15(01):1850011, 2017.

\bibitem{Lei:2020clg}
Yu-Qi Lei, Xian-Hui Ge, and Cheng Ran.
\newblock {Chaos of particle motion near a black hole with quasitopological electromagnetism}.
\newblock {\em Phys. Rev. D}, 104(4):046020, 2021.

\bibitem{Lei:2021koj}
Yu-Qi Lei and Xian-Hui Ge.
\newblock {Circular motion of charged particles near a charged black hole}.
\newblock {\em Phys. Rev. D}, 105(8):084011, 2022.

\bibitem{Chen:2025xqc}
Deyou Chen, Chuang Yang, and Yongtao Liu.
\newblock {Lyapunov exponents as probes for a phase transition of a Kerr-AdS black hole}.
\newblock {\em Phys. Lett. B}, 865:139463, 2025.

\bibitem{Ali:2025znb}
Riasat Ali, Terkaa~Victor Targema, Xia Tiecheng, and Rimsha Babar.
\newblock {Evaluation of circular geodesics and thermodynamics of a Bumblebee gravity-like black hole}.
\newblock {\em High Energy Dens. Phys.}, 55:101189, 2025.

\end{thebibliography}

\end{document}